\pdfoutput=1
\documentclass[useAMS,usenatbib,fleqn]{mn2e}
\usepackage{savesym}
\usepackage{amsmath}
\savesymbol{iint}
\usepackage{txfonts}
\restoresymbol{TXF}{iint}
\usepackage{graphicx}
\usepackage{grffile}
\usepackage{lineno}
\usepackage{subfigure}
\usepackage{color}
\usepackage{listings}
\usepackage{deluxetable}
\usepackage[T1]{fontenc}
\usepackage{ae,aecompl}
\def\vFv{\nu F_{\nu}}
\def\Ep{E_{\rm p}}

\newcommand{\vect}[1]{\mbox{\boldmath ${#1}$}}
\newcommand{\prob}{{\rm Pr}}
\newcommand{\diff}{{\rm d}}
\newcommand{\ereg}{\varepsilon}
\newcommand{\erf}{{\rm erf}}

\title[BALROG]{Awakening the BALROG (BAyesian Location Reconstruction Of GRBs): A new paradigm in spectral and location analysis of gamma ray bursts}
\author[J. Michael Burgess et al.]
{J.\ Michael Burgess$^{1,2}$\thanks{E-mail: jamesb@kth.se (JMB)},
Hoi-Fung Yu$^{3}$,
Jochen Greiner$^{3}$
and
Daniel J.\ Mortlock$^{4,5,6}$
\\  
$^{1}$The Oskar Klein Centre for Cosmoparticle Physics,
SE-106 91 Stockholm, Sweden\\
$^{2}$Department of Physics, KTH Royal Institute of Technology,
AlbaNova, SE-106 91 Stockholm, Sweden\\
$^{3}$Max Planck Institute for Extraterrestrial Physics, 
  Giessenbachstrasse, D-85748 Garching, Germany\\
$^{4}$Statistics Section, Department of Mathematics,
  Imperial College London, London SW7 2AZ, UK \\
$^{5}$Astrophysics Group, Imperial College London, Blackett Laboratory,
  Prince Consort Road, London SW7 2AZ, UK \\
$^{6}$Department of Astronomy,
  Stockholm University, Albanova, SE-10691 Stockholm, Sweden
}

\date{Accepted XXXX December XX. Received XXXX December XX; in original form 2016 October XX}

\begin{document}
\label{firstpage}

\pagerange{\pageref{firstpage}--\pageref{lastpage}} \pubyear{2016}

\maketitle

\begin{abstract}
The accurate spatial location of gamma-ray bursts (GRBs) is crucial for both producing a detector response matrix (DRM) and follow-up observations by other instruments. The Fermi Gamma-ray Burst Monitor (GBM) has the largest field of view (FOV) for detecting GRBs as it views the entire unocculted sky, but as a non-imaging instrument it relies on the relative count rates observed in each of its 14 detectors to localize transients. Improving its ability to accurately locate GRBs and other transients is vital to the paradigm of multi-messenger astronomy, including the electromagnetic follow-up of gravitational wave signals. Here we present the BAyesian Location Reconstruction Of GRBs ({\tt BALROG}) method for localizing and characterizing GBM transients.   Our approach eliminates the systematics of previous approaches by simultaneously fitting for the location and spectrum of a source.  It also correctly incorporates the uncertainties in the location of a transient into the spectral parameters and produces reliable positional uncertainties for both well-localized sources and those for which the GBM data cannot effectively constrain the position. While computationally expensive, {\tt BALROG} can be implemented to enable quick follow-up of all GBM transient signals. Also, we identify possible response problems that require attention as well as caution when using standard, public GBM DRMs. Finally, we examine the effects of including the variance in location on the spectral parameters of GRB 080916C. We find that spectral parameters change and no extra components are required when these effects are included in contrast to when we use a fixed location. This finding has the potential to alter both the GRB spectral catalogs as well as the reported spectral composition of some well-known GRBs.
\end{abstract}


\section{Introduction}

The localization of gamma-ray bursts (GRBs) on the sky has played an important role in the history of understanding these extreme objects. The {\em Burst and Transient Source Experiment} ({\em BATSE}) helped to confirm the cosmological origin of GRBs by determining the isotropic sky distribution of over 3000 GRBs \citep{Briggs:1996}. The {\em Fermi} Gamma-ray Burst Monitor (GBM), which can monitor the entire sky (with the exception of the portion occulted by the Earth) is currently providing locations for hundreds of GRBs and other transient gamma-ray events every year. With the advent of gravitational wave astronomy and the potential for electromagnetic follow-up of associated events, GBM could play a crucial role in the new era of multi-messenger astronomy. 

Using differential rates in each of its 14 non-imaging detectors, GBM is able to provide approximate localization of transient events that can then be used by imaging observatories to search for counterparts at other wavelengths. Unfortunately, the standard GBM localization procedure, the Daughter Of Location (DOL) produces offsets of 8-13$^{\circ}$ from the subsequently confirmed true locations for many GRBs localized by GBM \citep{Connaughton:2015}. The exact cause of this systematic is unknown and could be attributed to several factors such as imperfections in the simulated spacecraft response and/or model.

Here we hypothesize that the systematics in the standard GBM localization algorithm are a byproduct of the assumption that all GRB spectral energy distributions (SEDs) are alike, and of inaccuracies in the DRM modeling. With the DOL, a set three of so-called template spectra are assumed in order to calculate a spatial grid of putative observed counts on the sky. This grid is then compared with the real observed counts via a $\chi^2$-minimization on source position to produce a best fit location. The template spectra are built from the ubiquitous \cite{Band:1993} function with three sets of fixed spectral parameters that allow for only the amplitude to be varied. It GRBs exhibit a variety of spectral shapes as well as temporal spectral evolution (e.g., \citet{Burgess:2014} ) so the assumption of a fixed set of spectral shapes could influence inferred location of a GRB if its true SED differs from the templates.

To explore this possibility we have developed a method for the location and spectrum are determined simultaneously via Bayesian sampling with a new tool: the BAyesian Location Reconstruction Of GRBs ({\tt BALROG}\footnote{Our {\tt BALROG} tool is completely unrelated, except in name, to the BaLROG project to examine the dynamical influence of bars in galaxies described by \citet{seidel_etal_2015}.}). First, we detail our methodology in Section~\ref{sec:method} then demonstrate the viability of the method via simulations in Section~\ref{sec:sims} and real data in Section~\ref{sec:real}. Finally, we investigate the effect on spectra introduced by including the variance of location in Section~\ref{sec:spec} where we use GRB 080916C as a case study.

\section{Inference methodology}
\label{sec:method}

We analyze GBM data by using Bayesian inference to obtain the posterior distribution of the angular position and emission spectrum of a GRB.  This requires a model of the GBM instrument (Section~\ref{sec:instrument}), a GRB source (Section~\ref{sec:source}), and the sky background (Section~\ref{sec:background}), which can then be combined to obtain the full likelihood for the data-set (Section~\ref{sec:likelihood}).  Combining the likelihood with the parameter priors (Section~\ref{sec:priors}), the implied constraints are then obtained by generating samples from the posterior distibution (Section~\ref{sec:sampling}).  This allows for easy marginalization over nuisance parameters, most obviously integrating out the parameters describing the GRB's spectrum to focus on localization.

\subsection{GBM instrumental response}
\label{sec:instrument}

Any method of localizing and characterizing sources using GRM must account for the complicated nature of its detectors, which have a broad angular response and suffer from energy dispersion (i.e., there is a significant probability that only a fraction of a detected photon's energy is registered).  These effects are accounted for using forward-folding, described in detail by \citet{Vianello:2015}, which provides a probabilistic connection between the actual properties of incident photons and the data recorded by the GBM detectors.  

A critical simplifying approximation that helps make this problem tractable is that the orientation of the $N_{\rm d} = 14$ GBM detectors is both perfectly known and constant for the duration of the burst being analyzed.  This means that the rotation angles (represented as quaternions) which define each detector's orientation can be left implicit, although the formalism presented below could be extended to include the changing orientation of the detectors with the satellite's motion.

During the on-source period each detector is exposed to a radiation field which here is described by the number flux of photons (i.e., the rate per unit area, per unit energy, per steradian) at the detector, $n(E, \psi)$, which depends on both photon energy, $E$, and angular position, $\psi$, but is assumed to be time-independent.  The rate at which photons are registered by detector $d$ (with $d \in \{1, 2, \ldots, N_{\rm d}\}$) is $n(E, \psi) \, A_d(E, \psi)$, where $A_d(E, \psi)$ is the effective area of the detector.  The angular dependence of the effective area can largely be understood in terms of the geometrical area presented to a putative source, and so it decreases approximately with the cosine of the angle between $\psi$ and nominal detector orientation, although the output of a full numerical simulation is used in practice.

Each photon that is registered then deposits an energy $\ereg \la E$, a process which can be fully described by the conditional distribution $\prob(\ereg | E, \psi, d, {\rm detected})$.  The most important aspect of this energy dispersion is that there is a significant probability that $\ereg$ is considerably lower than $E$, due to photons escaping before depositing their entire energy in the scintillation crystal as the result of Compton scattering. This aspect of the detector response has an angular dependence, inextricably linking the probability that a given photon is detected with the energy energy recorded from it.

In the GBM analysis formalism photon detection and energy registration are combined into a single function,
\begin{equation}
R_d(\ereg, E, \psi)
 = A_d(E, \psi) \, 
  \prob(\ereg | E, \psi, d, {\rm detected}),
\end{equation}
which is known as the DRM and fully characterizes each detector.

The measured energy of an individual photon is not recorded per se, but rather each photon is recorded as being in one of $N_{\rm c} = 128$ channels that have a nominal correspondence with energy.  The rate at which photons are recorded by channel $c$ (with $c \in \{1, d, \ldots, N_{\rm c}$\}) is
\begin{equation}
\label{eq:sourcerate}
S_{d,c} =
  \int_{E_{{\rm min}, c}}^{E_{{\rm max}, c}} \diff \ereg
  \int_{0}^\infty \diff E
  \int \diff \Omega
  \, n(E, \psi) \, R_d(\ereg, E, \psi)
\end{equation}
where $E_{{\rm min}, c}$ and $E_{{\rm max}, c}$ are the minimum and maximum energies for this channel and the inner integral extends over all photon arrival directions.

\subsubsection{DRM generation}
\label{sec:rspgen}

GBM DRMs are built via the General Response Simulation System (GRESS) GEANT4 Monte Carlo code \citep{Kippen:2007gx} and separated into two components consisting of a direct DRM and an atmospheric scattering DRM. A grid of direct DRMs created for 272 points on the sky for each of the 12 GBM sodium-iodide (NaI) detectors and 2 bismuth-germinate (BGO) detectors is used as the base response library for DRM generation. These responses exist in an energy compressed format which must be decompressed and added to the atmospheric scattering response via a series of transformations that depend on GBM's position and orientation with respect to the Earth \citep{Pendelton:1988}.  The directional dependence of the response adds in an extra complication, as it is too computationally expensive to calculate $R_d(\ereg, E, \psi)$ for all values of $\psi$ over the sphere, so this operation is undertaken only for the source positions proposed during the posterior sampling process (Section~\ref{sec:sampling}).  Therefore, we designed efficient DRM generation software based of the original GBM tool that allows for vectorized and parallel generation of DRMs. The final direct DRM is added to the atmospheric scattering DRM to account for scattering of photons off the Earth's atmosphere \citep{Meegan:2009}.

{\tt BALROG} has the ability to correctly reject source positions that are occulted by the Earth even though these positions are accounted for in the base DRM database. The area of the sky that is occulted is calculated in real time using the current altitude of the Fermi satellite and finding the horizon line from the spacecraft zenith.  But it is more computationally efficient to do this before generating the full detector response, so the DRM is immediately set to $R_d(\ereg, E, \psi) = 0$ for occulted directions.

\subsection{Source model}
\label{sec:source}

For an astronomically-distant point-source (as GRBs are assumed to be) all photons arrive at the detector from the direction of the source, $\psi_{\rm s}$.  This allows the number flux defined in Section~\ref{sec:instrument} to be factorized as
\begin{equation}
n(E, \psi) = f(E, \vect{\phi}_{\rm s}) \, \delta_{\rm D}(\psi , \psi_{\rm s}),
\end{equation}
where $\vect{\phi}_{\rm s}$ is the vector of parameters describing the source's spectral energy distribution (SED) and $\delta_{\rm D}(\psi , \psi_{\rm s})$ is the delta function distribution on the sphere centered on $\psi_{\rm s}$.  The presence of the delta function means the predicted count rate from the source given in Eq.~\ref{eq:sourcerate} simplifies to be
\begin{equation}
\label{eq:sourceratesimple}
S_{d,c} (\psi_{\rm s}, \vect{\phi}_{\rm s}) = 
  \int_{0}^\infty \diff E \, f(E, \vect{\phi}_{\rm s})
  \int_{E_{{\rm min}, c}}^{E_{{\rm max}, c}} \diff \ereg
  \, R_d(\ereg, E, \psi_{\rm s}),
\end{equation}
where the dependence on the source parameters has been made explicit.  The order of integration has been swapped to emphasize the possibility of pre-calculating the inner integral independent of the source emission parameters.

The SED models that are used here are: Band; cut-off power law (CPL); and smoothly-broken power law (SBPL).  The SED parameters $\vect{\phi}_{\rm s}$ and the source position $\psi_{\rm s}$ represent the full set of GRB parameters to be constrained using the GBM data.

\subsection{Background estimation}
\label{sec:background}

GBM data include an isotropic sky background and an anisotropic Earth-albedo background, both of which are time- and energy-dependent.  Accounting for the combined background rate from these two effects is a necessary step to inferring the properties of the GRBs detected by the instrument.  The ideal approach to this problem would be to simultaneously model the background and source (cf \citealt{loredo:1992}), although the fact that the background counts are generally fairly high means that it is a good approximation to estimate the background rate separately.  The background is taken to be a polynomial in time, and the coefficients are fit to the off-source light curve in each energy channel, with a Poisson likelihood used to ensure valid results even for high-energy channels with low counts. The result of this process is a full joint posterior distribution of the polynomial coefficients.  

This posterior is then propagated to the on-source period, the result of which is a data-driven estimate of the background rate, $\hat{B}_{d,c}$, and its uncertainty, $\sigma_{d,c}$, in channel $c$ of detector $d$.  Only the former would be used if a simple background-subtraction were sufficient; but, as has been shown by \cite{Greiner:2016}, failure to include the background uncertainty can lead to qualitatively incorrect inferences about the source of interest.  

Here the implied posterior distribution for the background is taken to be a truncated normal of the form 
\begin{eqnarray}
\label{eq:truncated}
\prob(B_{d,c} | \hat{B}_{d,c}, \sigma_{d,c})
  & = & \frac{\Theta(B_{d,c}) }
    {\left[1 + \erf(\hat{B}_{d,c} / 2^{1/2} \, \sigma_{d,c}) \right] / 2}
\\
  & \times &
    \frac{1}{(2 \pi)^{1/2} \, \sigma_{d,c}}
     e^{-\left(B_{d,c} - \hat{B}_{d,c}\right)^2 
     / \left(2 \, \sigma^2_{d,c} \right)}
    ,
  \nonumber
\end{eqnarray}
where $\erf(.)$ is the standard error function and the Heaviside step function $\Theta(.)$ imposes the absolute prior constraint that the background rate cannot be negative.

\subsection{Likelihood}
\label{sec:likelihood}

The likelihood, $\prob(\{N_{d,c}\} | \psi_{\rm s}, \vect{\phi}_{\rm s})$, is the probability, conditional on a given source position $\psi_{\rm s}$ and emission model $\vect{\phi}_{\rm s}$, of obtaining the measured counts, $\{N_{d,c}\}$, during the period (of duration $T$) that the GBM was observing the GRB.  As such, it encodes all the relevant information that can be used to localize and characterize the source.  Assuming that each incident photon is only registered once (i.e., in a particular channel of a particular detector), the likelihood has the form
\begin{equation}
\prob(\{N_{d,c}\} | \psi_{\rm s}, \vect{\phi}_{\rm s})
  = \prod_{d = 1}^{N_{\rm d}}
  \prod_{c = 1}^{N_{\rm c}}
   \prob\left[N_{d,c} | S_{d,c} (\psi_{\rm s}, \vect{\phi}_{\rm s})\right],
\end{equation}
where $S_{d,c} (\psi_{\rm s}, \vect{\phi}_{\rm s})$ is the predicted rate at which source photons are registered in channel $c$ of detector $d$ is given in Eq.~\ref{eq:sourceratesimple}.

\subsubsection{Single channel contribution}

As the contribution from each channel and detector combination has the same mathematical form, it is simplest to consider this without reference to detector number $d$ or channel index $c$, which are hence left implicit in what follows.

If the background rate, $B$, were known precisely then the probability of $N$ counts being registered during the on-source period would be given by a simple Poisson distribution as 
\begin{equation}
\label{eq:poisson_1}
\prob(N | S, B)
  = \frac{[(S + B)\,T]^{N}
   \, e^{- (S + B)\,T}}{N!},
\end{equation}
where $S$ is the rate of photons being registered from the source.  This depends on $\psi_{\rm s}$ and $\vect{\phi}_{\rm s}$, but it can be considered as the one model parameter of interest here.

The fact that the background is not known perfectly must be incorporated, which formally should be done by including $B$ as an extra parameter to be subsequently marginalized out.  Fortunately, however, almost all the available information about $B$ comes from the polynomial fitting to the off-source intervals described in Section~\ref{sec:background}, and not the single on-source measurement.  It is hence a very good approximation to simply average the Poisson distribution given in Eq.~\ref{eq:poisson_1} over the plausible values of the background rate.  Given the background estimate $\hat{B}$ and associated uncertainty $\sigma$, and adopting the truncated normal defined in Eq.~\ref{eq:truncated}, the contribution to the likelihood from a single channel is modified to be 
\begin{eqnarray}
\prob(N | S, \hat{B}, \sigma)
  & = &
  \int_0^\infty \diff B \,
    \prob(B | \hat{B}, \sigma)
    \,  
   \prob(N | S, B) 
\nonumber \\
  & \propto &
  \int_0^\infty \diff B \,
  (S + B)^{N} 
  e^{- [(B + S)\,T + (B - \hat{B})^2 / (2 \, \sigma^2)]},
\nonumber \\
\end{eqnarray}
where only the terms with that have a contribution from the source parameters have been retained in the second expression.  

While the above integral could be evaluated numerically, the fact that it will have to be evaluated numerous times during the posterior sampling process motivates using a closed form approximation instead.  The option taken here is to evaluate the profile likelihood, in which, for a given value of $S$, the integral is taken to be the value of the integrand evaluated at its peak (i.e., for the value of the background rate, $\tilde{B}$, that maximizes the integrand).  This is the approach implemented by \cite{Arnaud:2015} to obtain the pgstat statistic that is an option in {\tt XSPEC}.  This is equivalent\footnote{The actual expressions given in the pgstat section of \cite{Arnaud:2015} are $-2$ times the logarithm of the likelihoods given here.  There are also some terms with no $S$ dependence that have been omitted here.} to assuming that
\begin{equation}
\prob(N | S, \hat{B}, \sigma)
  \propto
  \left\{
  \begin{array}{@{}ll@{}}
    e^{-[ (S + \hat{B})\, T - \sigma^2 \, T^2 / 2 ]},
    & \!\!\!\! \text{if}\ N = 0, \\
  \\
    (S + \tilde{B})^{N} \,
    e^{- [(S + \tilde{B}) \, T + (\tilde{B} - \hat{B})^2 / (2 \, \sigma^2)]},
    & \!\!\!\! \text{if}\ N > 0,
  \end{array}\right.
\end{equation}
where in the $N > 0$ case
\begin{equation}
\tilde{B}
   = \frac{1}{2} \! \left\{
   \hat{B}  \!-\!S  \!-\!\sigma^2 T 
  + \! \left[
  (S \!+\! \sigma^2 T \!-\! \hat{B})^2 
  - 4
  (\sigma^2 T S \! - \! \sigma^2 N \! - \! \hat{B} S)
  \right]^{1/2}
  \right\} \!
  .
\end{equation}

Independent of whether numerical integration or the profile likelihood is used to evaluate $\prob(N | S, \hat{B}, \sigma)$, there is an increase in the range of values of $S$ for which the likelihood is appreciable.  This can lead to the apparently undesirable result that the parameter constraints are similarly broadened, but in fact this is simply correctly reflecting the range of source positions that are consistent with the data and hence which should be considered as potential GRB locations.

\subsection{Parameter priors}
\label{sec:priors}

A necessary component of Bayesian parameter estimation is the specification of priors for all parameters.  The joint prior in $\psi_{\rm s}$ and $\vect{\phi}_{\rm s}$ can reasonably be assumed to factorize (i.e., the plausible SED parameters are independent of position on the sky), so that $\prob(\psi_{\rm s}, \vect{\phi}_{\rm s} | {\rm GRB}) = \prob(\psi_{\rm s} | {\rm GRB}) \, \prob(\vect{\phi}_{\rm s} | {\rm GRB})$.   

For $\psi_{\rm s}$ we assume a prior that is uniform over the celestial sphere, so that $\prob(\psi_{\rm s} | {\rm GRB}) = 1 / (4 \pi)$, although there is no operational need to normalize this distribution correctly.

For $\vect{\phi}_{\rm s}$ we assume uniform improper priors with appropriate bounds determined for the spectral model at hand (e.g., to ensure that the SED is not negative for positive $E$).  The choice of prior can of course be more physically motivated, however, for our purposes these choices are sufficient because, as we show in Section~\ref{sec:real}, the marginalized positional posteriors do not have significant prior dependence.

\subsection{Posterior distribution and sampling}
\label{sec:sampling}

Given the likelihood described in Section~\ref{sec:likelihood} and the parameter priors given in Section~\ref{sec:priors}, the desired posterior distribution has the form
\begin{equation}
\prob(\psi_{\rm s}, \vect{\phi}_{\rm s} | \{N_{d,c}\}, {\rm GRB})
\propto 
\prob(\psi_{\rm s}, \vect{\phi}_{\rm s} | {\rm GRB})
\,
\prob(\{N_{d,c}\} | \psi_{\rm s}, \vect{\phi}_{\rm s}).
\end{equation}
There is no analytic way of determining the combinations of parameter values for which the posterior is high, nor of obtaining the normalizing constant which is needed to marginalize over $\vect{\phi}_{\rm s}$. As the posterior can be multi-modal, we use the form of nested sampling \citep{Skilling:2004} as implemented in {\tt MULTINEST} \citep{Feroz:2009} to obtain samples from the posterior.  

A particular utility of representing the posterior distribution as a set of samples is that marginalization is straightforward.  In our case, depending on the situation, either $\vect{\phi}_{\rm s}$ or $\psi_{\rm s}$ are nuisance parameters and marginalizing over one or other allows us to examine the marginal distribution of the parameter(s) of interest, $\prob(\psi_{\rm s} | \{N_{d,c}\}, {\rm GRB})$ or $\prob(\vect{\phi}_{\rm s} | \{N_{d,c}\}, {\rm GRB})$, respectively. 
	
\begin{figure*}
\includegraphics[width=14cm]{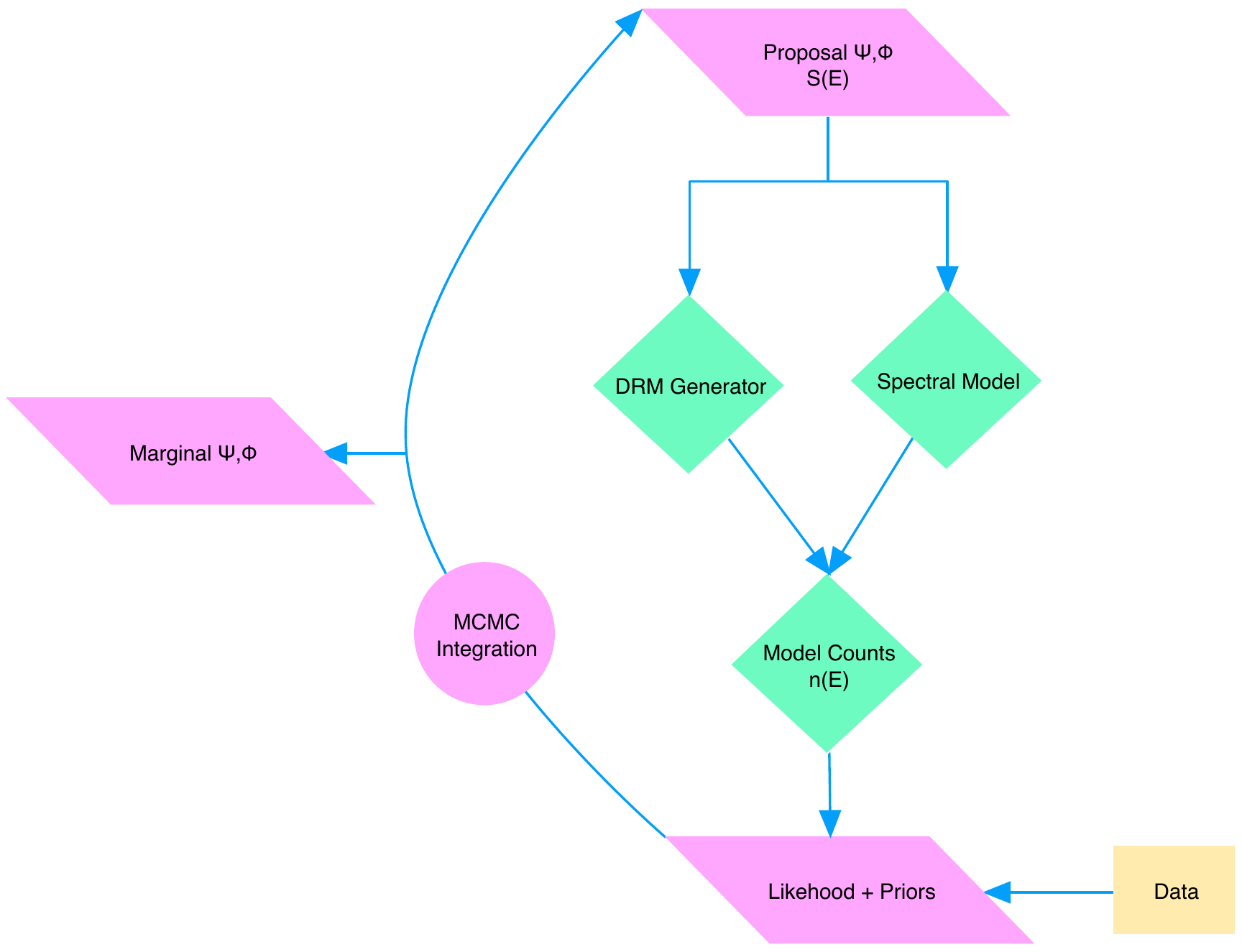}
\caption{The control flow of the {\tt BALROG} algorithm.}
\label{fig:flow}
\end{figure*}

Figure \ref{fig:flow} demonstrates the design of the {\tt BALROG} algorithm. In order to be useful for timely follow-up, the {\tt BALROG} must be able to localize a transient swiftly. The current implementation can be run on GBM data types including the quick-look TRIGDAT which consists of light curves for each of the 14 GBM detectors each with 8-channel energy resolution, as well as GBM time-tagged event (TTE) data which consists of photon time tags with 128-channel energy resolution \citep{Meegan:2009}. Localization on a multicore desktop workstation takes $\sim$ 5-10 minutes for TRIGDAT data and $\sim$ 10-20 minutes for TTE data due to its finer energy resolution. Faster speeds can be achieved with a computing cluster based installation which is used in this study. We discuss the differences between results with TRIGDAT and TTE data in the following sections. 

\section{Synthetic location analysis}
\label{sec:sims}

To access the validity of our method, we generate synthetic GRBs with known location and spectrum on a 25-point grid in the sky in celestial coordinates for a given spacecraft orientation. The CPL SED is used for the synthetic photon spectrum with power law index of $-1$ and $\vFv$ peak of 300 keV. For each point in the grid, a DRM is generated for the given sky position and the photon model is convolved with the DRM and given Poisson variability to produce an 8-channel synthetic data spectrum resembling TRIGDAT data. A simple Poisson background is added to the data obeying a power law energy spectrum.

\begin{figure}
  \includegraphics[scale=1]{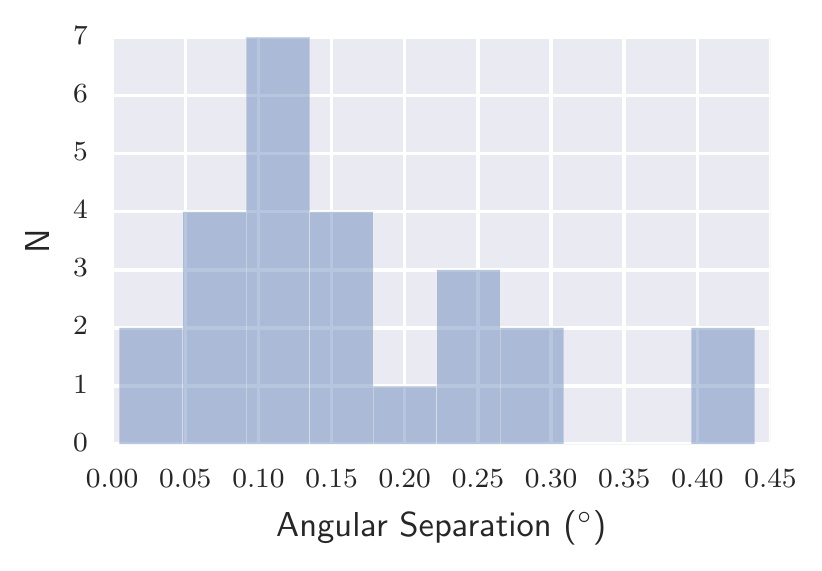}
  \caption{Distribution of angular separations from the simulated positions to the best fit {\tt BALROG} position using the CPL photon model for the 25-point simulated grid. }
  
  \label{fig:cpl_sim}
\end{figure}

Using {\tt BALROG}, we fit each simulated GRB in the grid with the CPL model to determine the location and spectrum using all 12 NaI detectors. Figure \ref{fig:cpl_sim} shows the distribution of angular separations between the true simulated position and the best fit {\tt BALROG} position.  This is always less than 1 deg and has a mean of $\sim 0.1$ deg. For real GRBs, the error could be much larger and depends on the intensity of the GRB. These results depend on knowing the true model of the spectrum \emph{and} the calibration of the NaI detectors being perfect. As we will see in the following sections, imperfections in the response will lead to systematic deviations from the true location when all 12 NaI detectors are used.
	
\begin{figure}
  \includegraphics[scale=1]{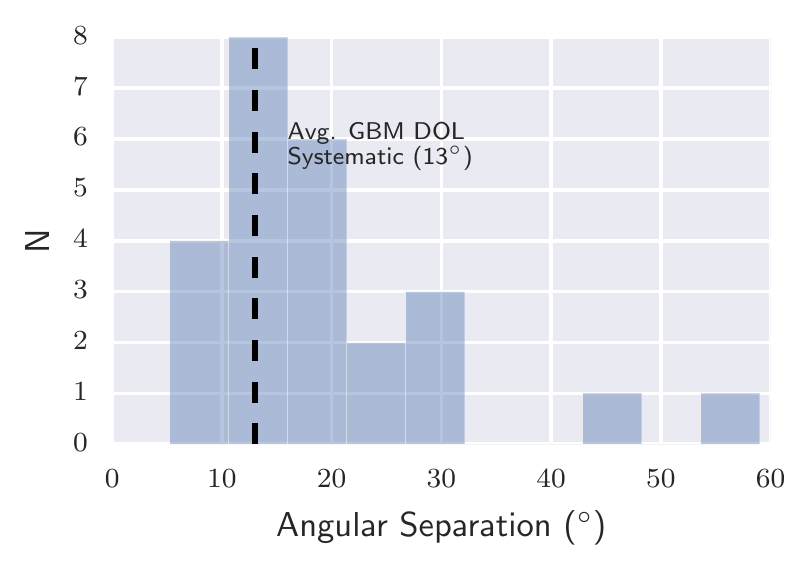}
  \caption{Distribution of angular separations from the simulated positions to the best fit {\tt BALROG} position using the hard template photon model for the 25-point simulated grid. }
  
  \label{fig:hard_sim}
\end{figure}

\begin{figure}
  \includegraphics[scale=1]{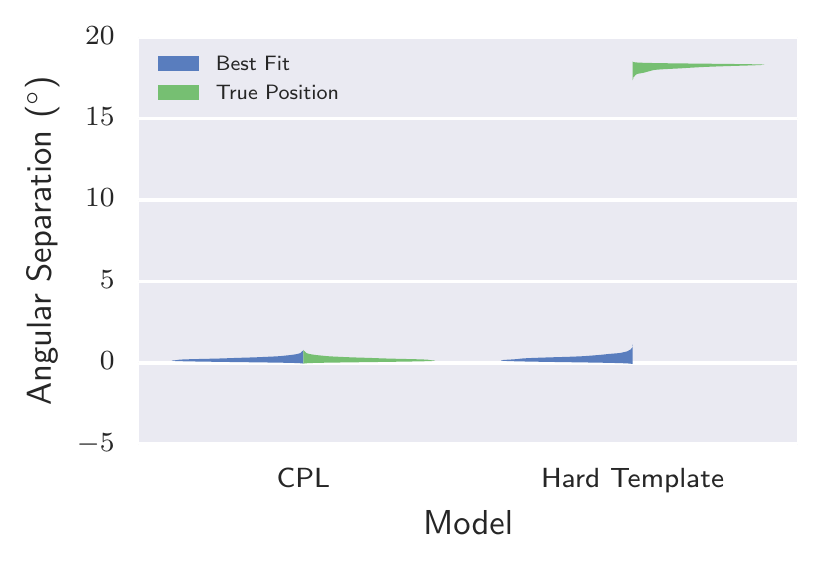}
  \caption{Comparison between distributions of the angular separation of the location posterior and the best fit location (blue) and true location (green) for the CPL and hard template photon models.}
  
  \label{fig:comp_sim}
\end{figure}

In order to demonstrate the effect of using fixed templates to determine locations, we fit the simulated grid with a Band function with spectral parameters $\Ep = 1 \, {\rm MeV}$, $\alpha = 0.$ and $\beta = -1.5$ (i.e., the hard template). The amplitude and location are the only parameters allowed to vary in the fit. Figure \ref{fig:hard_sim} shows that using the hard template introduces a systematic offset in the location of $\sim 13$ degrees, which is comparable to that identified by \cite{Connaughton:2015}. Moreover, we can examine the angular separation between the best fit positions and the location posterior (essentially the error region of the location) for one point on the grid and compare it with the angular separation between the true position and the location posterior. Figure \ref{fig:comp_sim} shows that, when fitting with the CPL model, these two distributions are the indistinguishable, which indicates that the error region of the fit is encompassing the true variance in the data. However, while the error region obtained from the hard template is very precise, it is not accurate (i.e., does not encompass the true error region). 
	
\begin{figure}
  \includegraphics[scale=.8]{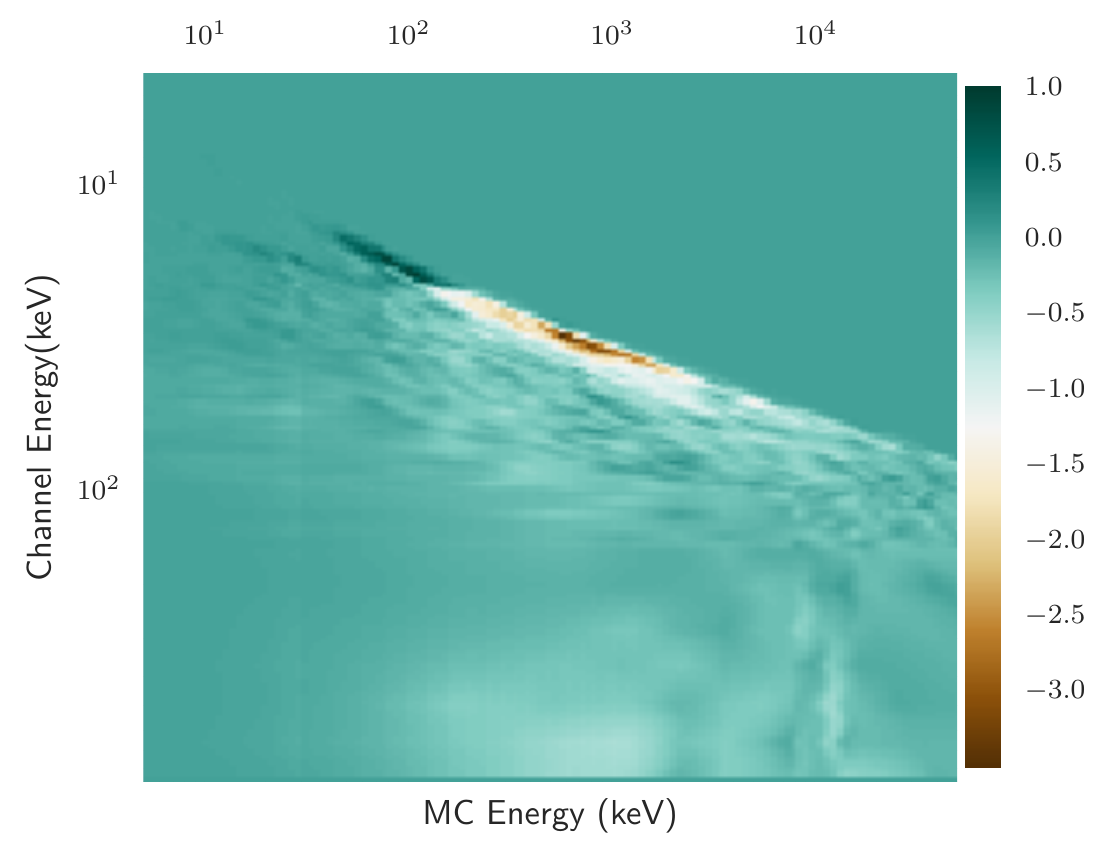}
  \caption{A probabilistic difference between the DRMs generated from the location posterior suing the hard template and the CPL photon model. The color scale indicates the the difference in effective area in arbitrary units. This serves to illustrate the differences that can arise in the normal public DRMS when an improper model is used to generate DRMs.}
  
  \label{fig:sim_drm}
\end{figure}

		We can look at how the differences in the locations found by the CPL and hard template models affect the DRM used for spectral analysis. As a toy example, we propagate the location posterior through the DRM generator to create a probabilistic DRM for both models. Figure \ref{fig:sim_drm} illustrates the difference between these two DRMs. The main differences are at high and low energy, which can cause different slopes and hence different spectral parameters to be recovered. However, in the {\tt BALROG} scheme, the DRM is no longer a fixed quantity, but rather a probabilistic quantity weighted by the uncertainty in location.

\subsection{Posterior checking}
\label{sec:ppc}
	
To assess whether {\tt BALROG} provides GRBs position estimates with realistic uncertainties, we need to perform model checking. The standard Bayesian approach is a posterior predictive check (PPC, see, e.g., \citealt{Gelman_etal:2014}), in which simulated data-sets obtained from parameters drawn from the posterior distribution for a GRB are compared to the actual data on that source.  Here, however, we are in the unusual situation of knowing the true locations of the GRBs from other observations.  That allows us to bypass the data-simulation step and simply assess whether our posterior distributions are consistent with the GRBs' actual positions.

For a single GRB, with known sky position $\psi_{\rm true}$, the key statistical quantity is the posterior density at this location, $\rho_{\rm true} = \prob(\psi_{\rm s} = \psi_{\rm true} | \{ N_{d,c} \})$.  In cases where the posterior is singly peaked and symmetric this reduces to a measure based on the angular separation between the peak of the posterior and the true position, but is also valid if the posterior has significant degeneracies or is multimodal.  The numerical value of $\rho_{\rm true}$ is not meaningful in isolation, as it is a probability density that is determined by, e.g., the scale of the positional uncertainties.  It must be evaluated in relation to the distribution of $\rho$ values that would be obtained by drawing random positions from the posterior.  A sample from this distribution can be obtained by drawing a position $\psi_{\rm sim}$ from the posterior distribution $\prob(\psi_{\rm s} | \{ N_{d,c} \})$ and then evaluating $\rho_{\rm sim} = \prob(\psi_{\rm s} = \psi_{\rm sim} | \{ N_{d,c} \})$ (i.e., the density at this simulated position).  Repeating this process yields a set of samples from the distribution of $\rho$ values from which from which $\rho_{\rm true}$ ought to have been drawn if the model were correct.  If the model is  badly wrong then the true position will lie outside the region identified by the posterior and $\rho_{\rm true}$ will be lower than (almost all of) the simulated values $\rho_{\rm sim}$.  Conversely, if the uncertainties are over-estimated then $\rho_{\rm true}$ will be higher than the simulated values.

This comparison can be made on an source-by-source basis, but for a set of $N_{\rm GRB}$ GRBs the $\rho$ values can be multiplied to obtain
\begin{equation}
  \rho_{\rm true} = \prod_{i = 1}^{N_{\rm GRB}} 
    \rho_{{\rm true},i},
\end{equation}
where $\rho_i = \prob(\psi_{\rm s} = \psi_{{\rm true},i} | \{ N_{d,c} \}, i)$ is evaluated from the posterior distribution for the $i$'th GRB.  Similarly, with $\rho_{{\rm sim},i,j}$ being the $j$'th simulated $\rho$ value drawn from the posterior of the $i$'th GRB, samples from the expected distribution are given by the product
\begin{equation}
\rho_{{\rm sim},j} = \prod_{i = 1}^{N_{\rm GRB}} \rho_{{\rm sim},i,j}.
\end{equation}

\begin{figure}
  \includegraphics{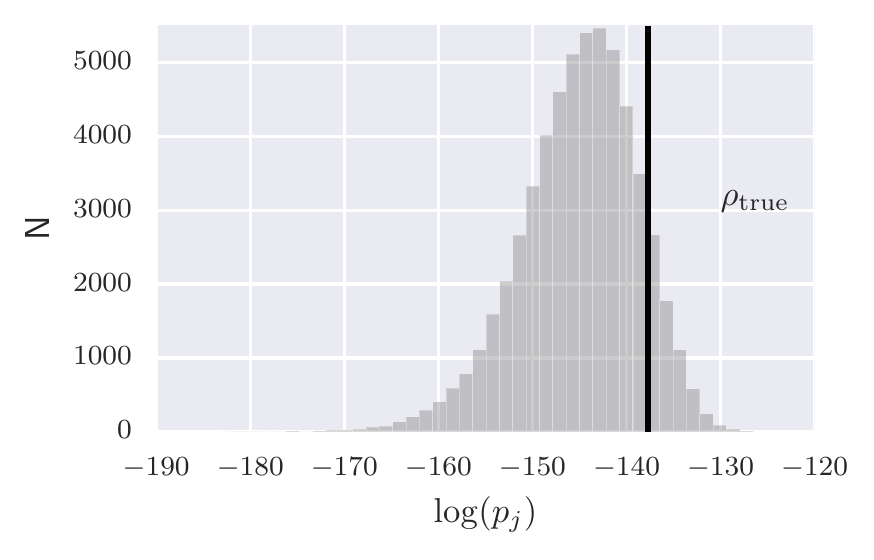}
  \caption{The PPC distribution for the simulated GRB sets using the full approach of the {\tt BALROG}.}
  \label{fig:ppc_sim}
\end{figure}

Applying this framework to the simulations, we obtain Figure \ref{fig:ppc_sim}. As expected from simulations assuming and {\emph ideal} detector response, the model is accurate at predicting future predictions ($\rho_{\rm true}$ is not too far to the left) without over-fitting the data with the increased variance of the {\tt BALROG} model ($\rho_{\rm true}$ is not too far to the right).	
	
\section{Analysis of real GRBs with known locations}
\label{sec:real}

Accessing the validity of our method depends strongly on our ability to precisely and accurately locate real GRBs that have been located by imaging instruments. We use the table of known GRB locations provided in \cite{Connaughton:2015} which additionally allows us to compare our method to GRBs localized by the standard GBM method. We divide the analysis into two sections: GRBs well localized by GBM and GRBs localized with a systematic deviation by GBM. First, we note a few details about fitting real data. In the above simulations, we assumed the DRMs are perfect and expect to recover the true location and spectral parameters. If there are calibration/simulation effects not properly modeled by the DRMs, then analysis of real data can produce sub-optimal results leading to offsets in location and spectral parameters. Similarly, if the spectral model we choose to fit the GRBs does not correspond to the true spectral model of the source, we can expect the location to be affected. 

\begin{figure}
  \includegraphics[scale=.5]{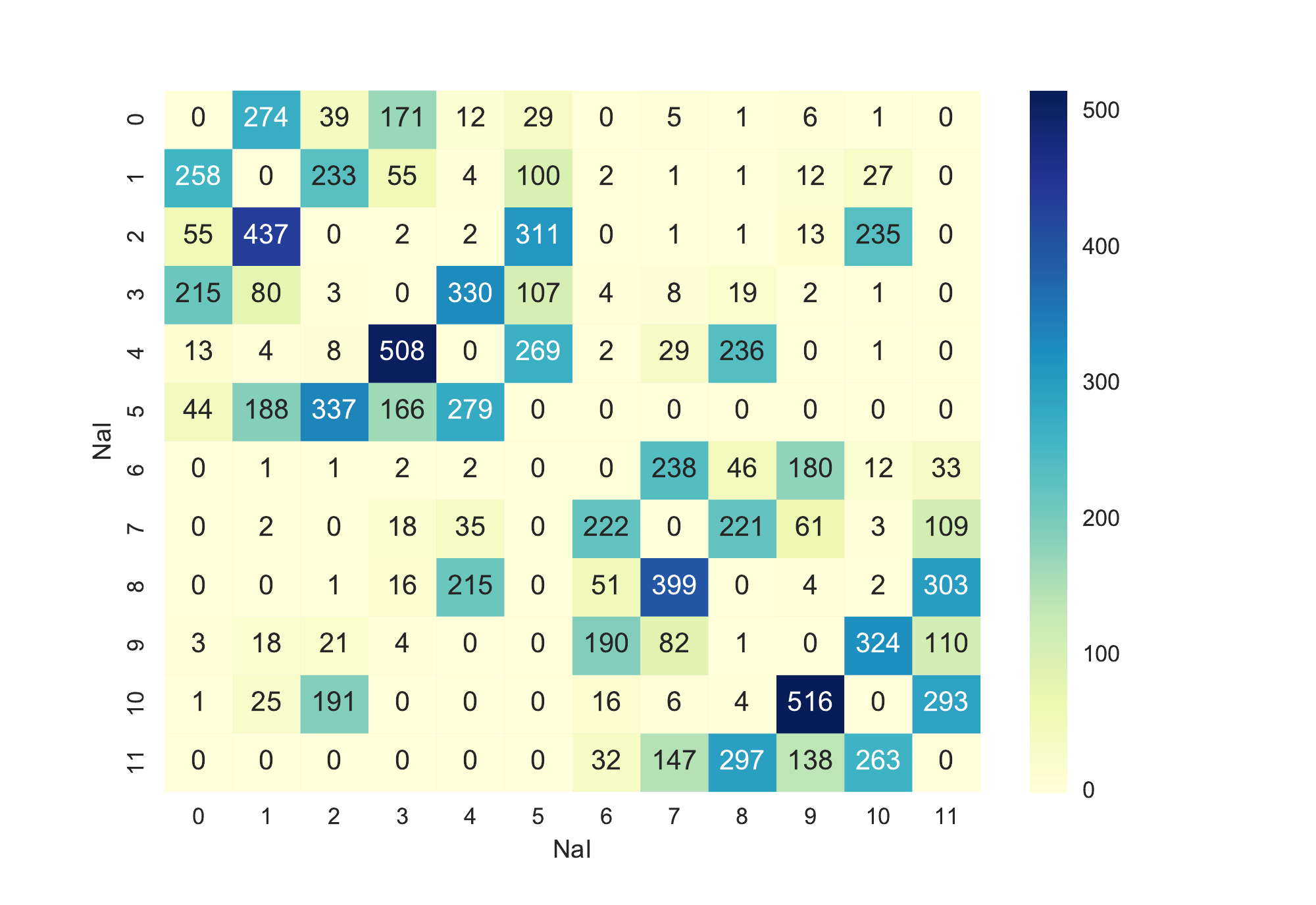}
  \caption{Chart of detectors which see the same source based off a simulation of 10,000 GRBs made by the Fermi GBM team. Color and number correspond to the number of coincident detections between a pair of detectors. It is intended to aid in choosing appropriate sets of detectors for spectral and location analysis.}
  \label{fig:legalpairs}
\end{figure}

	In realtime trigger situations, the user of {\tt BALROG} will not know the location a priori. As we will demonstrate in the following sections, the user must select a subset of detectors to obtain correct locations of sources. The GBM team has assembled a table of so-called legal detector sets to aid in choosing proper subsets of detectors for analysis (see Figure \ref{fig:legalpairs}). It will often be required to perform several location runs, with different spectral models and detector subsets to find the correct location. {\tt BALROG} provides the marginal likelihood or evidence which can be used to compare between these iterations and choose the best location. It is currently unlikely that an automated system can perform these actions due to the complicated interaction between background, source interval selection, model choice and detector selection. 

\subsection{GRBs localized with a systematic deviation by GBM}
\label{sec:sys}

We examine ten GRBs with systematic deviation in the DOL location from the true location to test if {\tt BALROG} can improve upon the result or provide appropriate location contours. We define a systematic deviation such that the angular separation between the DOL location and the true location exceeds the reported DOL $1\sigma$ error. We choose the ten worst cases from \citet{Connaughton:2015} which are listed in Table \ref{tab:sys_grbs}. For each of these GRBs, we try three spectral models (Band, CPL and SBPL) and both all detectors and a subset of detectors. 

\begin{figure*}
\centering
  \subfigure{\includegraphics[scale=.4]{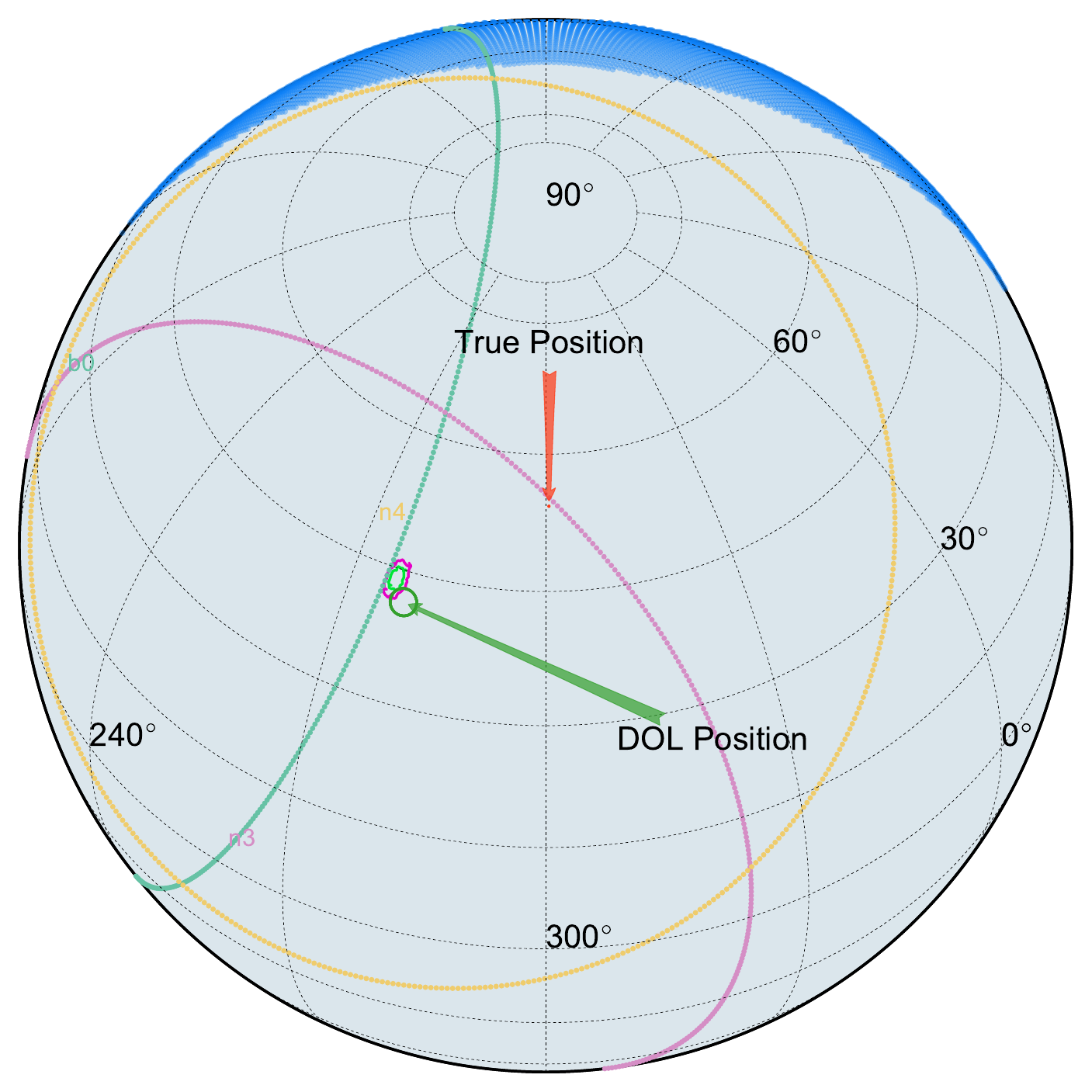}}\subfigure{\includegraphics[scale=.4]{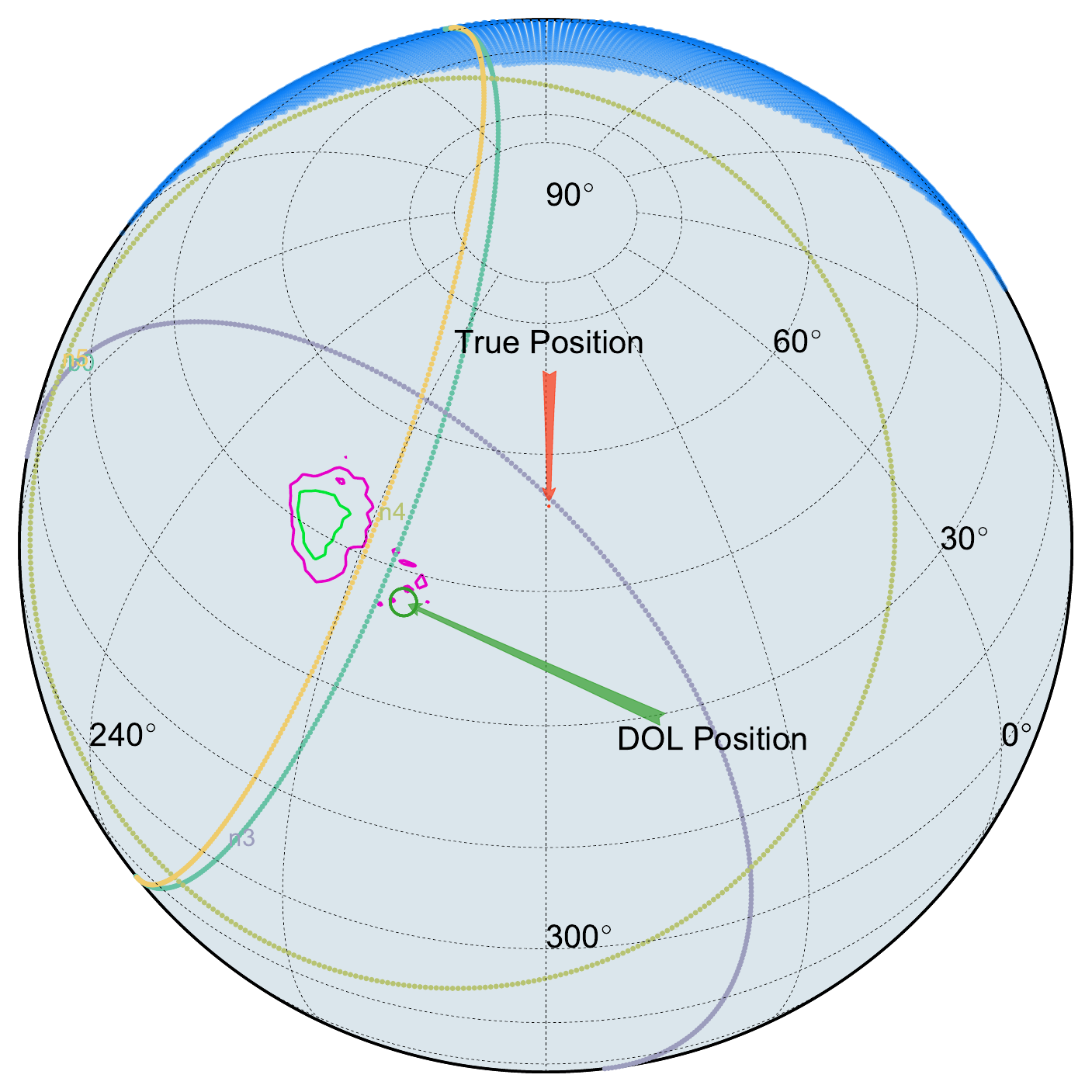}}\subfigure{\includegraphics[scale=.4]{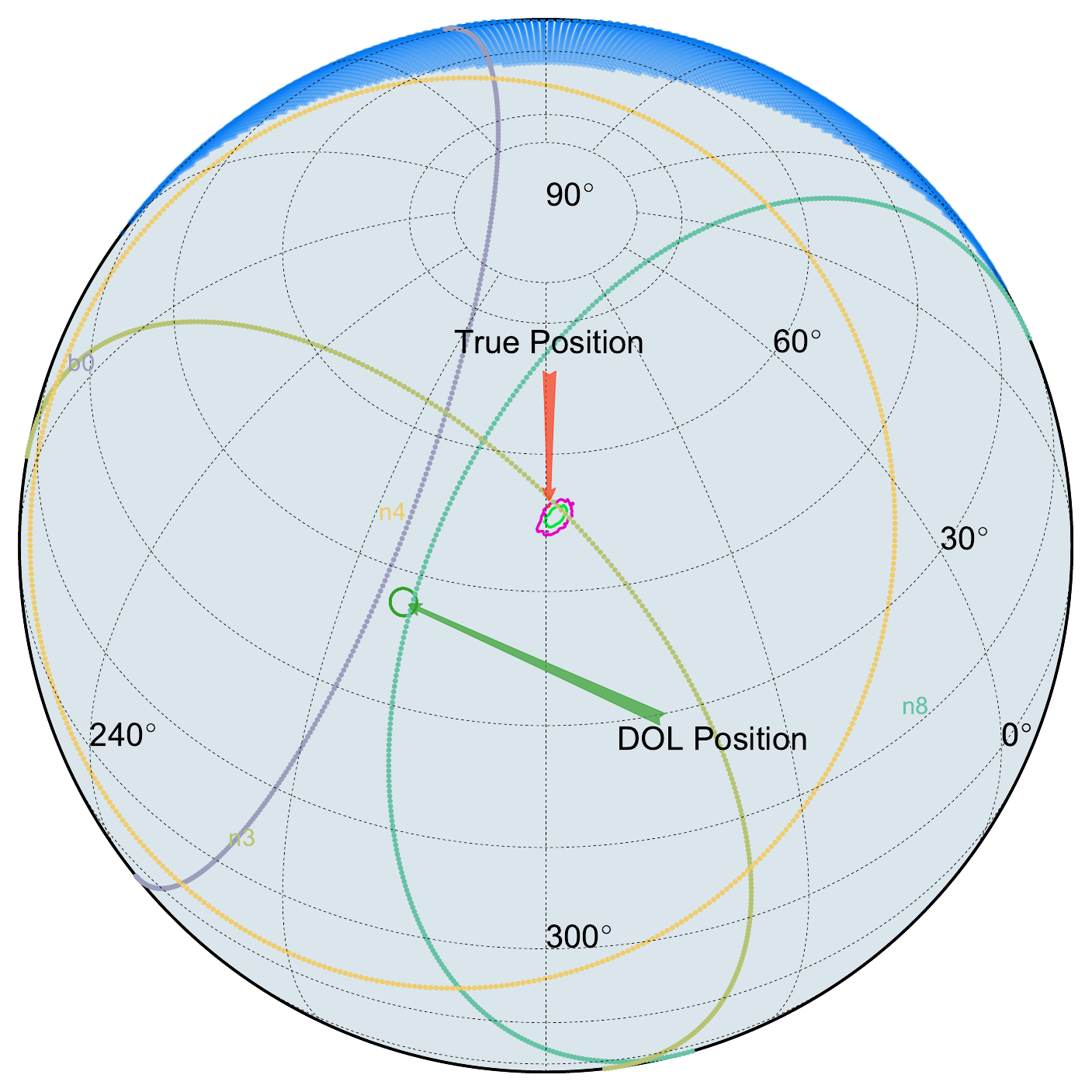}}
  \caption{The location plots from GRB 121128212. From left to right, the results of using all 14 NaI detectors, the result from assuming a template spectrum, and the {\tt BALROG} result. The $1\sigma$ and $2\sigma$ {\tt BALROG} contours are shown in green and purple respectively. The blue shaded region represents the portion of the sky occulted by the Earth. The various colored circles are the 60$^{\circ}$ FOV of the GBM detectors that view the best fit {\tt BALROG} position.}
  
  \label{fig:1211}
\end{figure*}

	Focusing on the worst case, GRB 121128212, we can specify a few important details about what provides a good location. For the case of using all detectors to fit the spectrum and location, we recover a position similar to the DOL (See Figure \ref{fig:1211}) . However, when a subset of detectors chosen by starting with the brightest NaI and then selecting detectors via the legal detector set table (see Figure~\ref{fig:legalpairs}), we find perfect agreement with the true source. Finally, when we mimic the so-called template spectra, we similarly find a position consistent with the DOL. This immediately points out two very important points:
	\begin{enumerate}
		\item obtaining a correct position requires the proper spectral model and not a fixed template
		\item using 12 NaI detectors for spectral and/or location analysis can lead to systematics.
	\end{enumerate}
	The last point indicates that the off-axis response of the GBM NaI detectors may contain inaccuracies, otherwise, using all 12 NaI detectors would result in an accurate location as shown in our simulations. It is well known \citep{Goldstein:2012} that for spectral analysis, only a subset of detectors with viewing angles less than 60$^{\circ}$ should be used. Similar care must be taken when fitting for location. Unfortunately, without knowing a location a priori, one must iterate the fitting process until a good match between the detectors viewing the location and those used is found.
	
	To demonstrate this, we look at the case of GRB 120624309 which has a true location that is not within the good FOV of any NaI, but has detectable signal in all NaI detectors. We iterate through the detector selection and use the value of the marginal likelihood ($Z$ or Bayesian evidence) to choose the best location. We find that certain detector combinations greatly alter the location while others simply change the contours slightly. This can be troublesome in real-time followup, and demands better modeling of GBM DRMs to be addressed. Essentially, different interval, background, and detector selections are different statistical models and highly non-nested. This makes using likelihood ratio tests (such as choosing the lowest $\chi^2$) for choosing locations improper. Still, it should be noted that the use of improper priors such as the uniform priors we assume for spectral parameters makes the marginal likelihood insensitive for model comparison.

	The performance of the {\tt BALROG} is demonstrated for the entire set of ten GRBs with the highest systematic in Table \ref{tab:sys_bal}. We show that we can accurately recover the true positions of all the GRBs within the 95\% highest density posterior (HDP) bayesian credible region. There are a few exceptions where the credible region misses the true location by fractions of a degree. This is most likely due to response deficiencies. Still, the {\tt BALROG} practically eliminates the systematics in GBM GRB locations. In some cases, the locations were made more accurate via larger error regions, while in many cases, the error region was simply shifted to the true location. In practice, this depends on the intensity of the source. For TRIGDAT data, we find that a CPL model is sufficient for determining a proper location. The medium and soft templates always resulted in a systematic offset unless the true spectral model was similar to the template. In a single case (GRB 120624309), a power law photon model provided the best location which was also indicated by having the highest marginal likelihood. The remaining figures demonstrating these fits are in shown in Figure \ref{fig:all_sys} of Appendix \ref{sec:appendix}.

\subsection{GRBs well localized by GBM}
\label{sec:well}
We examine ten GRBs that are well localized by the DOL, i.e., GRBs with angular separation between the DOL location and the true location smaller than the DOL $1\sigma$ error circle radius from \citet{Connaughton:2015}. Table~\ref{tab:good_bal} details the results. We discarded GRB 110625881 because it has no detector with $5\sigma$ signal within 60 degrees. The same models as are used in Section \ref{sec:sys} are applied to these GRBs as well. 

Table~\ref{tab:good_bal} shows that {\tt BALROG} is able to recover all the 10 well localized GRBs. We found that for GRB 091003191, {\tt BALROG} is able to recover the true location with sub-degree $1\sigma$ precision. Such accurate (i.e., the error contour covers the true location) and precise (i.e., small error region) gamma-ray localization of GRBs can be used for X-ray and optical afterglow follow-up observations.

It is observed that using all 12 NaI detectors in the localization results in less accurate and precise locations. The fact that the CPL gives more accurate locations in the 10 bursts examined in Section.~\ref{sec:sys}, as well as with this sample, indicates again that the fixed templates may result in inaccurate locations unless they are close to the intrinsic GRB spectrum.

\subsection{Summary of {\tt BALROG} localization}

  Using the PPC framework developed in Section \ref{sec:ppc}, we perform model checking of the {\tt BALROG} on real data. Once again, we are in the fortunate situation of knowing the true locations via follow-up X-ray and optical observations of the GRBs in our sample as referenced in \citet{Connaughton:2015}.

\begin{figure*}
  \subfigure{\includegraphics[scale=1]{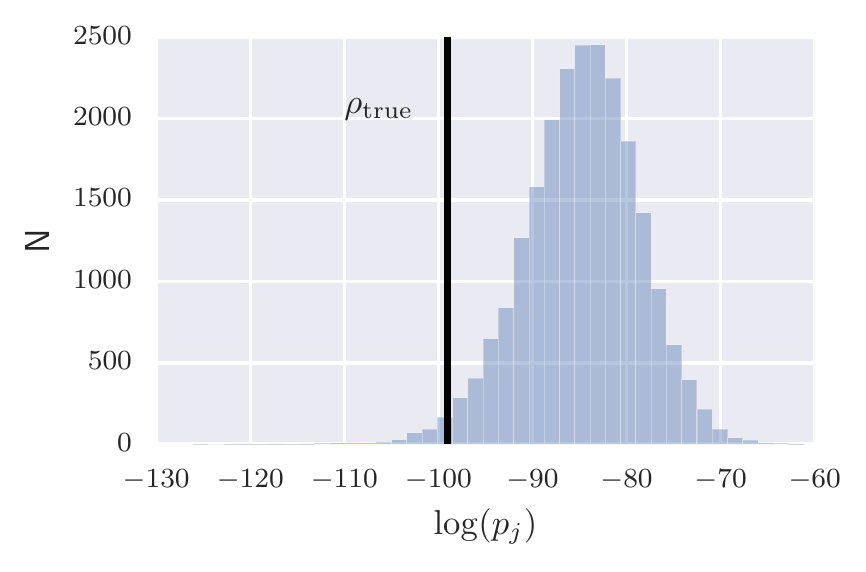}}\subfigure{\includegraphics[scale=1]{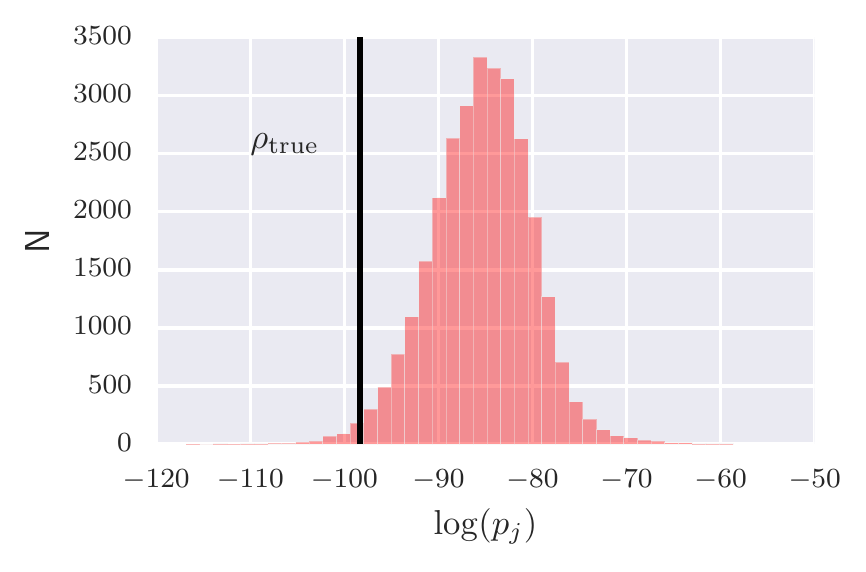}}
  \caption{The PPCs for the systematically off (left) and well-located sample (right) from the {\tt BALROG} posterior.}
  \label{fig:ppc}
\end{figure*}

\begin{figure*}
   \subfigure{\includegraphics[scale=1]{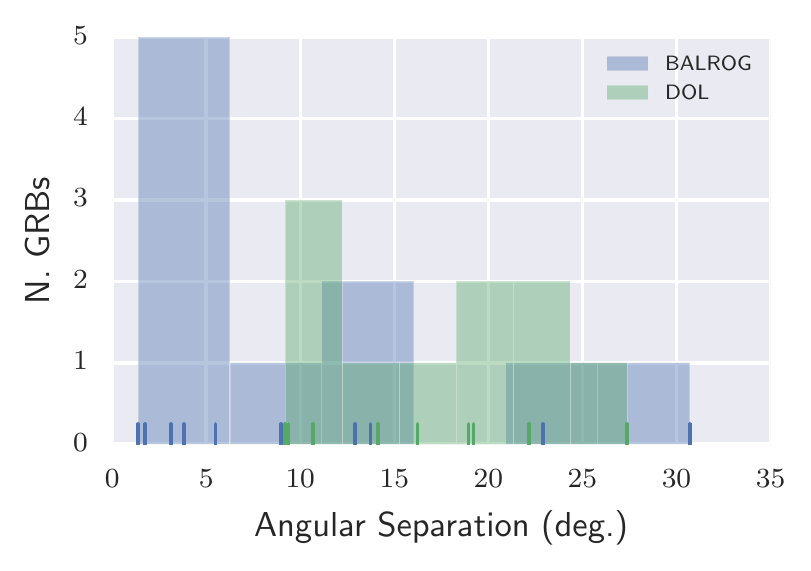}}\subfigure{\includegraphics[scale=1]{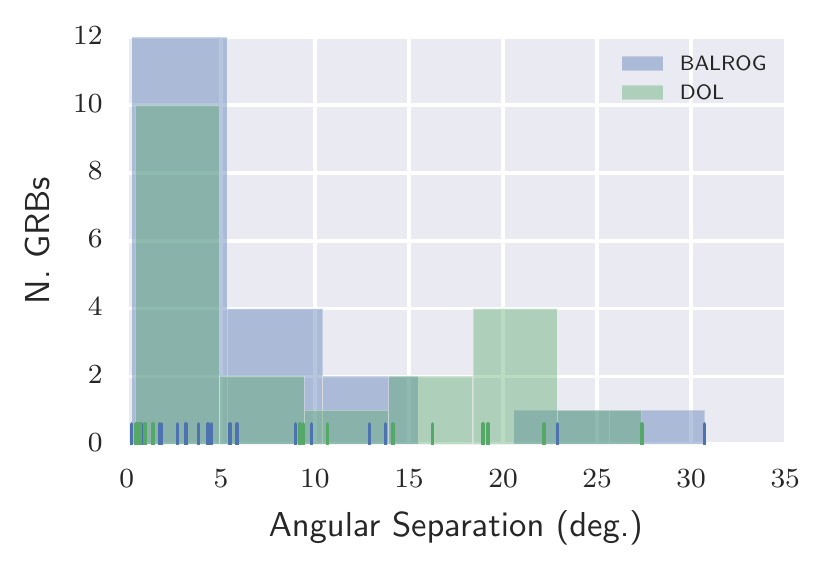}}
  \caption{(Left) Histogram of {\tt BALROG} and DOL location angular separation from the best fit location to the true location. Note that all the DOL locations systematically miss the true location, though {\tt BALROG} captures all within the 95\% posterior. (Right) The same as the left panel, but for all locations. The DOL's tail is longer and includes GRBs that were not located within its error contours. }
  \label{fig:hist_all}
\end{figure*}

Figure \ref{fig:ppc} compares $\rho_{\rm true}$ to the distribution of $\rho_{\rm sim}$ for both samples.  Both samples show perform equally well at predicting the true positions compared to the DOL was able to capture the true position on for half of this sample. Still, the model does not perfectly predict the true locations, but is within the 95\% region. Figure~\ref{fig:hist_all} shows the angular separation from the true position to the MLE estimate from {\tt BALROG} as well as GBM for the systematically off sample and the entire sample. It is important to point out that for the systematically off sample, while the {\tt BALROG} angular separation can be very large, it fully covers the \emph{true} location while the DOL error radius does not. 

\begin{figure*}
  \subfigure{\includegraphics[scale=1]{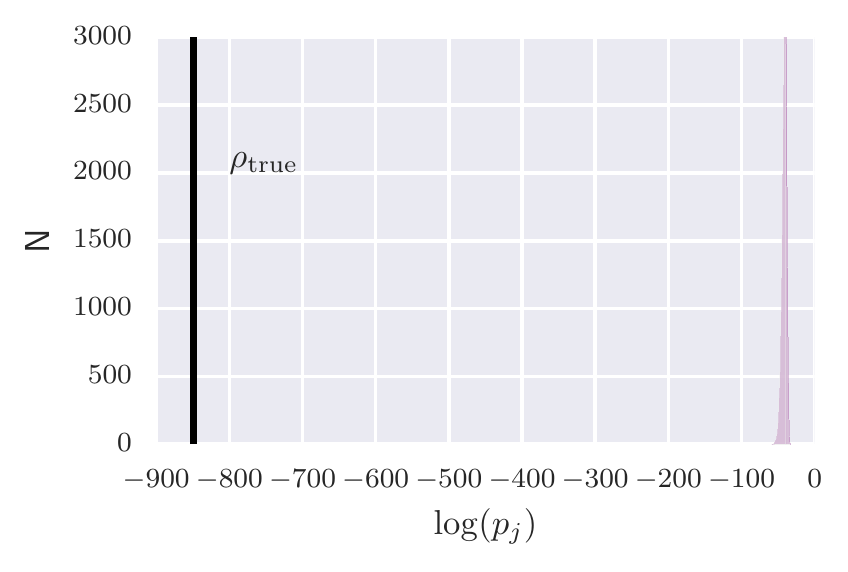}}\subfigure{\includegraphics[scale=1]{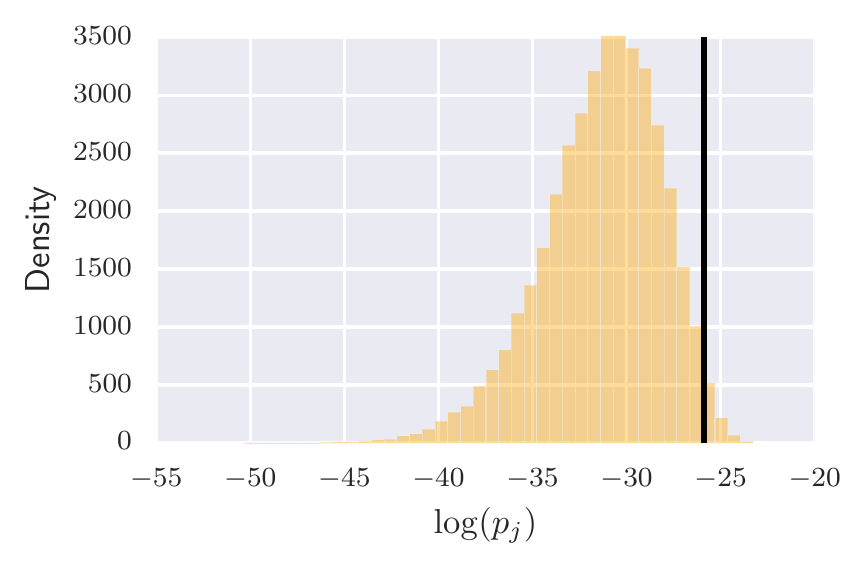}}
  \caption{The PPCs for the systematically off (left) and well-located sample (right) from the assumed DOL posterior.}
  \label{fig:ppc_dol}
\end{figure*}

To further emphasize the how {\tt BALROG} compares to the DOL, we perform the same check on the systematically off and good sample using the DOL locations. A major caveat is that the DOL locations accessible via \citet{Connaughton:2015} are not the result of Bayesian sampling, and do not possess the non-gaussian error regions calculated in the GBM location software. Therefore, we assume the reported location is the center of a multinomial Gaussian with standard deviation equal to the reported error radius. This forms our pseudo-posterior for each DOL location. We then perform the same PPC analysis as before for the two samples. Figure \ref{fig:ppc_dol} shows that the systematically off sample analyzed with the DOL performs significantly worse than the {\tt BALROG} at predicting the true locations. Moreover, the tightly-peaked shape of the DOL posterior contrasts with the broader distribution of the {\tt BALROG}. This translates to the fact that while the {\tt BALROG} does not perfectly reconstruct the location in all cases, (for reasons relating to source geometry, detector calibration, etc.) the method admits its ignorance with wider error regions (See Figure \ref{fig:all_sys}). On the other hand, the DOL performs very well for those GRBs which it located accurately as expected. These GRBs make are part of the so-called core distribution of the DOL location accuracy distribution as noted in \citet{Connaughton:2015}.

\section{Effect on GRB spectra}
\label{sec:spec}

The variance in location can be viewed as variance in the known DRM for each detector used in a spectral analysis. Just as the background should not be fixed with no variance in spectral analysis, so should the variance in the DRM be left free. Therefore, we investigate how this added freedom can affect the spectral analysis of GRBs. For this study, we use GBM TTE data which has 128-channel resolution. The {\tt BALROG} can produce DRMs for an arbitrary number of channels. We focus on the well-known GRB 080916C which has an external and precisely measured location (RA=119.8$^{\circ}$ Dec=-56.6$^{\circ}$). We break the study into two parts: time-integrated spectra and time-resolved spectra. In both, we focus on detector selection and purported claims of extra spectral components \citep[e.g.][]{Guiriec:2015}. While the GRB was also seen by the Fermi Large Area Telescope (LAT), we neglect these data as they are outside the scope of GBM locations and spectra.

\subsection{Time-integrated analysis} 

\begin{figure}
  \includegraphics[scale=1]{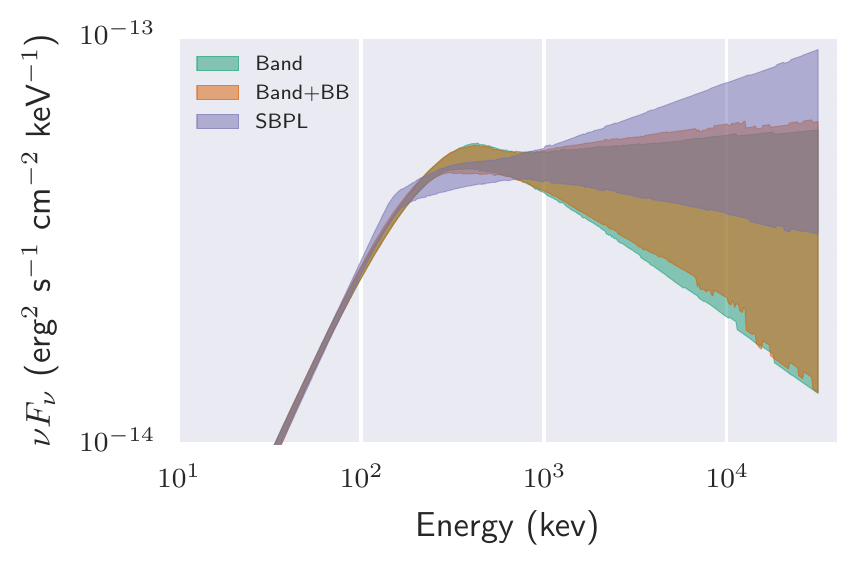}
  \caption{A $\vFv$ comparison of photon models used for the integrated spectrum of GRB 080916C. The Band and Band+blackbody spectra are very similar due to the fact that the blackbody component contributes little to the spectrum. This indicates that the blackbody's contribution in \citet{Guiriec:2015} may be due to neglecting the variance of the DRMs and location.}
  \label{fig:spec-int}
\end{figure}

The interval T0+0.-71. s. is chosen for time-integrated analysis. The detectors with favorable viewing angles are NaI detectors 0,3,4,6,7 and BGO 0. First we fit the Band function with these selections. Alternative photon model choices have no effect on correcting the location. We find that the optimal detector selection is NaI detectors 0,3,4 and BGO 0 which result in a correct location with large errors (see Table \ref{tab:080916C-int}). With this detector selection, we additionally fit Band+blackbody (BB) and SBPL law photon models. We find that all models have the same marginal likelihood and similar locations. Therefore, the simpler Band model is preferred. We note that the Band+blackbody model found in this analysis differs from that found in \citet{Guiriec:2015} with the blackbody's temperature highly unconstrained. This is likely due to the blackbody acting to modify the low-energy slope to accommodate the localization parameters. It has been shown that blackbodies in time-integrated spectra may arise from spectral evolution \citep{Burgess:2014}. Yet, this analysis shows that the narrow blackbody spectrum may be absorbing response deficiencies in the GBM DRMs when their variance is not accounted for as is done in the standard GRB spectral analysis. Figure \ref{fig:spec-int} displays the spectra of all three photon models fit. 

We now fit the same data assuming no variance in location, i.e, with a fixed DRM corresponding to the known location. Table \ref{tab:080916C-fix} shows the recovered parameters corresponding to the fixed DRM fits. There are two important observations: the fitted parameters change when using a fixed DRM resulting in softer low energy slopes and higher $\Ep$ and the Band+BB fit results in the well-known $kT\sim 40$ keV blackbody with preferential marginal likelihood ($\Delta \log Z \simeq 6$) for the model over the Band function. This demonstrates that without the included variance of location, extra spectral components may be serving to artificially correct the DRM for this missing variance.

\subsection{Time-resolved analysis}

The high intensity of GRB 080916C caused Fermi to repoint \citep{Abdo:2009}. The {\tt BALROG} assumed one spacecraft orientation when fitting spectra and locations though we have designed it with the ability to fit multiple spectra simultaneously all sharing the same GRB location. We do not test this capability here and leave it for a future study. However, the 71 s duration used in the time-integrated analysis may introduce systematics as the spacecraft is repointing and a single spacecraft orientation is not valid. We can instead look at time-resolved spectra and use the spacecraft orientation corresponding to individual intervals to analyze the spectra. Ideally, all intervals should give similar locations. Using the same detector selection as was used in the time-integrated analysis, we choose the first eight intervals identified in the intriguing fine-time spectroscopy of \citet{Guiriec:2015} and fit them with Band and CPL+BB photon models. We additionally fit with the new GRB model identified in \citet{Guiriec:2015} which consists of three spectral components, a CPL, a blackbody and a power law where the CPL and PL have their spectral indices fixed to -0.7 and -1.5 respectively. 
	
\begin{figure*}
  \includegraphics[scale=0.5]{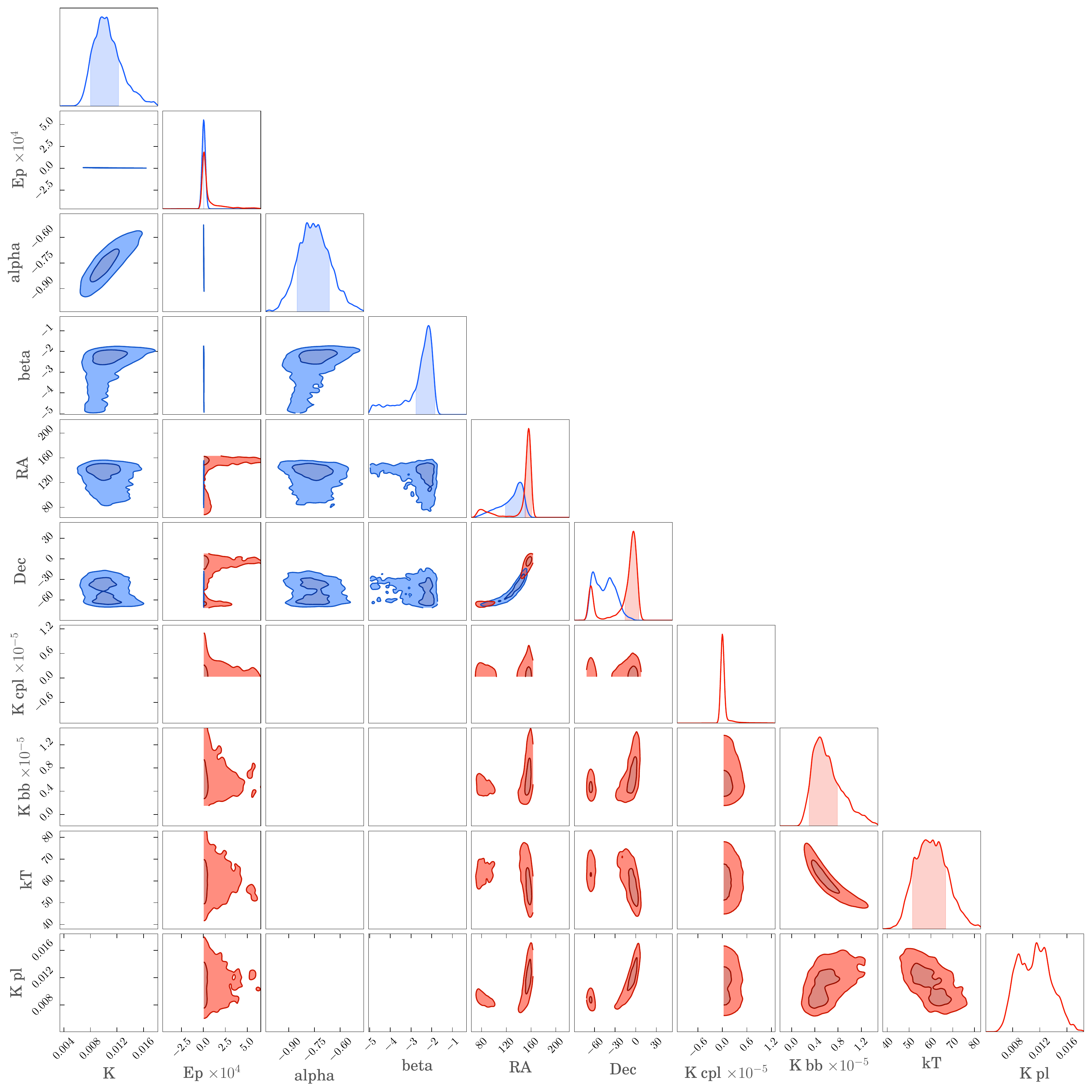}
  \caption{The marginal distribution for the Band (blue) and CPL+BB+PL (red) fit for the time interval T0+3.8-4.3 for GRB 080916C. K represents normalizations for the various spectral components. It is evident that for the CPL+BB+PL model, the CPL component is consistent with zero, allowing the BB to dominate the spectral peak. The differences in the location parameters are also clearly demonstrated. }
  \label{fig:band_marg}
\end{figure*}

\begin{figure*}
\centering
  \subfigure{\includegraphics[scale=.4]{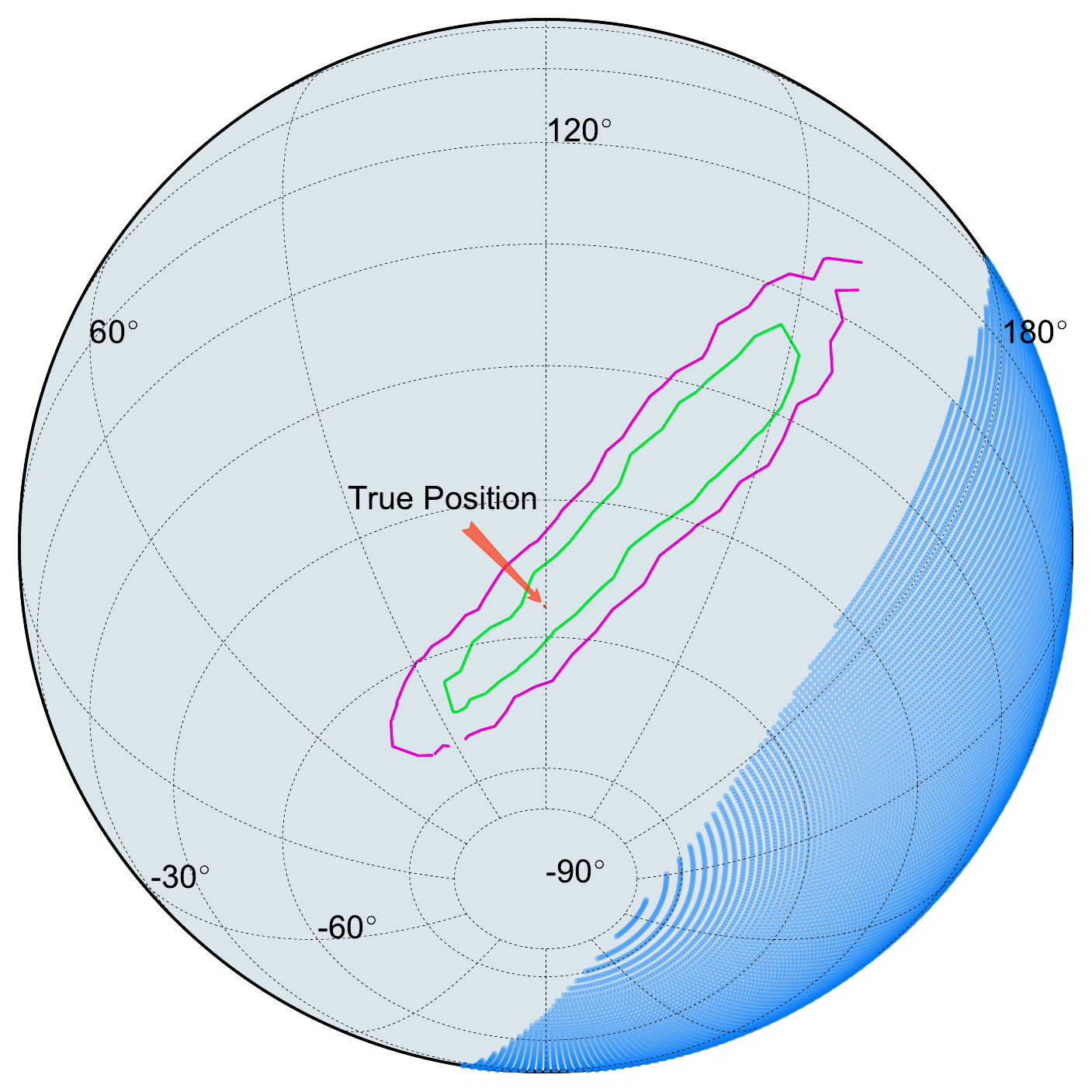}}\subfigure{\includegraphics[scale=.4]{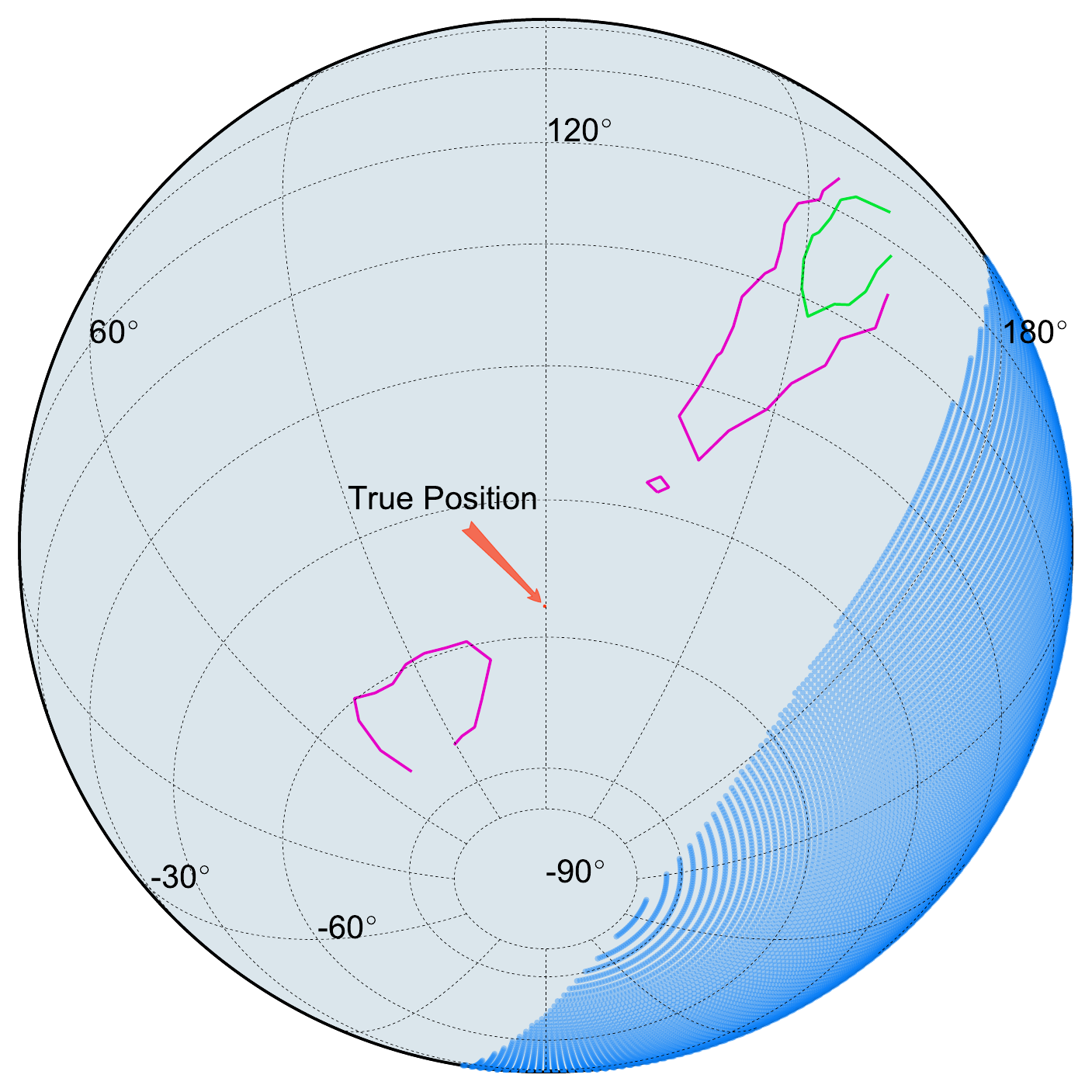}}
  \caption{Locations produced from fitting the Band function (right) and the CPL+BB+PL model (left) to the interval T0+3.8-4.3 of GRB 080916C. The CPL+BB+PL model provides an inaccurate location.}
  \label{fig:sgcomp}
\end{figure*}

The results from each time interval are detailed in Table \ref{tab:080916C}. As with the time-integrated analysis, there was no significant Bayesian evidence for extra components in the spectra and a Band function fit the spectra sufficiently. The first interval provided an incorrect location as NaI 0 was pointed close enough to the true location until after Fermi repointed. Therefore, we did not use NaI zero in the first interval. Interestingly, while the marginal likelihood did not support the three component model of \citet{Guiriec:2015}, the model resulted in inaccurate locations in several time intervals (see Figure \ref{fig:sgcomp}). Due to its fixed low-energy slope, the model's flexibility to fit the data is passed to the freedom in the location parameters which adjusts to the poor-fitting of the spectral model by giving an incorrect location. The marginal distributions of the Band fit and the CPL+BB+PL (Figure \ref{fig:band_marg}) demonstrate that while for the Band function proper constrained distributions are obtained, the more complex CPL+BB+PL model's parameters fill out their priors and hence lead to lower Bayesian evidence. One could argue that the incredibly small time-intervals do not allow for more complex models to be statistically significant, however, the inability of the model to locate the GRB also makes it difficult to explain it as the true spectrum. However, recent work in \citet{Guiriec:2016} shows that the model does help to explain the optical data of some GRBs, and hence, further investigation is required. 

Therefore, this model is not able to explain the data of GRB 080916C. Moreover, this raises an interesting paradigm for model selection in GRBs, when the true location is known externally, models, such as the new physical models being fit to data \citep[e.g.][]{Burgess:2014a,Vurm:2015}, can be rejected if they do not provide proper location parameters. In future work we will address this with the synchrotron model of \citet{Burgess:2014a}, but the computationally expensive work is beyond the scope of the project at hand.

\begin{figure}
  \includegraphics[scale=1]{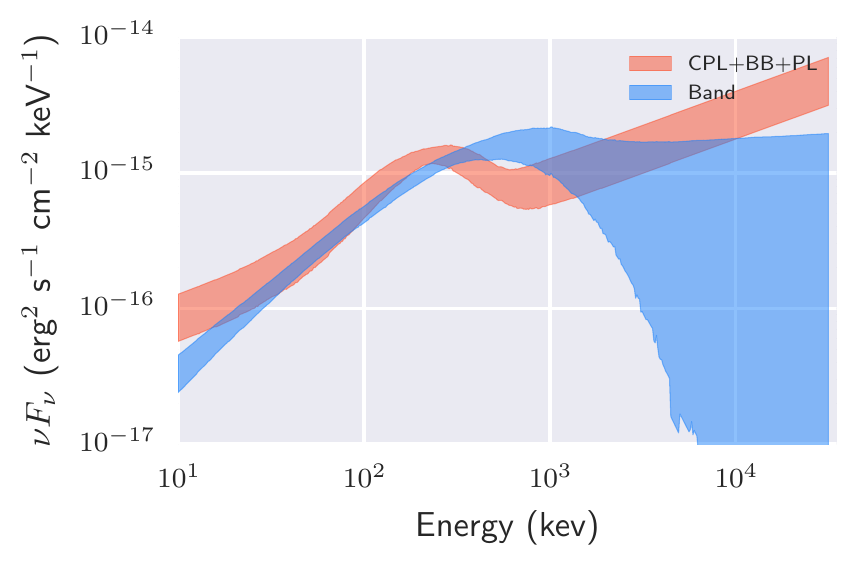}
  \caption{A $\vFv$ comparison between the Band and CPL+BB+PL for the interval T0+3.8-4.3. The width of the curves demonstrate the 95\% HPD marginal distributions including the error associated with location.}
  \label{fig:speccomp}
\end{figure}

\section{Comment on the case of GW150915-GBM}

The purported detection of electromagnetic radiation by GBM of LIGO event GW-150914 \citep{Connaughton:2016,Abbott:2016,Abbott:2016ki} was in contrast with upper limits from the INTEGRAL-SPI/ACS \citep{Savchenko:2016} and an independent analysis of the GBM data \citep{Greiner:2016}. While in \citet{Greiner:2016} we used an alternative statistical method to arrive at the conclusion that the data were statistically preferred to be background, another key difference to the analysis of \citet{Connaughton:2016} was the use of only two detectors (NaI 5 and BGO 0) in the analysis rather than all 14 GBM detectors. We have found in our analysis herein that the use of all NaI detectors can result in systematically off locations owing to possible deficiencies in the GBM DRMs. Here, we use the {\tt BALROG} with two data sets: NaI 5 and BGO 0 as well as all 12 NaI and 2 BGO detectors. We use a PL and CPL photon model for the spectral function as they were used in \citet{Connaughton:2016} to describe the spectra. 

\begin{figure*}
\centering
  \subfigure{\includegraphics[scale=.3]{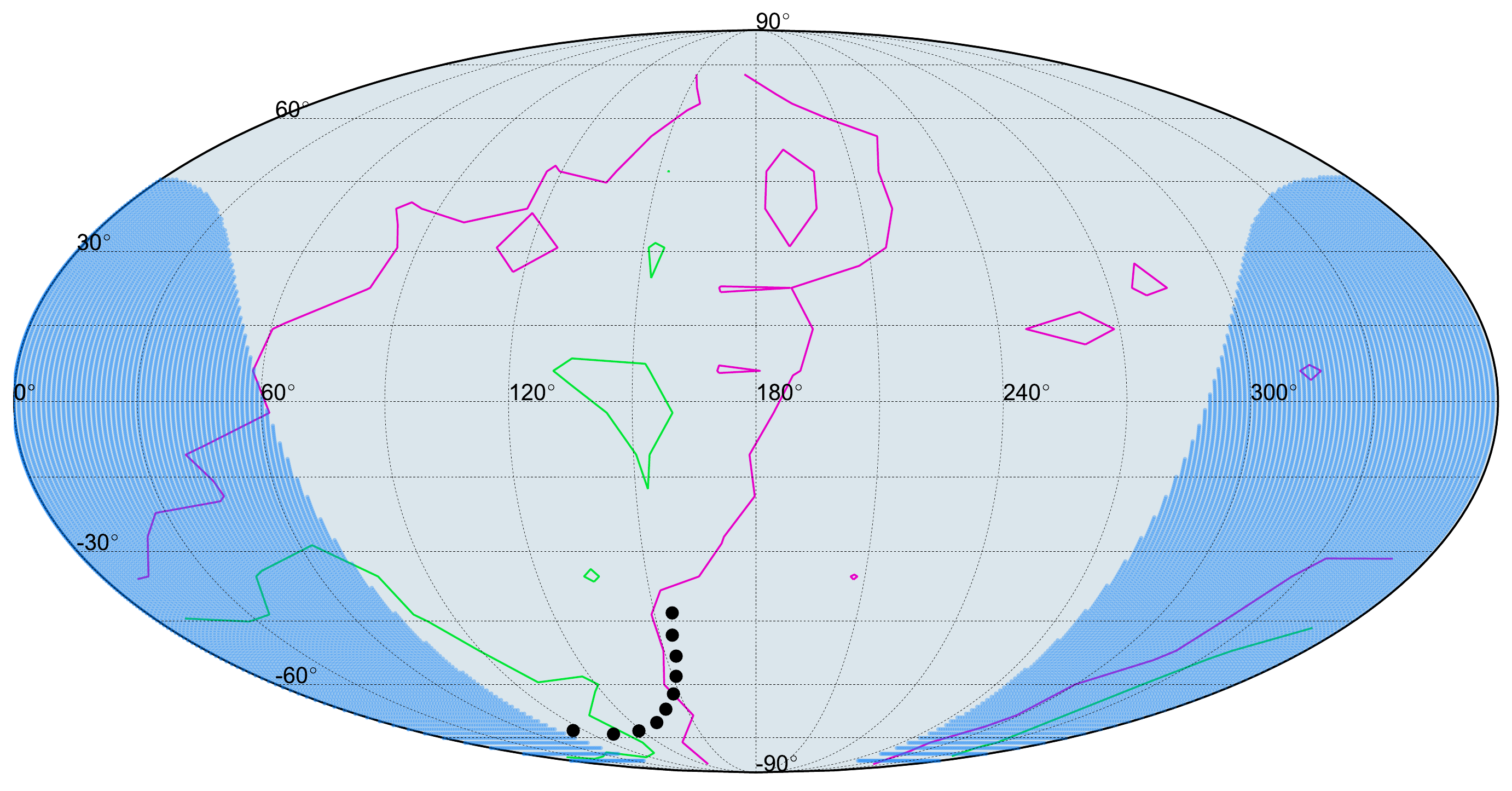}}\subfigure{\includegraphics[scale=.3]{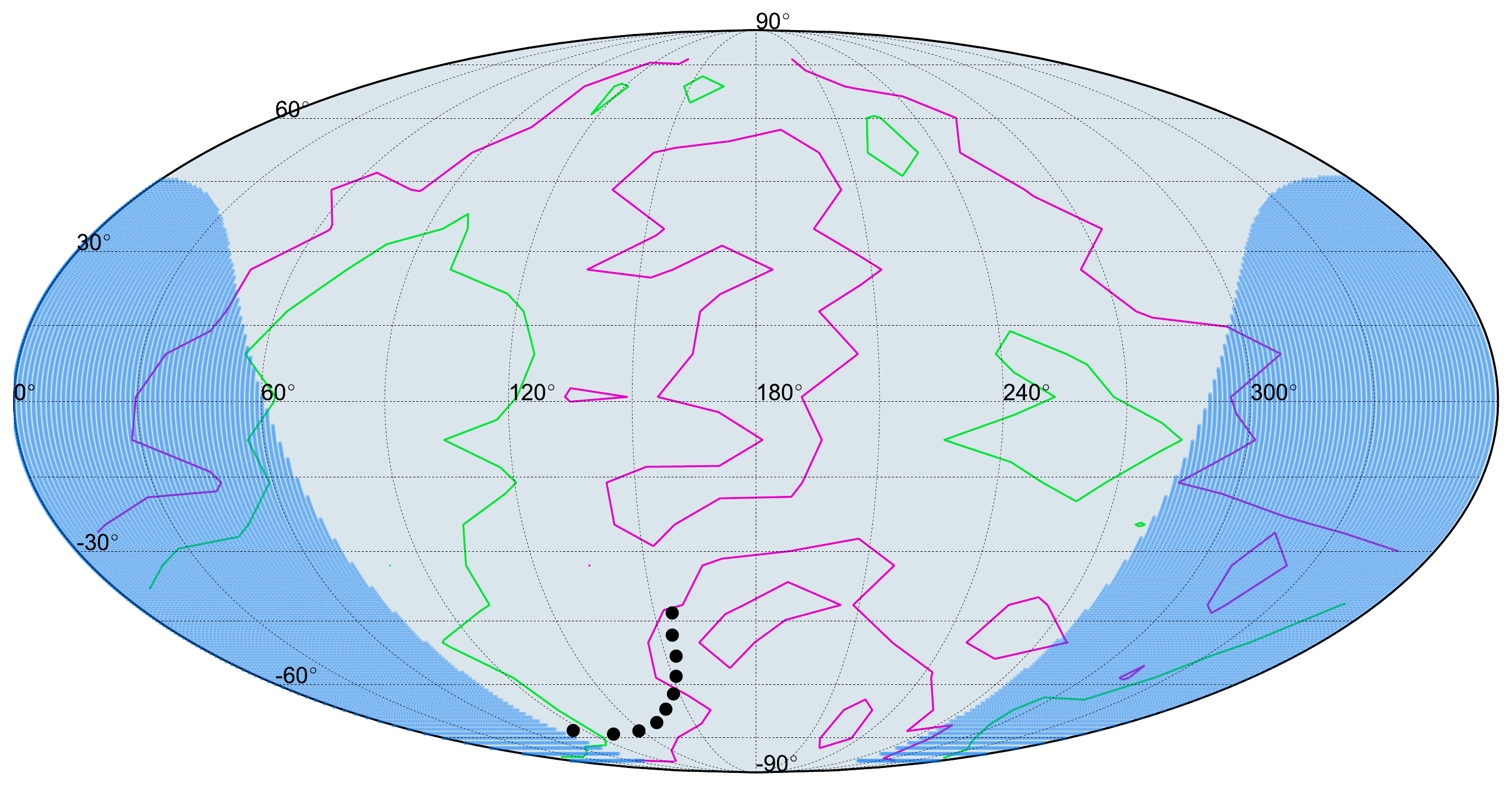}}
  \caption{The location of the event associated with GW-150914 all GBM detectors and a PL photon (left) and CPL photon model (right). The black points represent positions along the LIGO arc.}
  \label{fig:gw_all}
\end{figure*}

\begin{figure}
  \includegraphics[scale=1]{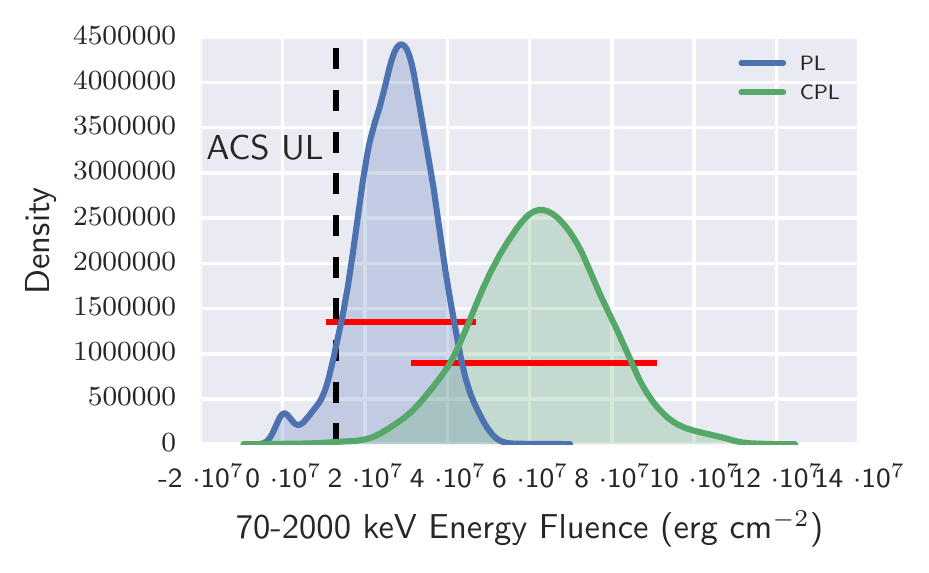}
  \caption{The marginalized fluence distributions resulting from fits using all GBM detectors to the event associated with GW-150914. The red lines indicate the 95\% HDP for each distribution. The distributions are at odds with the ACS upper limits \citep{Savchenko:2016}.}
  \label{fig:fluence}
\end{figure}

Figure \ref{fig:gw_all} shows the results of the analysis when using all detectors with both photon models. Both one 1 and 2$\sigma$ regions of the GBM location intersect the LIGO arc. Disregarding the fact that using all 14 GBM detectors has the potential to introduce systematics, we can compare the fluences derived from these fits with the upper limits derived in \citet{Savchenko:2016}. Marginalizing over the location, Figure \ref{fig:fluence} shows the marginal fluence distributions over the interval 70-2000 keV for both spectral functions. The PL function does not have a high-energy cut off and therefore has less believable limits though the 95\% HDP does intersect the ACS upper limit. The CPL does have a high-energy cut off and is clearly violating the ACS upper limits.

\begin{figure*}
\centering
  \subfigure{\includegraphics[scale=.3]{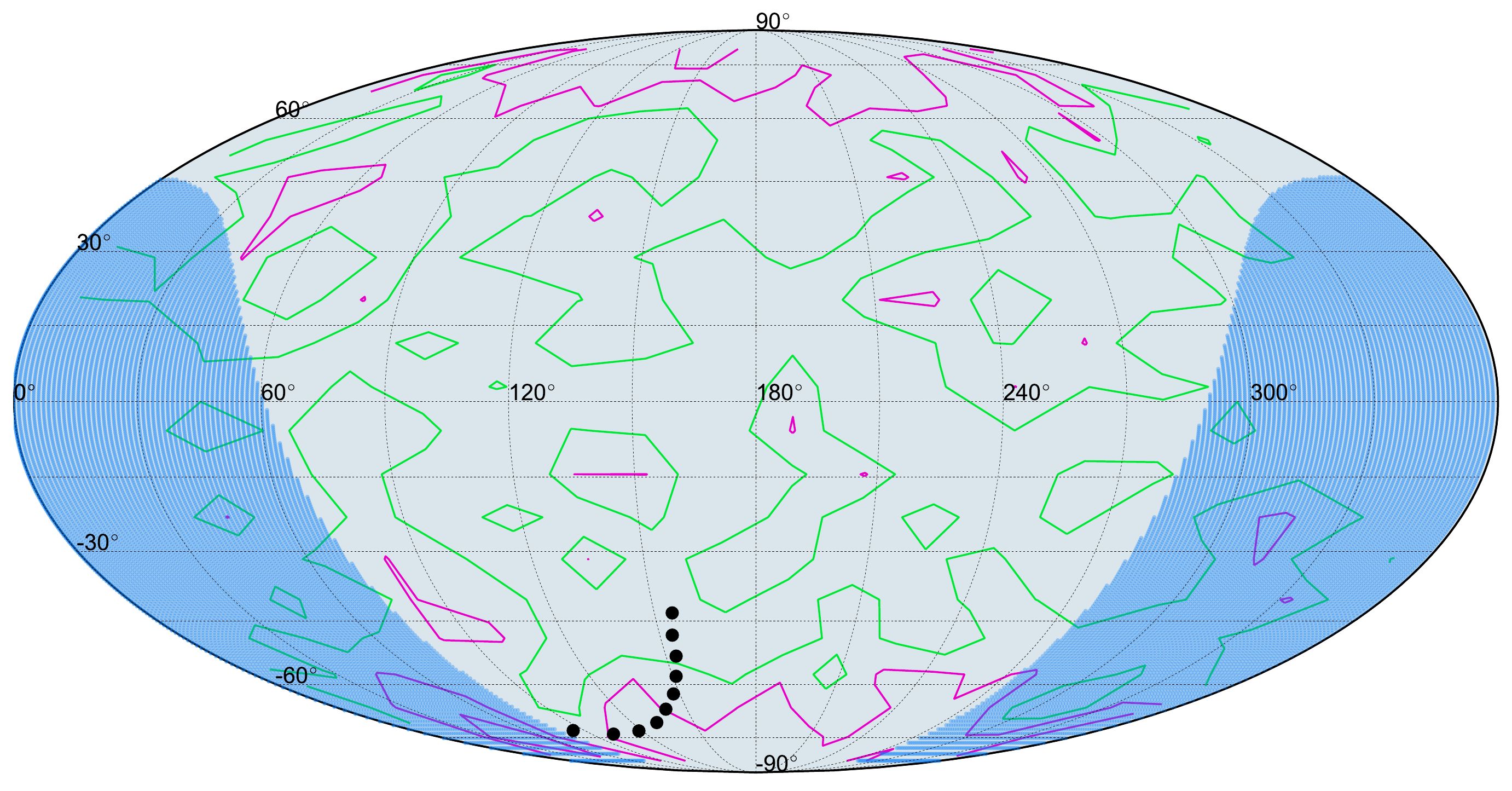}}\subfigure{\includegraphics[scale=.3]{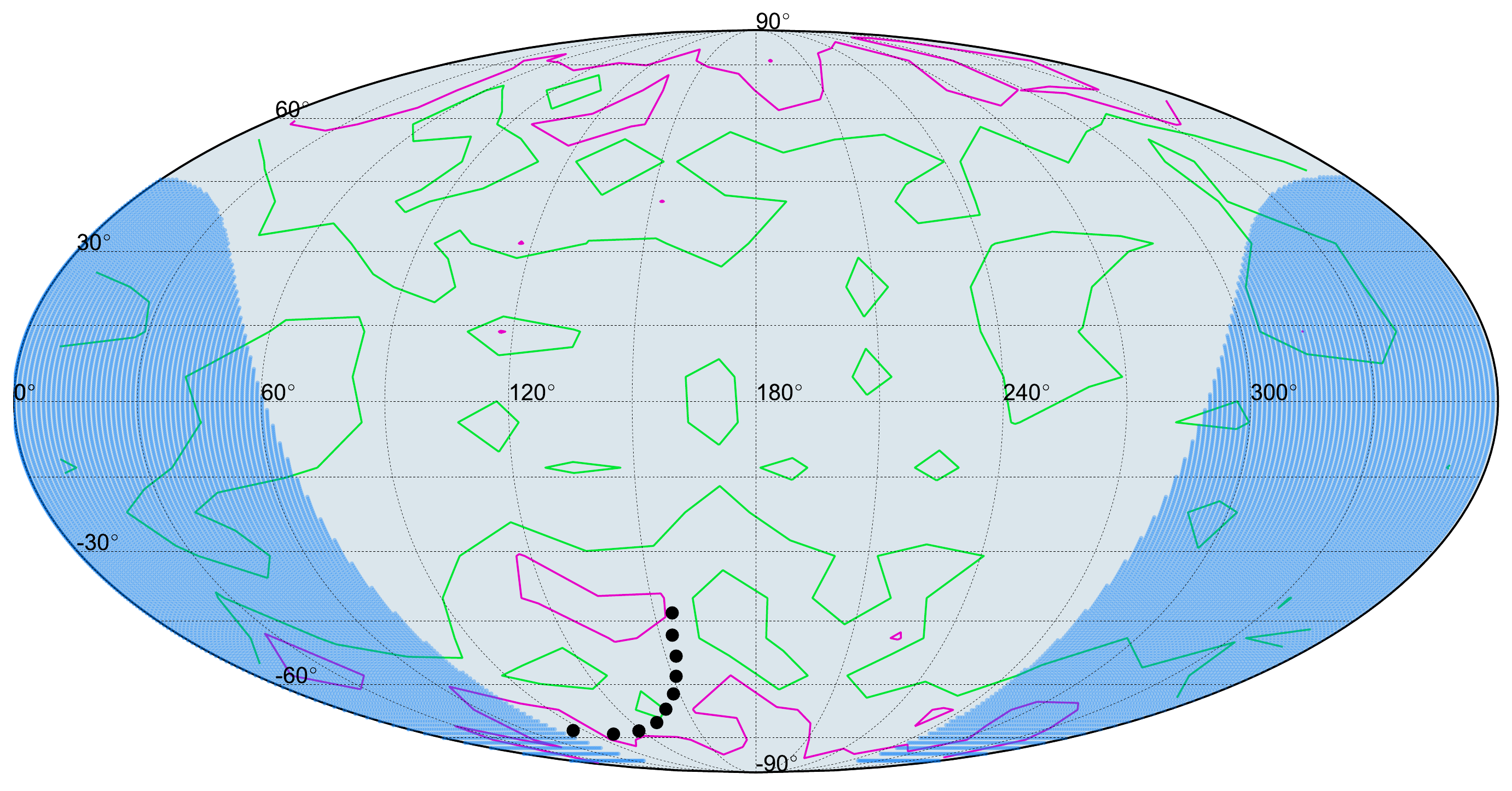}}
  \caption{The location of the event associated with GW-150914 only NaI 5 and BGO 0 and a PL photon (left) and CPL photon model (right). The black points represent positions along the LIGO arc.}
  \label{fig:gw_some}
\end{figure*}

 We examine the fits where only NaI 5 and BGO 0 are used. Figure \ref{fig:gw_some} shows the location contours resulting from these fits. The location is clearly unconstrained. These results combined with the non-detection results of \citet{Greiner:2016} point to a background (non-astrophysical) origin of the GBM counts data at the time of the GW-150914. In \citet{Greiner:2016}, we used two different methods to access the spectra with NaI 5 and BGO 0. We used a Poisson-Gaussian likelihood for fixed positions with maximum likelihood estimation and a Bayesian method where the background and source are fit simultaneously. Both methods found that the signal was consistent with background, the method presented here is similar to the maximum likelihood method of \citet{Greiner:2016} except that the location parameters are free. 

\begin{figure*}
\centering
  \subfigure{\includegraphics[scale=.3]{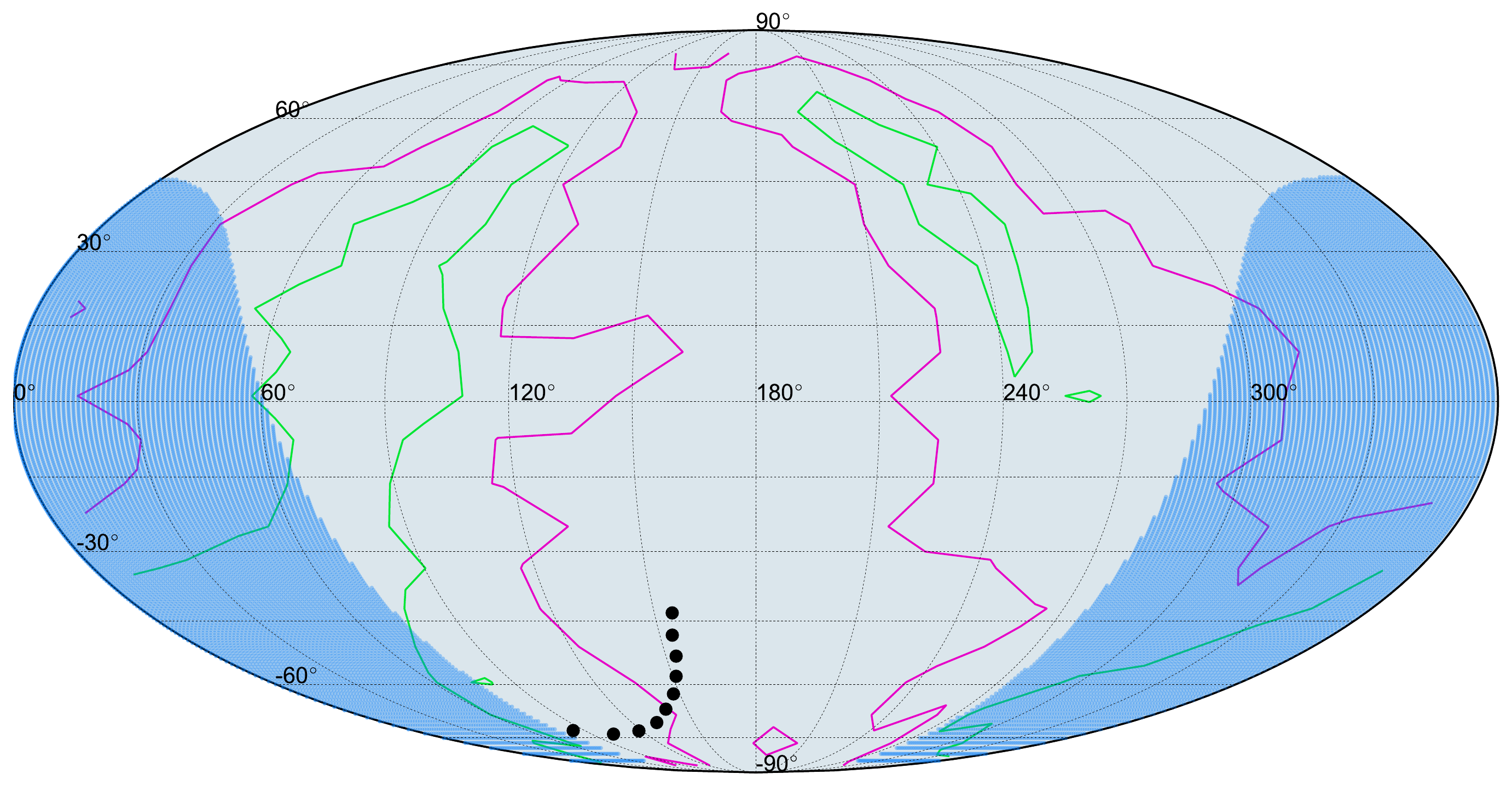}}\subfigure{\includegraphics[scale=.3]{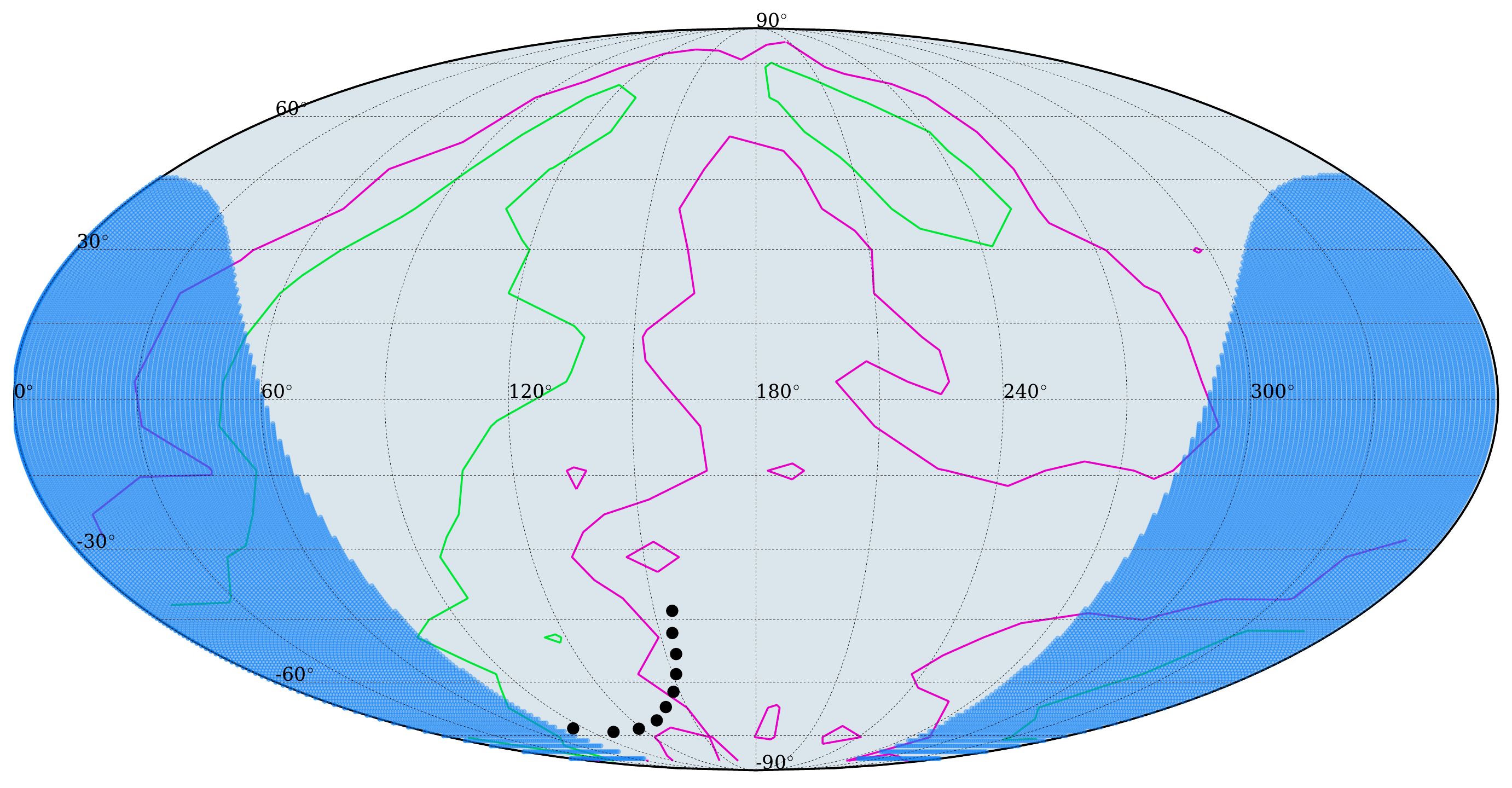}}
  \caption{The location of the event associated with GW-150914 all GBM detectors repeating the analysis of \citet{Veres:2016ve} where the we hold $\Ep=$ 1 MeV fixed (left) and $\alpha=-0.42$ fixed (right). The locations are marginally consistent with the LIGO arc at the 2$\sigma$ level.}
  \label{fig:gw_veres_map}
\end{figure*}

\begin{figure*}
\centering
	\includegraphics[scale=.6]{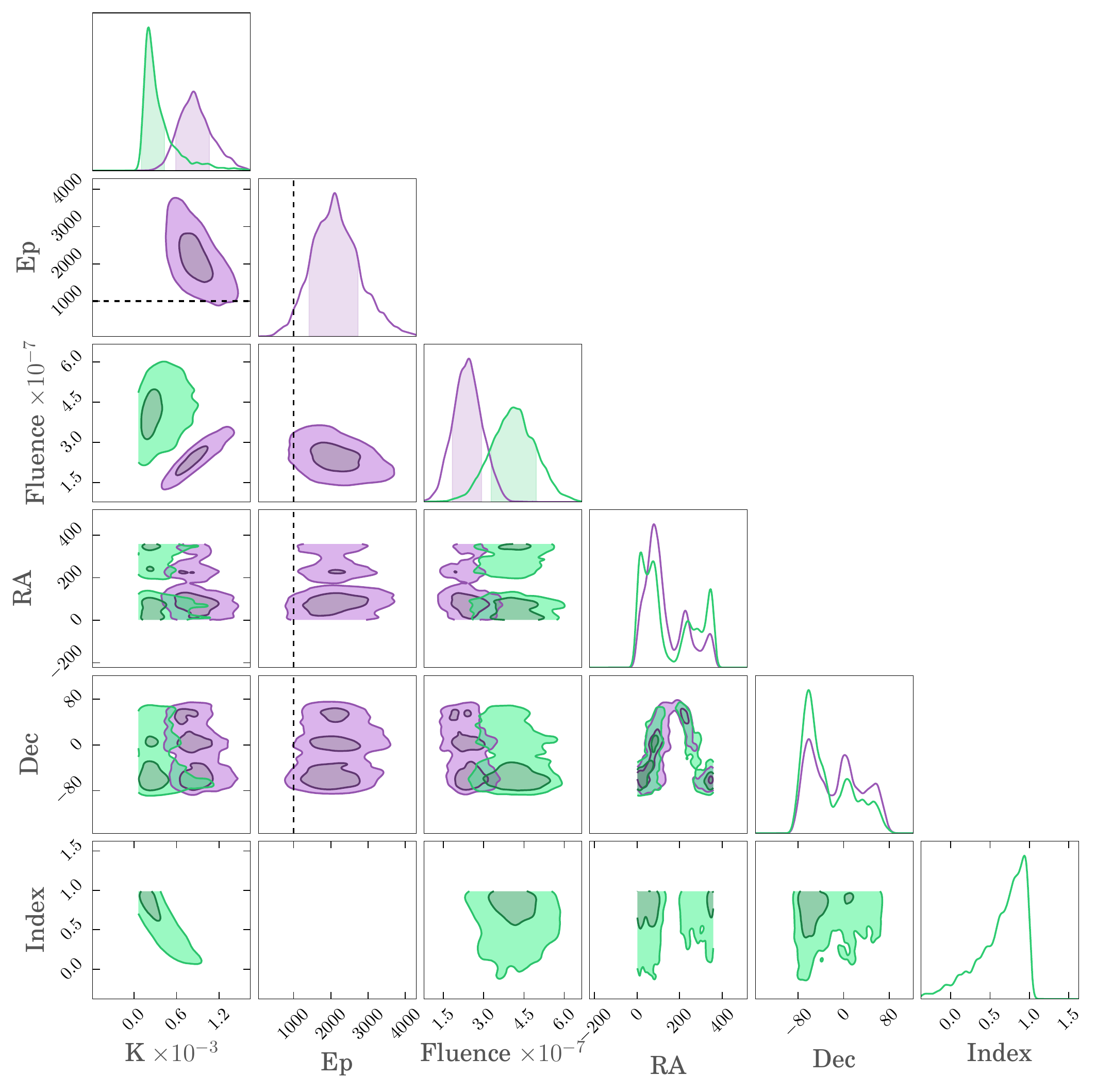}

   \caption{The marginal distributions of the fixed low-energy index (purple) and fixed $\Ep$ (green) from the repeated analysis of \citet{Veres:2016ve}. The dotted lines represent the fixed values of each fit which show that the two different fits are inconsistent with each other. The fluence (erg cm$^{-2}$) is integrated over 10-1000 keV.}
  \label{fig:gw_veres_param}
\end{figure*}

\begin{figure}
  \includegraphics{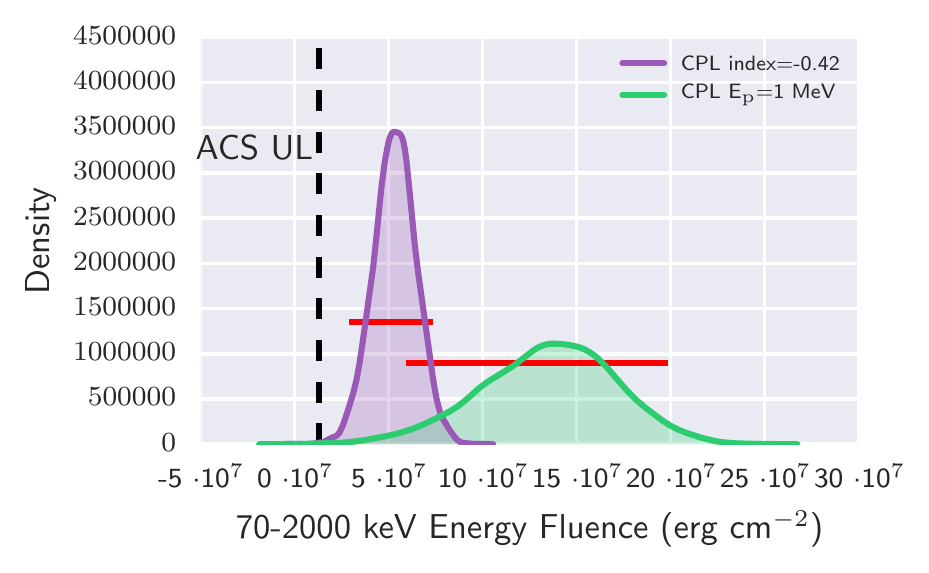}
  \caption{The fluences from the fits of the CPL models following the analysis of \citet{Veres:2016ve} compared with the ACS upper limit derived in in \citet{Savchenko:2016}.}
  \label{fig:veres_ul}
\end{figure}

In light of a new study by \citet{Veres:2016ve} where a CPL was fit to the 14 detector GBM data to understand the plausible physical GRB scenarios that could explain the signal. The authors use an a method which is not described to incorporate multiple locations along the LIGO arc and monte carlo the maximum likelihood profile of the parameters. Two sets of these monte carlos are created each hold either the $\vFv$ peak constant as 1 MeV or the low-energy index at $\alpha=-0.42$. We repeat this analysis to examine the marginal distributions under the {\tt BALROG} scheme. The locations of each fit are consistent with the LIGO arc at the $2\sigma$ level (See Figure \ref{fig:gw_veres_map}). we arrive at slightly different marginal distributions (Figure \ref{fig:gw_veres_param} ) most likely due to the fact that we fully marginalize over all locations, though it is difficult to compare the methods without a detailed description from \citet{Veres:2016ve}. Additionally, both fits violate the ACS upper-limits shown in Figure \ref{fig:veres_ul}.

\section{Discussion}

Localizing GRBs is the most important function of GBM in the era of multi-messenger astronomy. With the possibility of joint LIGO-GBM observations in the near future, GBM localizations have the potential to drastically constrain the broad localization arcs provided by LIGO leading to higher probability that imaging instruments can zero in on the location of a LIGO trigger. With this goal in mind, we have proposed a modern method to localize GBM GRBs that simultaneously fits the spectrum and location of GRBs via Bayesian nested sampling. The technique eliminates the systematics inherent in the standard GBM localization technique reported in \citet{Connaughton:2015} though some issues still remain that may be the result of poor spacecraft modeling or improper spectral model choice. Proper spacecraft modeling can be addressed with better simulations but proper model choice is a problem that plagues GRB spectroscopy for even the very wise cannot see all ends. In attempting to find the true location of GRBs with systematic offsets via {\tt BALROG}, we found that special care must be taken with both model and detector selection. The following methodology was developed:
\begin{enumerate}

  \item A subset of detectors with $5\sigma$ significance above background \citep{Li:1983} and less than 60$^{\circ}$ angular separation from the GBM flight software position are initially used 
  \item once a location has been determined, it is checked that the detectors selected view the final location. If not, the selection is modified and repeated
  \item this processes is repeated while also iterating model choice until the highest marginal likelihood is found which resulted in the best location. 
\end{enumerate}
\noindent
We emphasize that the fact that proper location depends heavily on detector selection is indicative of 
deficiencies in the detector response. This could be due to various effects, among others
(i) the $>$60$^{\circ}$ off-axis response of the individual detectors has some calibration uncertainties,
(ii) the Bayesian fitting scheme does not perfectly cover the noise-only detectors,
(iii) the simulations include deficiencies because they are based on imperfect mass models of the detectors and the spacecraft.
The effect of these deficiencies has been noted already at the pre-launch source survey,
when the full Fermi spacecraft with all GBM detectors mounted
was irradiated with a radioactive source. The simulations were never
able to reproduce the measurements, in particular for detectors which
are at large angles relative to the source beam. Since these mismatches did not occur
for the calibration measurements of individual detectors, we presently suspect the above effect (iii)
to be the dominant.
A deeper investigation into these deficiencies is beyond the scope of this work, but may alleviate the need for careful detector selection which currently makes using the {\tt BALROG} difficult in realtime follow-up. It is worth noting that this study only analyzes a small fraction of the GRBs listed in \citet{Connaughton:2015}. However, these are the tail cases and {\tt BALROG} performs well in both. It is conceivable that for the median of the distribution, {\tt BALROG} will perform equally well. We advocate for a deeper study to access this, but feel that the introduction of the concept herein provides the required impetus to modernize GRB location algorithms. It is also worth restating that this method is a superset of the original DOL method.

{\tt BALROG} is computationally expensive and fast locations will require its implementation on high-performance computing clusters. A standard location on TRIGDAT data takes $\sim 1-5$ minutes on a 48-core node without hyper-threading. The main computational bottleneck is the DRM generation. In the future, it may be possible to pre-compute a finer base DRM grid, but the multiplicity of spacecraft orientations that require recomputing of the atmospheric scattering will make such a task difficult. 
	
A secondary finding of this study is the effect of including the location errors in spectra. The spectral shape of GRB 080916C for both time-integrated and time-resolved analysis favor a simple Band function rather than multiple components. The new model proposed in \citet{Guiriec:2015} is not favored by  and results in incorrect locations due to its fixed low-energy slope. To reject a model in this way, it is required that a GRB is localized externally by an instrument such as  Swift. In the future, optical/radio counterparts will be much
 easier to identify given the few square-degree error regions as compared
 to the hundred square-degree error regions (where afterglow identification was possible 
 already in about 50\% of the cases \citep{Lipunov:2015bv,Singer-L.-P.:2015}.  While these findings focus on a single GRB, they provide impetus to study these effects on a broader sample which will be the subject of a future work. We reiterate that this opens a new path for GRB model selection. While the Band function's flexibility is likely free enough to allow for proper location, the push for using physical models in GRB spectroscopy which have less flexibility will allow for quick rejection of these models if they result in inaccurate locations. 

The systematics identified when using all NaI detectors raises further issues associated with the claims of \citet{Greiner:2016} regarding the purported signal associated with GW-150914 \citep{Connaughton:2016}. While including all NaI detectors results in a location that touches the LIGO arc, it also includes most of the Earth. When using a detector selection that corresponds to the position of the LIGO, the location contours encompass the entire sky. This in combination with the lack of signal found in \citet{Greiner:2016} place serious constraints on the astrophysical origins of the event. Furthermore, the {\tt BALROG} can be used for future LIGO events as it provides accurate locations for GBM transients that can be employed in optical follow-up strategies.

With these findings, we advocate for adopting the {\tt BALROG} for the location of GBM GRBs as it provides realistic and accurate locations suitable for followup strategies. Additionally, the shifts in spectral properties found in GRB 080916C warrant an adoption of this scheme for spectral analysis of GBM GRBs but could also be used for similar instruments. We will investigate the GBM GRB catalogs in a forthcoming publication. Such a scheme could be implemented in modern analysis tools such as the Multi-Mission Maximum Likelihood Framework (3ML) \citep{Vianello:2015} which would allow the {\tt BALROG} to be used to fit data simultaneously with other instruments while still incorporating the uncertainty in location. 
	  
\section*{Acknowledgements}

This work made use of several open source Python packages and we are thankful to the authors of matplotlib \citep{Hunter:2007}, astropy \citep{astropy}, seaborn \citep{michael_waskom_2016_45133}, and pymultinest \citep{Buchner:2014}. The authors are thankful to the authors of the original GBM DRM generation code which helped in designing {\tt BALROG} as well as to  Gudlaugur J{\'o}hannesson for insight into making the code more efficient. Additional statistical discussion with  Brandon Anderson and Giacomo Vianello helped steer this work in the right direction. We thank Andreas von Kienlin (MPE) and Marc Kippen (Los Alamos) for discussion on the GBM response calibration and simulation.

\bibliographystyle{mn2e}
\bibliography{bib}

\appendix

\section{Location plots}
\label{sec:appendix}

\begin{figure*}
\centering
  \subfigure{\includegraphics[scale=.4]{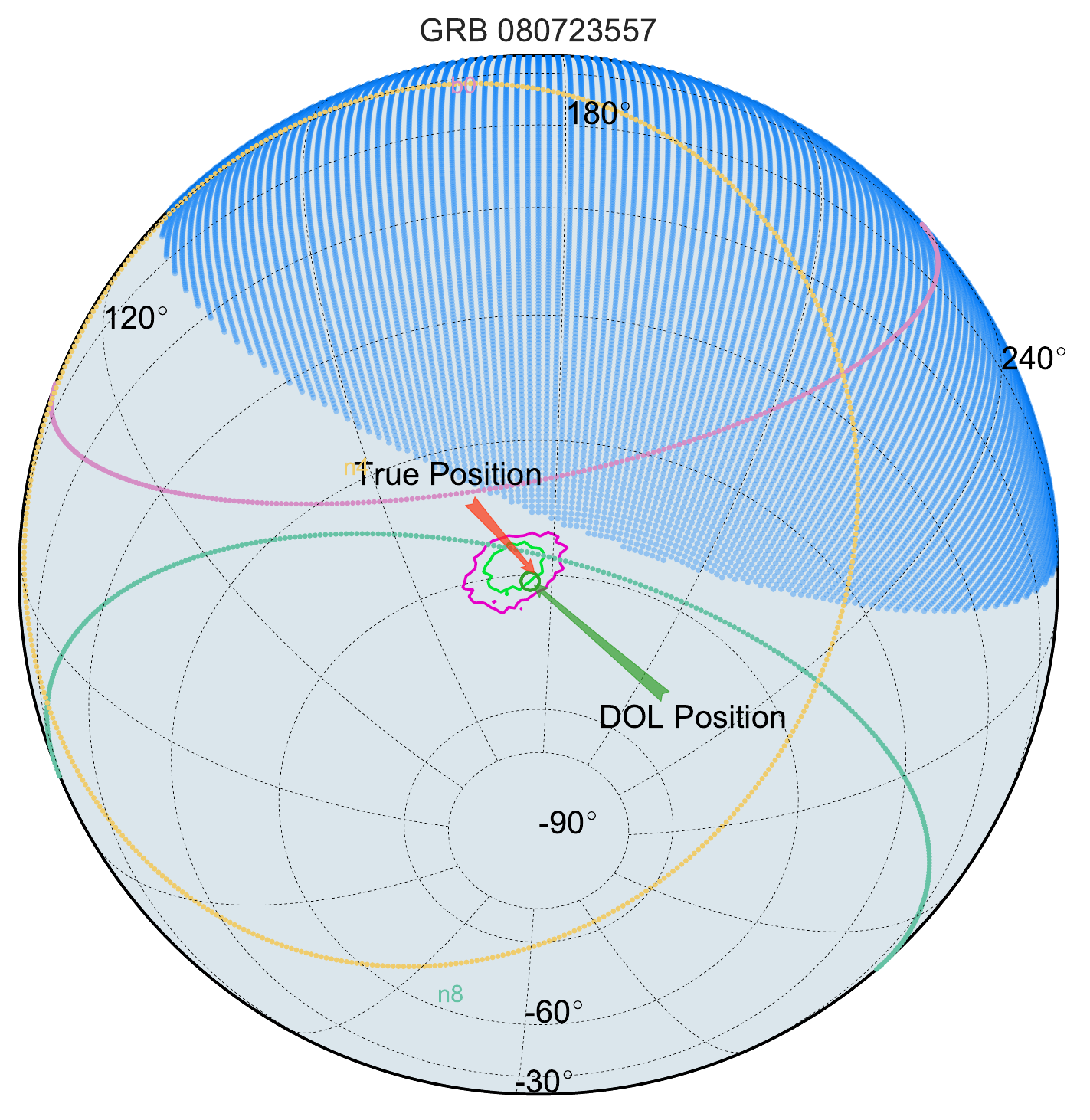}}\subfigure{\includegraphics[scale=.4]{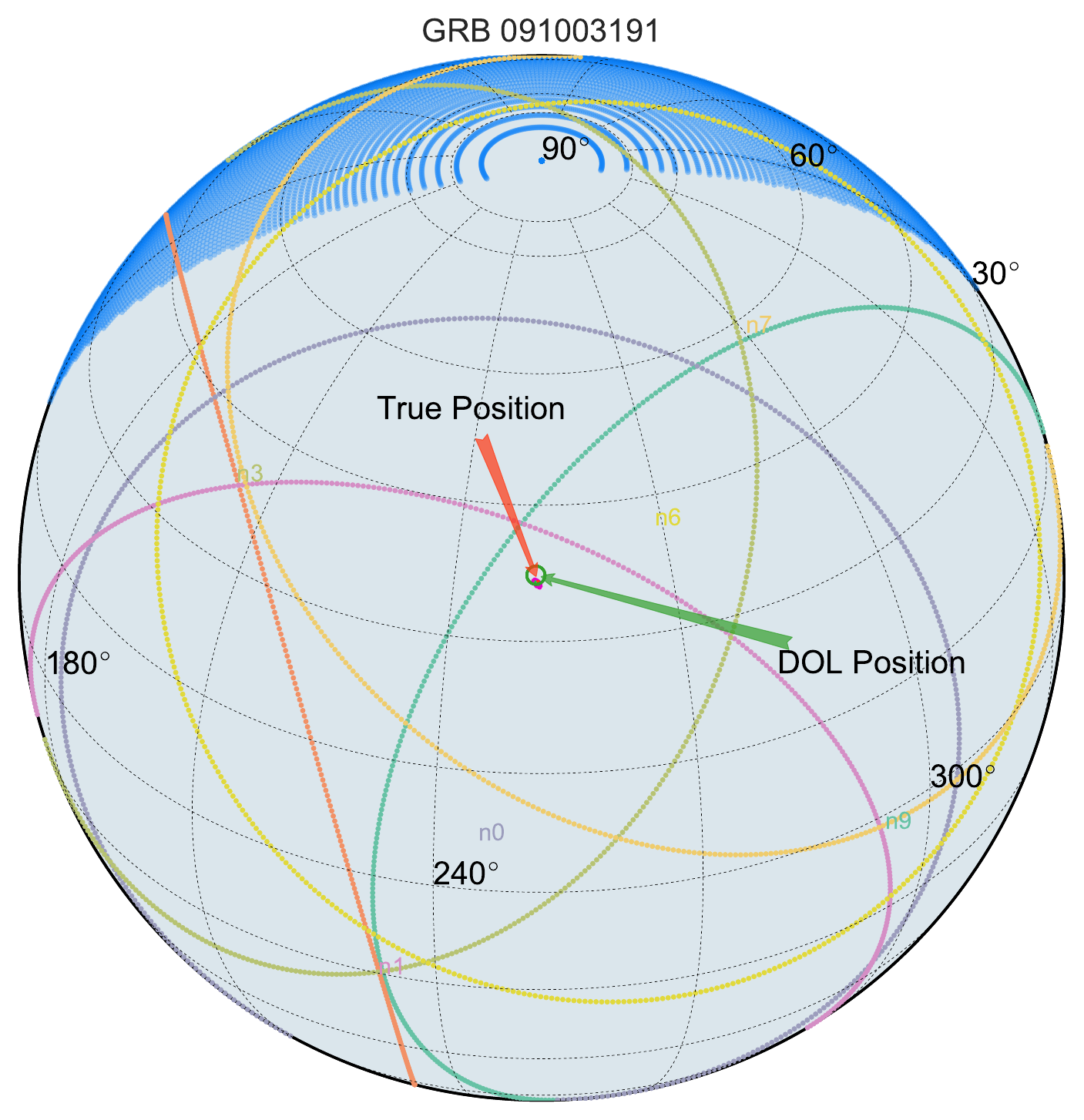}}\subfigure{\includegraphics[scale=.4]{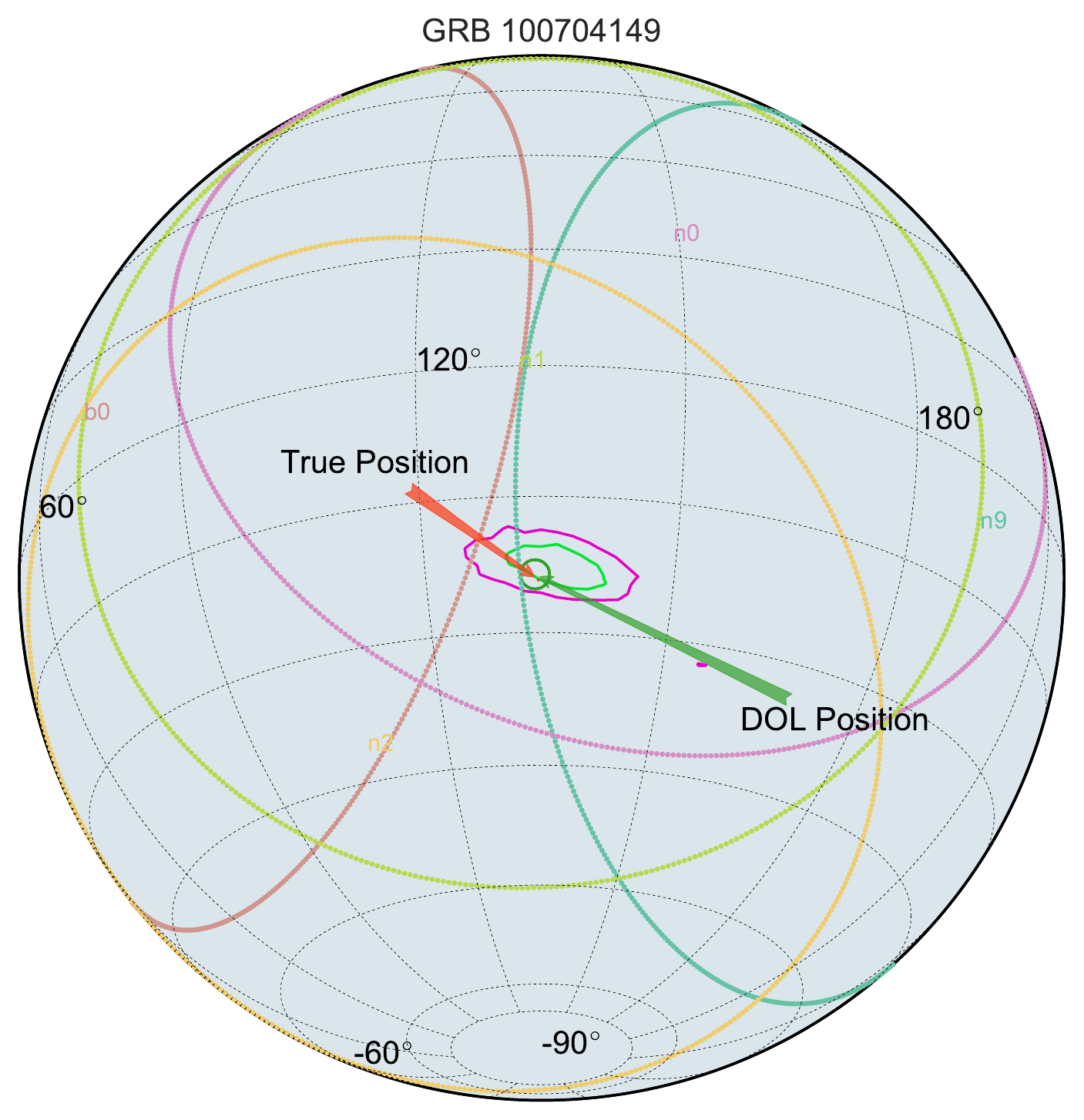}}
  \subfigure{\includegraphics[scale=.4]{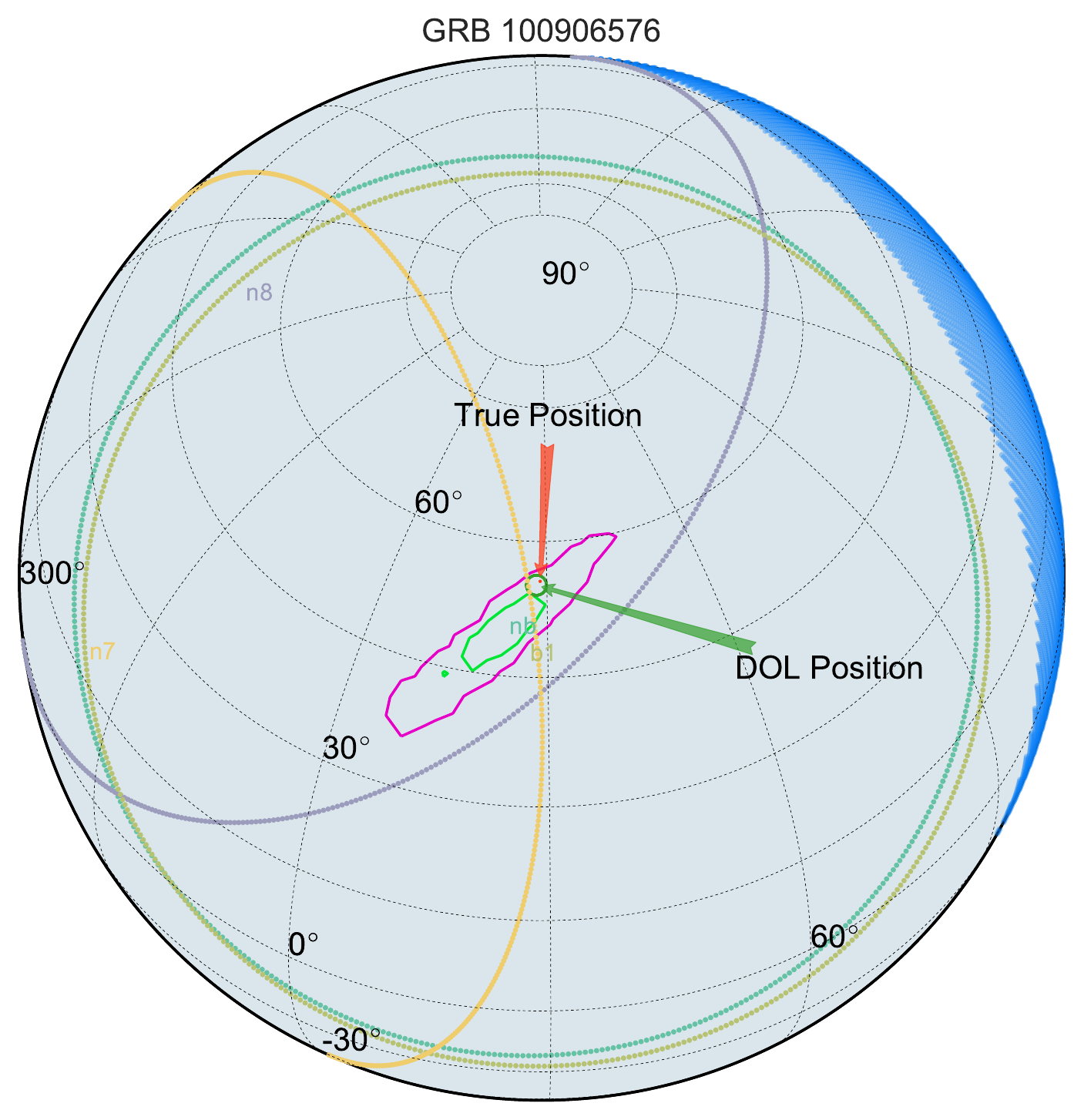}}\subfigure{\includegraphics[scale=.4]{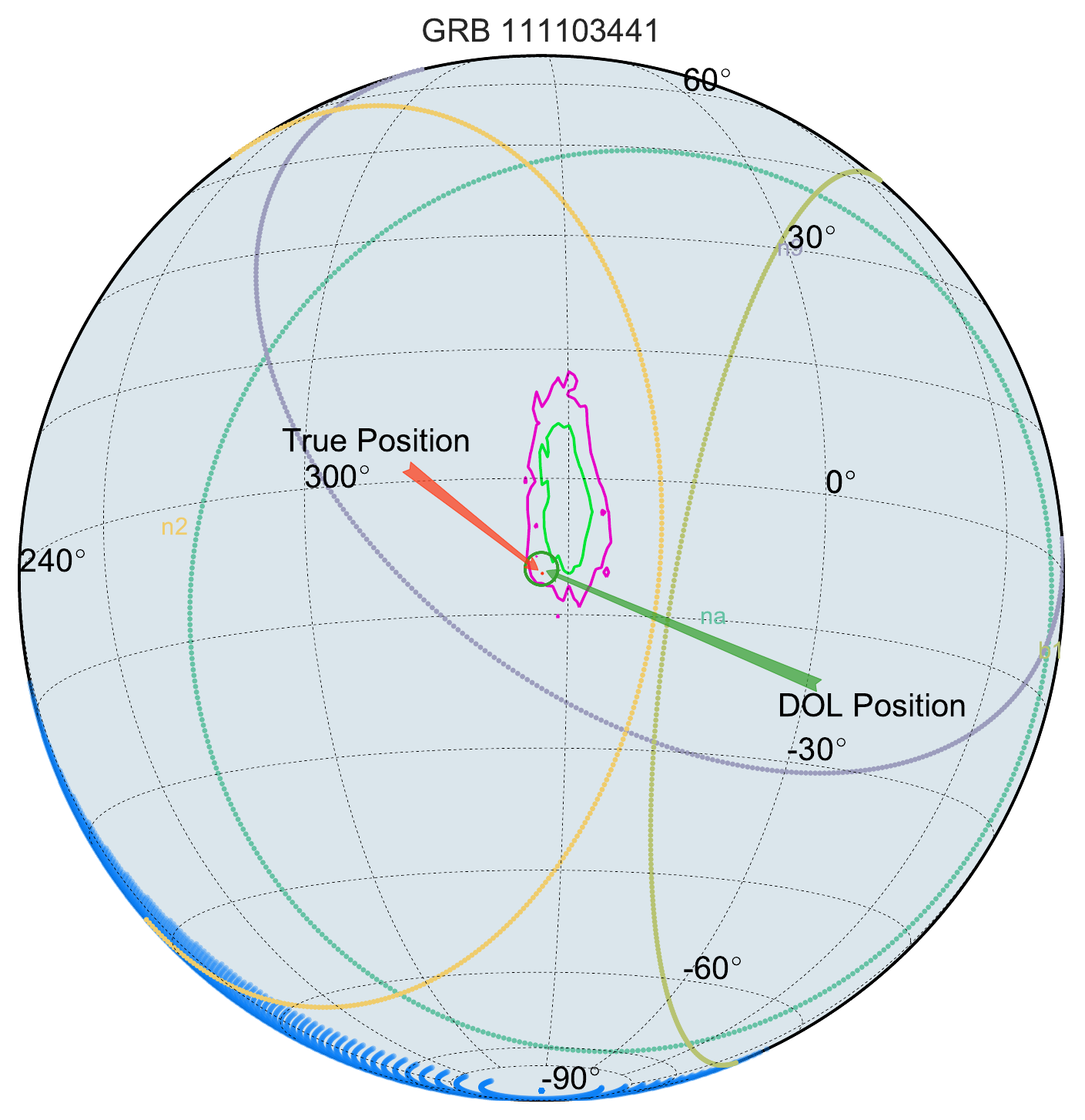}}\subfigure{\includegraphics[scale=.4]{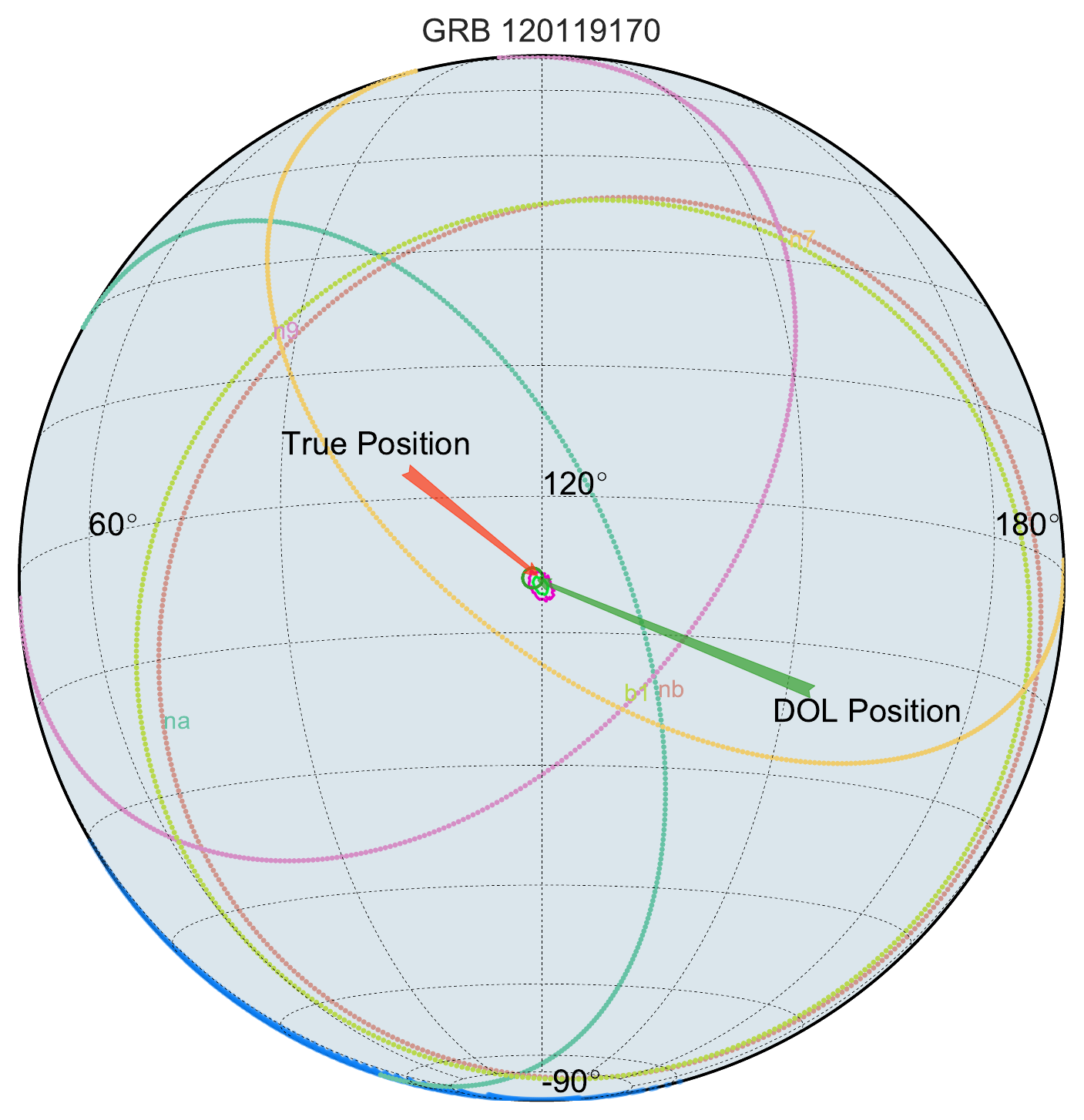}}
  \subfigure{\includegraphics[scale=.4]{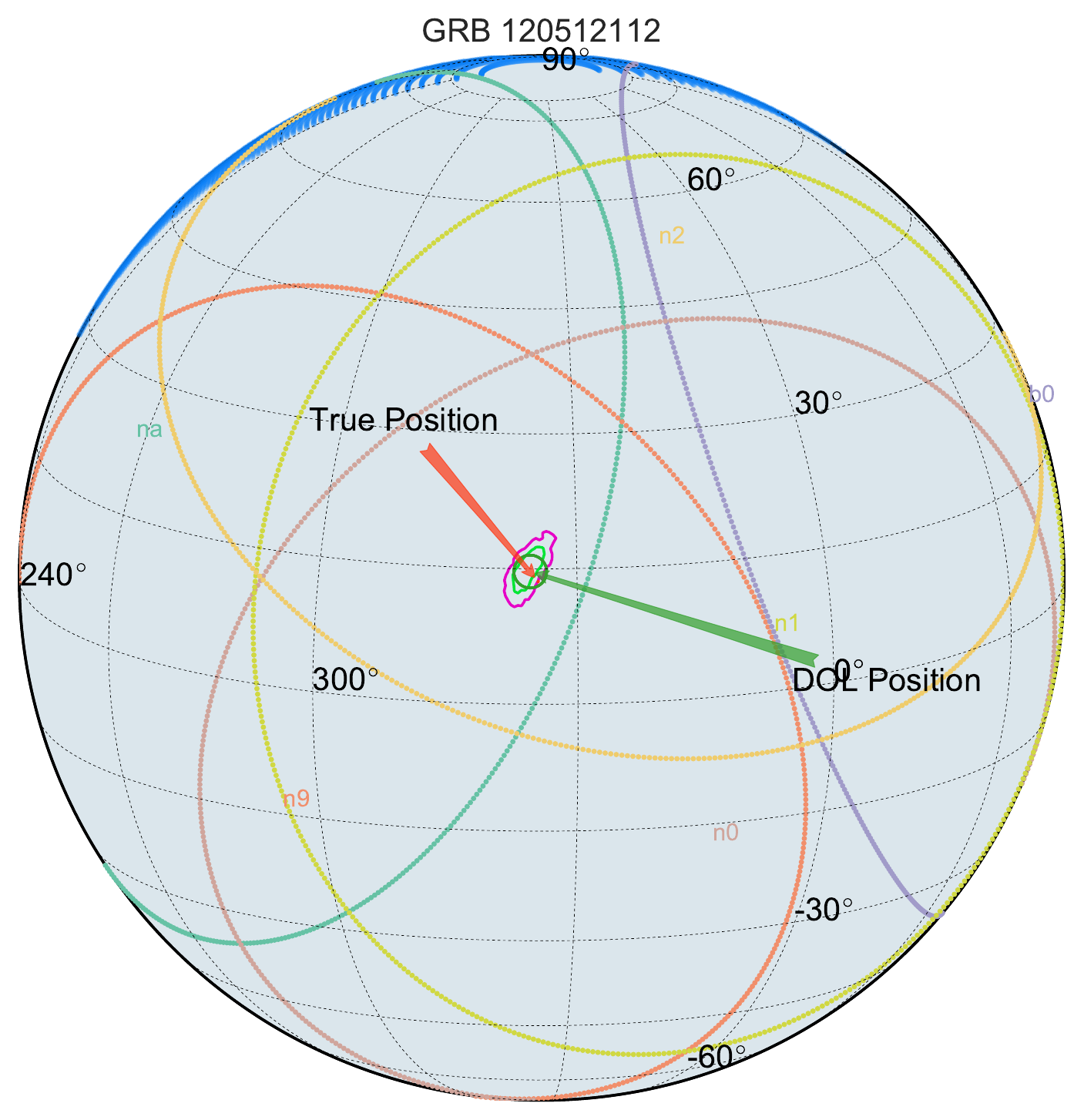}}\subfigure{\includegraphics[scale=.4]{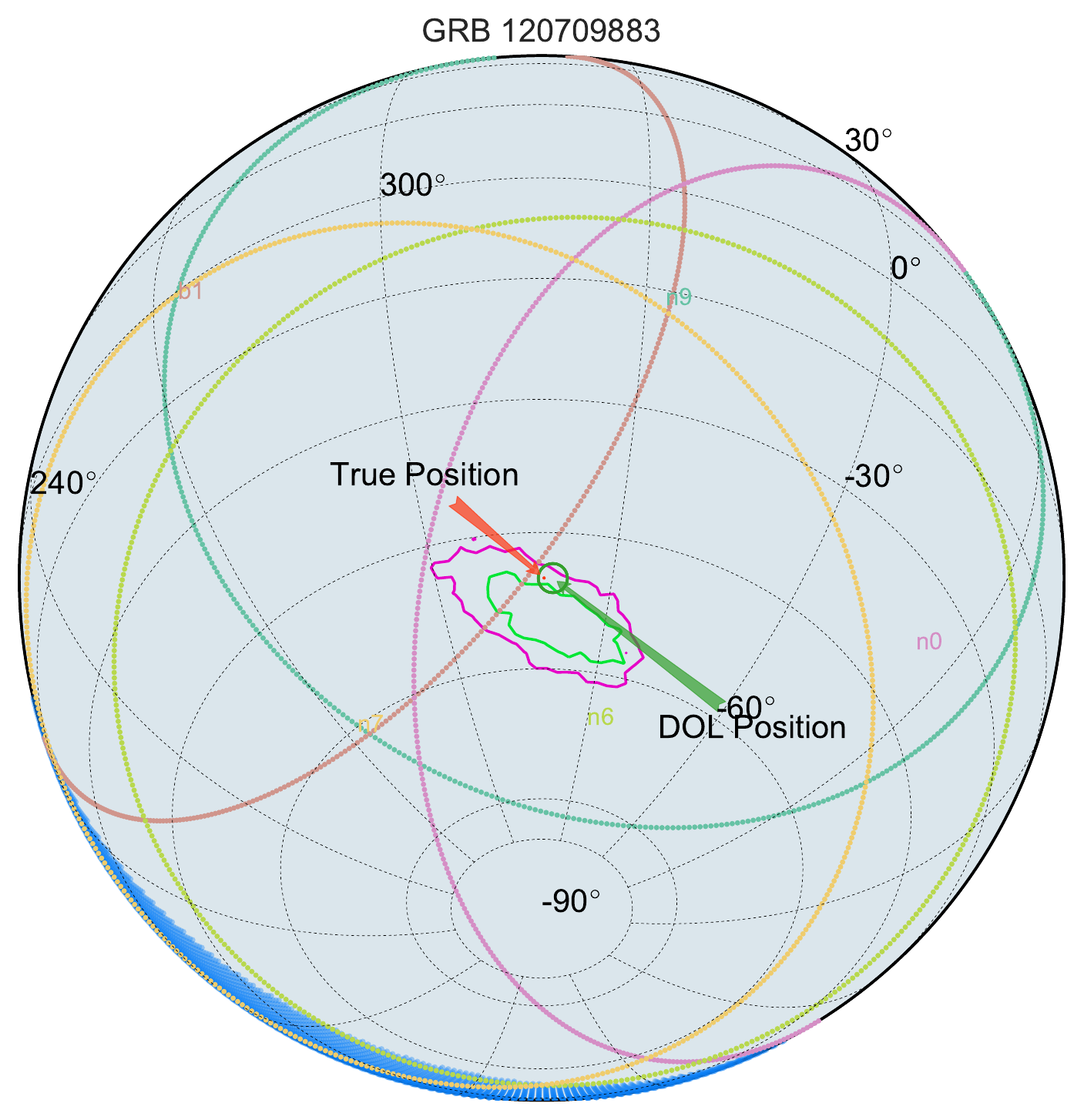}}\subfigure{\includegraphics[scale=.4]{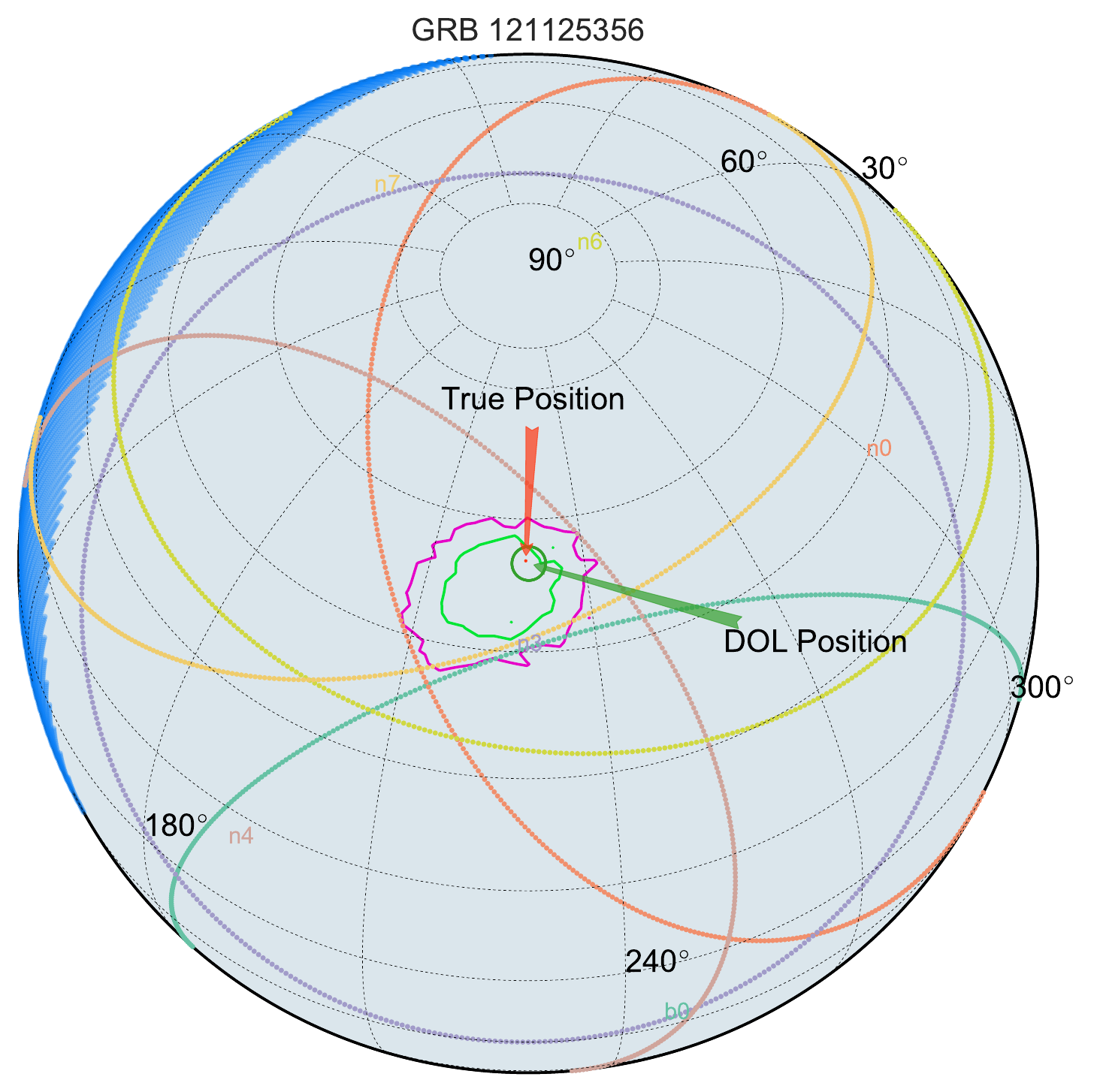}}
  \caption{The plots of the remaining nine GRBs well located by the DOL and {\tt BALROG}.}
  \label{fig:all_good}
\end{figure*}

\begin{figure*}
\centering
  \subfigure{\includegraphics[scale=.4]{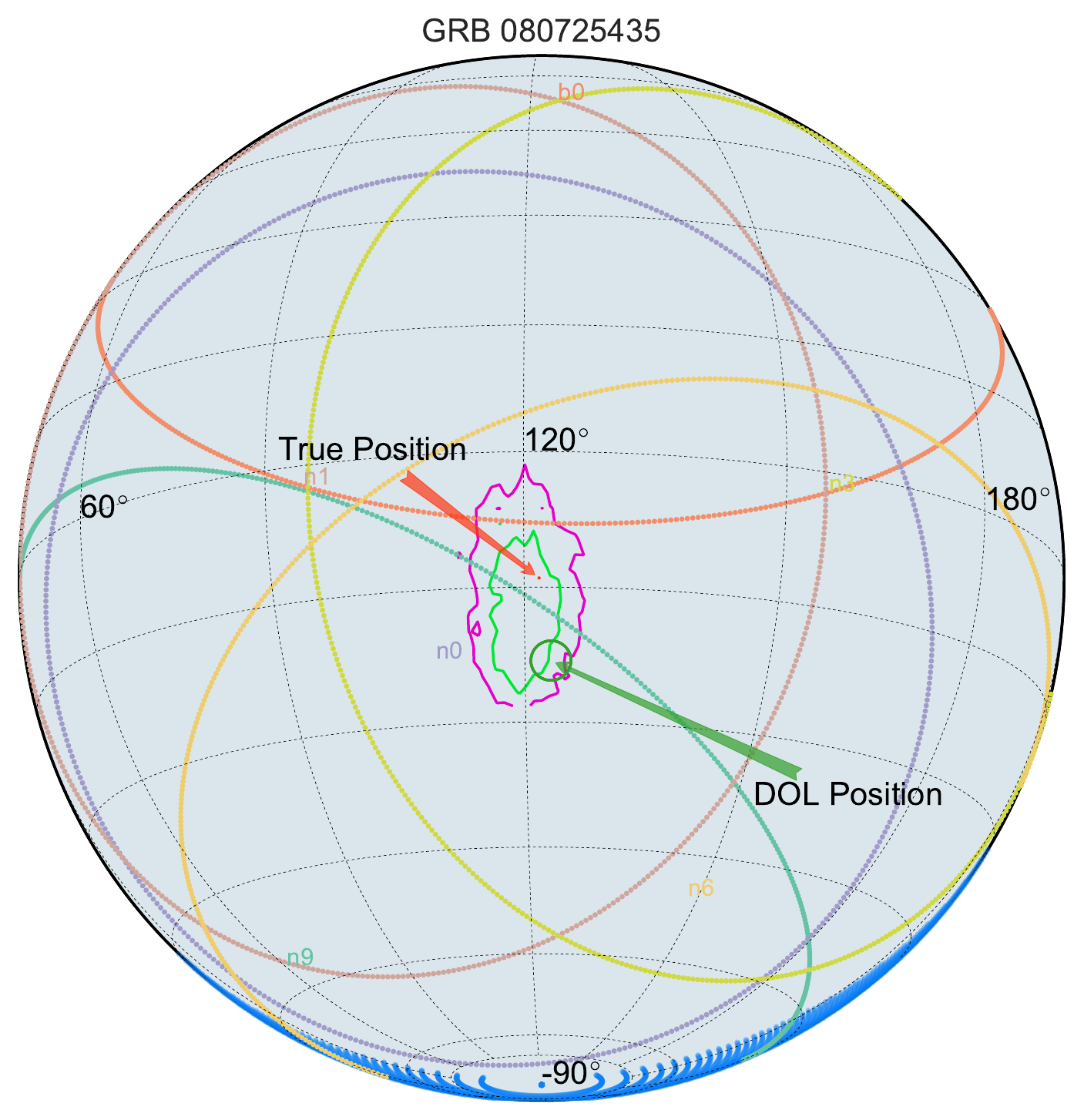}}\subfigure{\includegraphics[scale=.4]{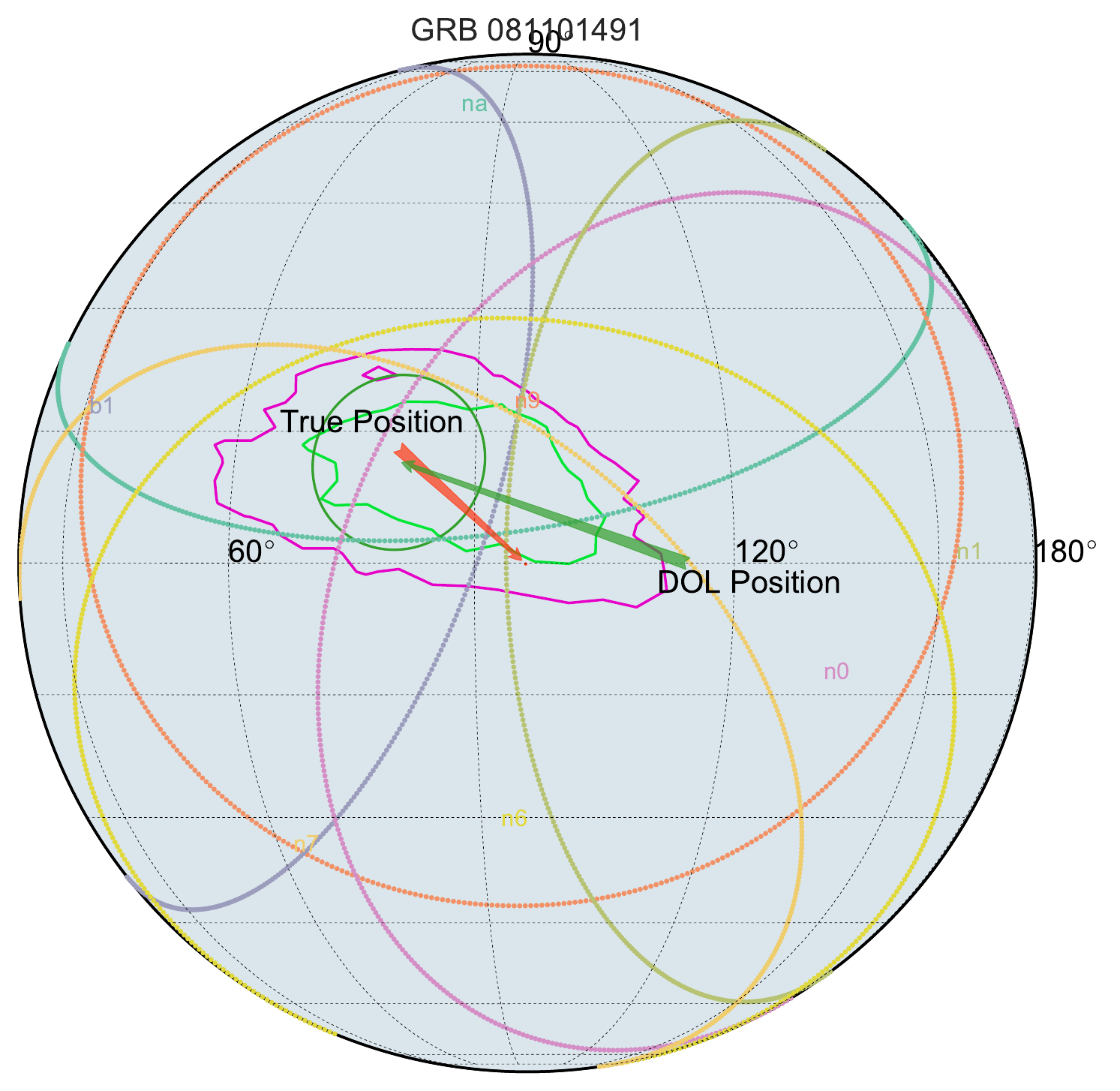}}\subfigure{\includegraphics[scale=.4]{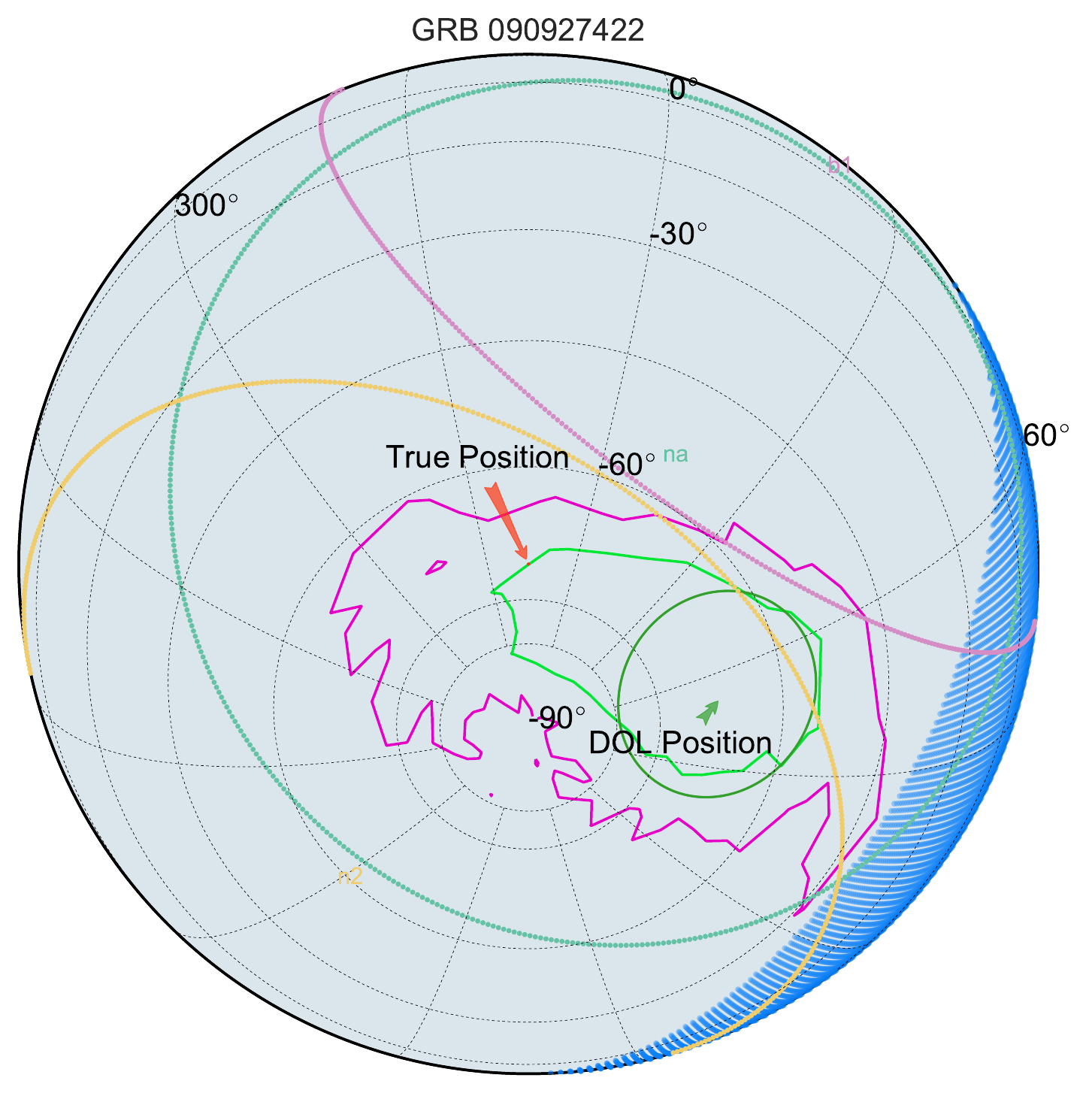}}
  \subfigure{\includegraphics[scale=.4]{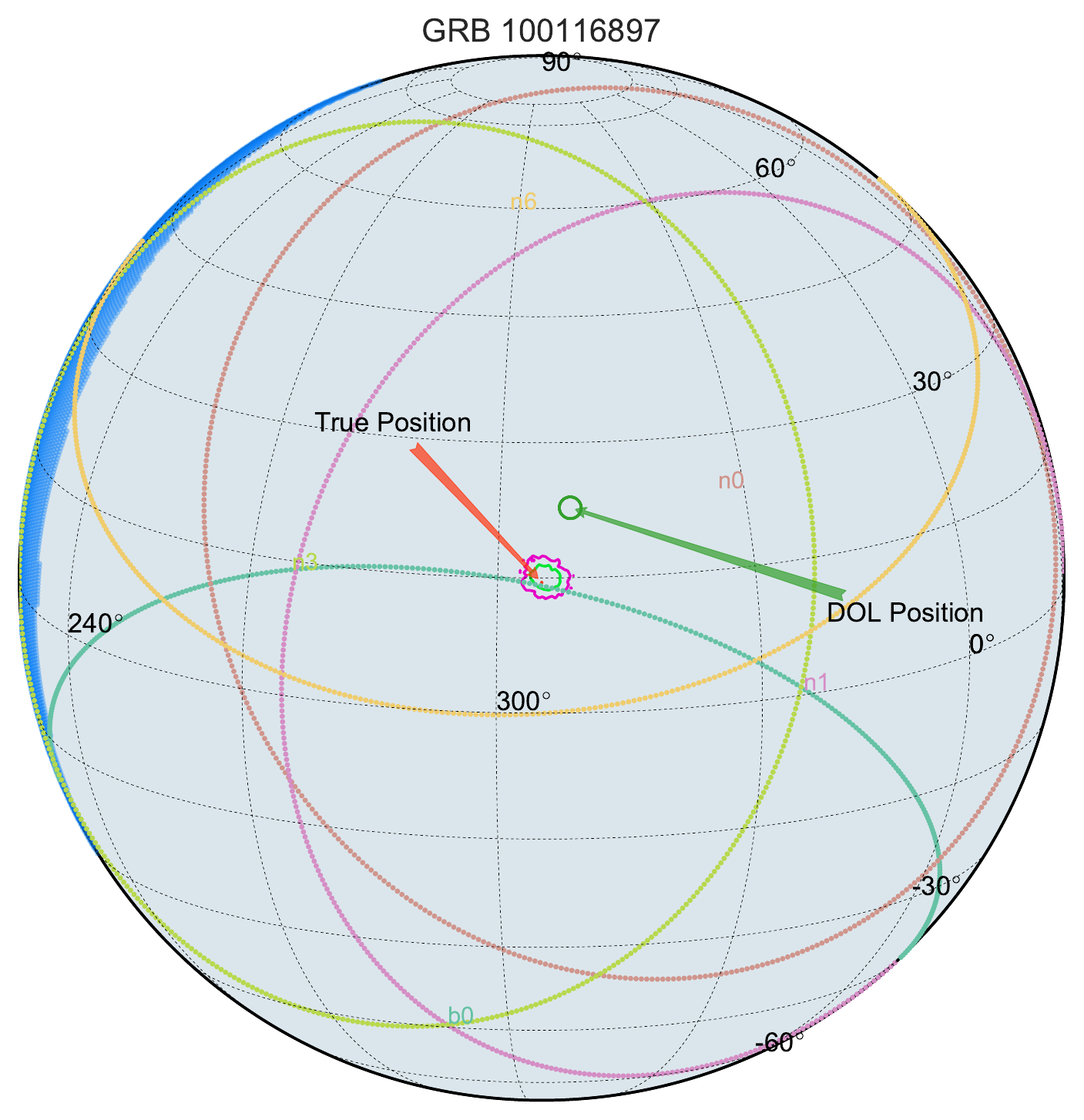}}\subfigure{\includegraphics[scale=.4]{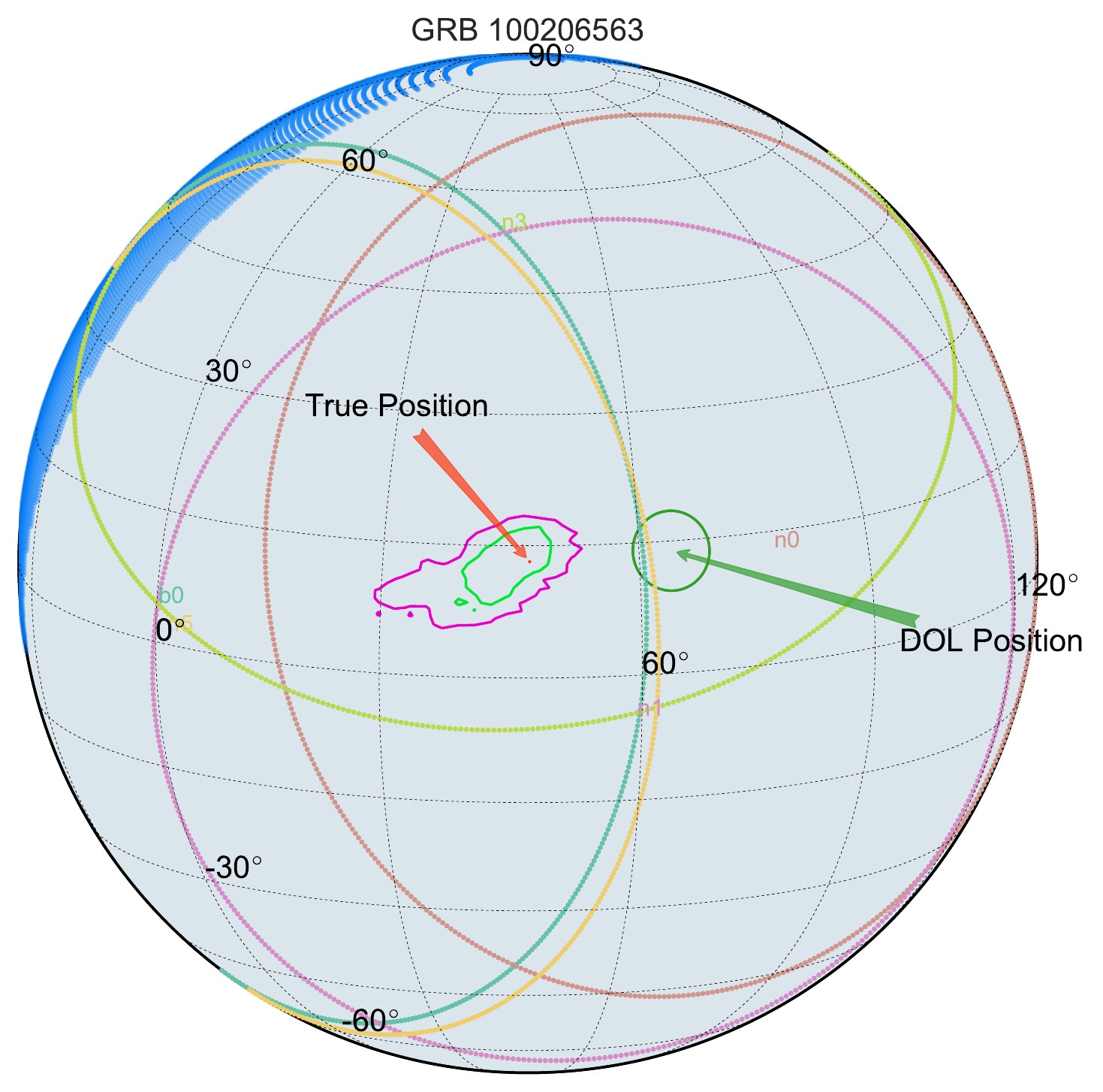}}\subfigure{\includegraphics[scale=.4]{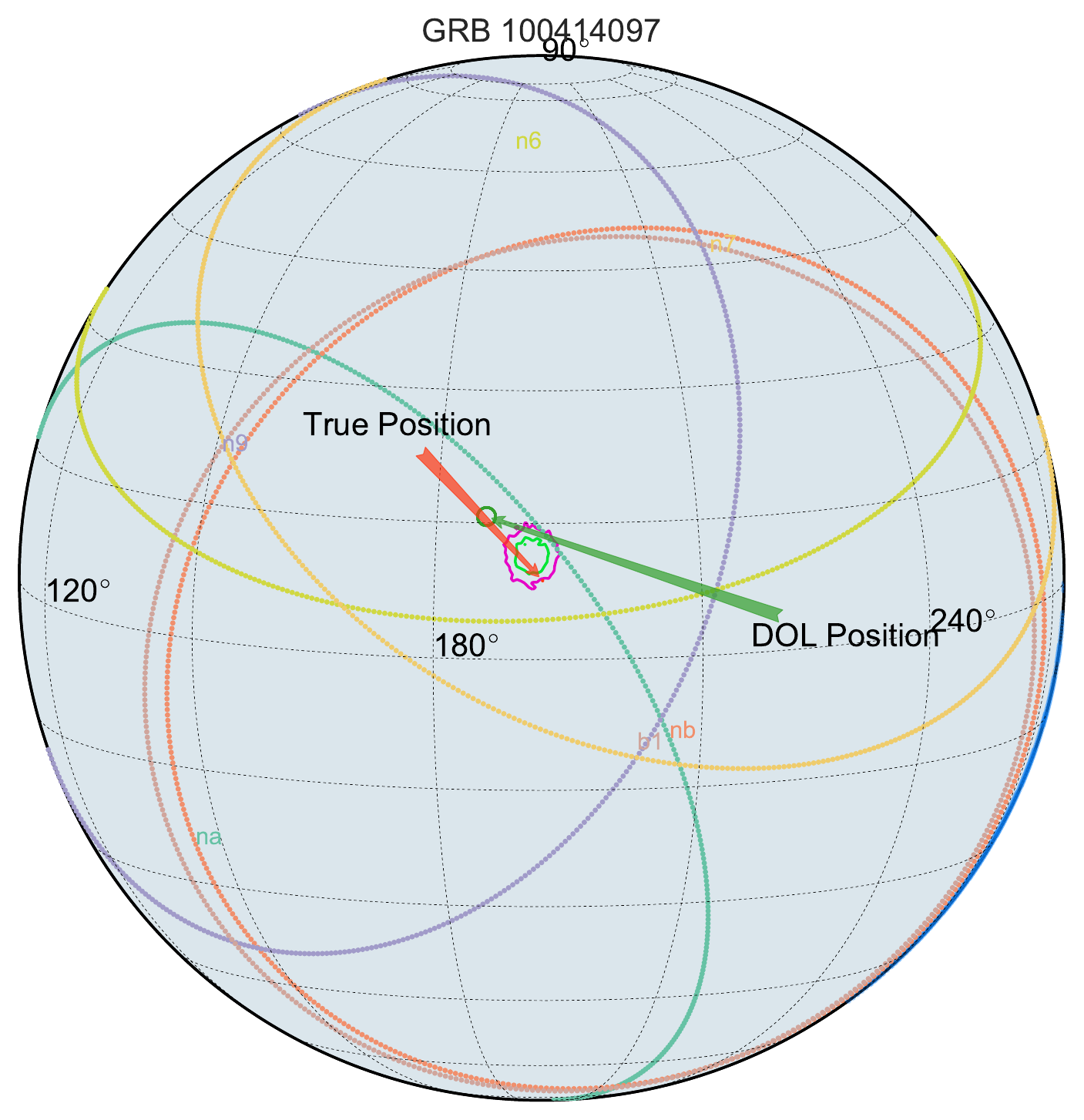}}
  \subfigure{\includegraphics[scale=.4]{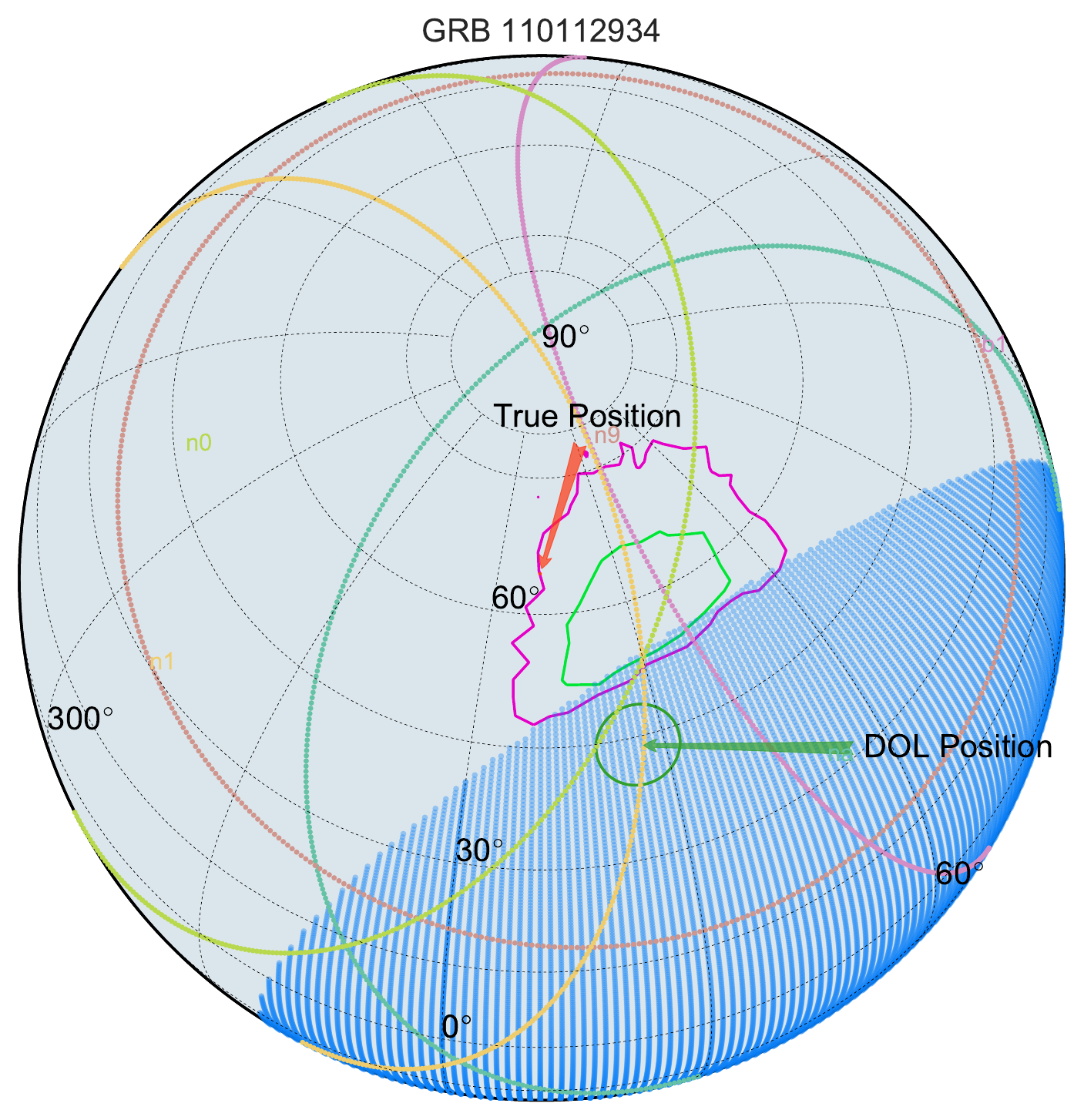}}\subfigure{\includegraphics[scale=.4]{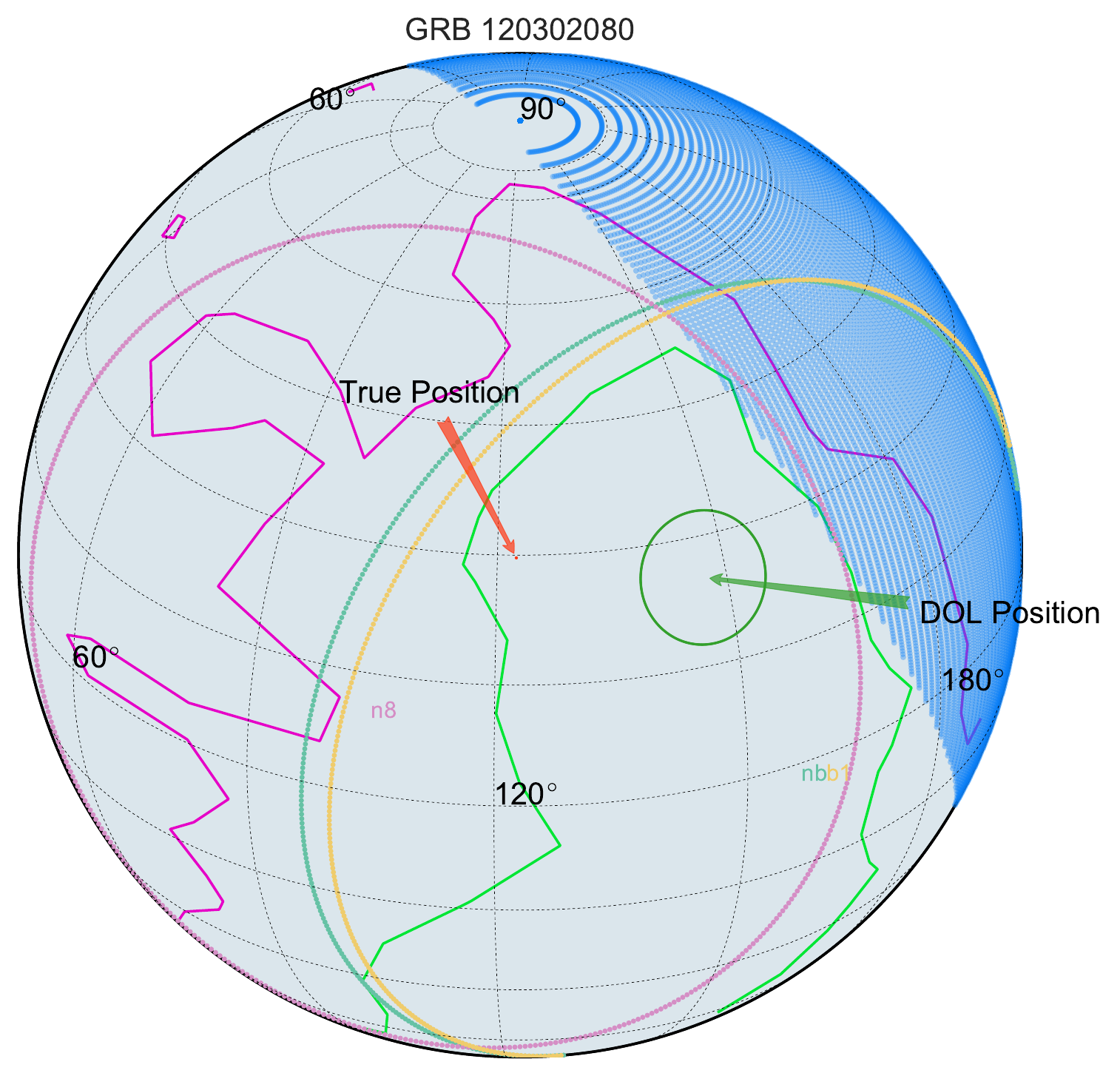}}\subfigure{\includegraphics[scale=.4]{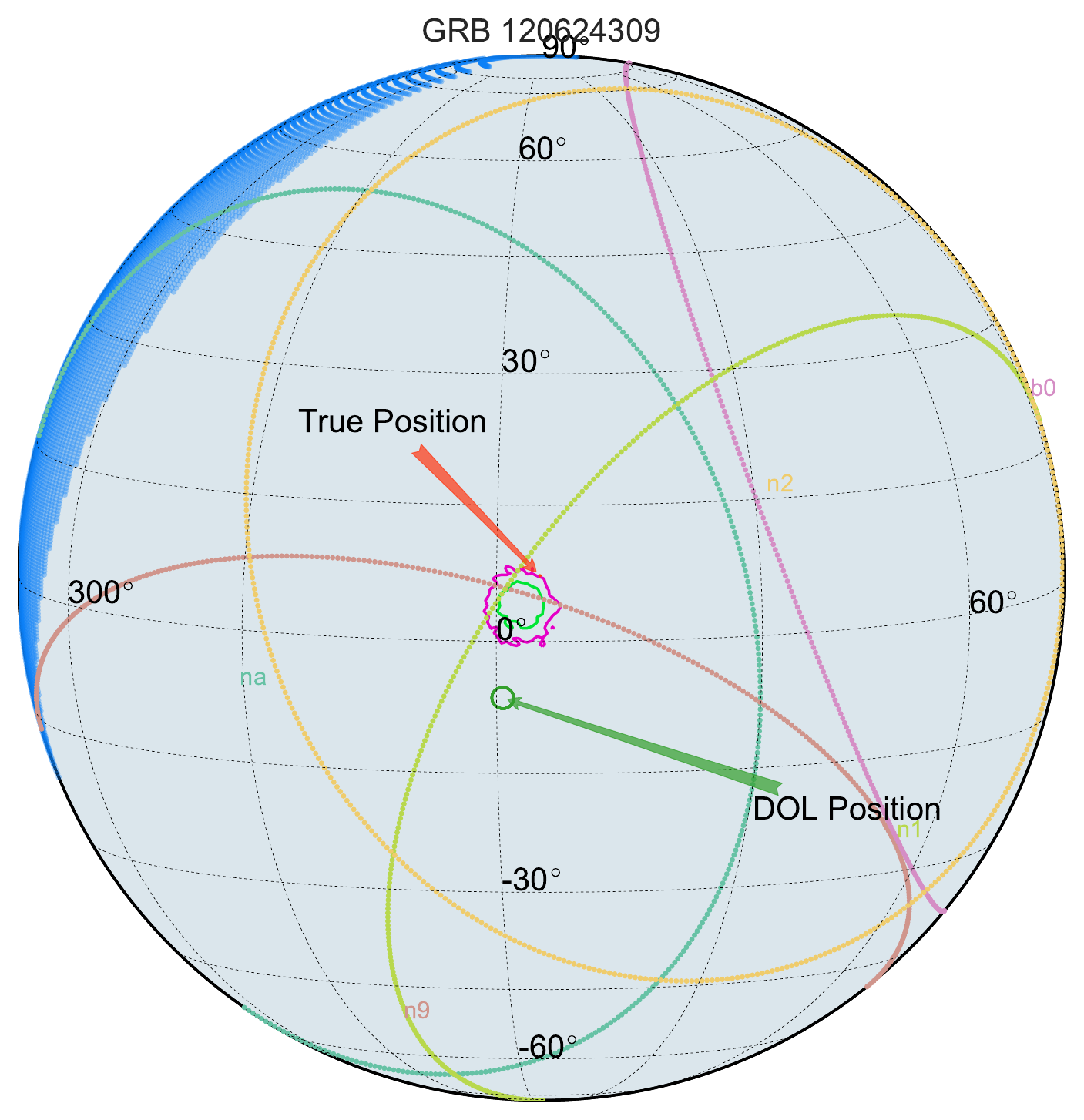}}
  \caption{The plots of the remaining nine GRBs with large DOL systematic error.}
  \label{fig:all_sys}
\end{figure*}

\clearpage

\begin{deluxetable}{ccccccccc}
\tabletypesize{\scriptsize}  
\tablecaption{Systematic Location GRB Sample \label{tab:sys_grbs}}
\tablewidth{0pt} 
\tablehead{\colhead{GRB} & \colhead{DOL RA} & \colhead{DOL Dec} & \colhead{DOL Err} & \colhead{True RA} & \colhead{True Dec} & \colhead{Separation Angle}}
\startdata
080725435 & 123.1 & -23.1 & 2.2 & 121.7 & -14.0 & 9.19 \\
081101491 & 80.8 & 11.6 & 10.1 & 95.8 & -0.1 & 18.94  \\
090927422 & 67.6 & -67.6 & 12.1 & 344.0 & -71.0 & 27.37  \\
100116897 & 308.4 & 22.7 & 1.2 & 305.0 & 14.5 & 8.80 \\
100206563 & 63.9 & 13.9 & 4.5 & 47.2 & 13.2 & 16.24  \\
100414097 & 185.7 & 15.7 & 1.0 & 192.1 & 8.7 & 9.38  \\
110106893 & 155.3 & 38.6 & 9.3 & 134.2 & 47.0 & 17.53 \\
110112934 & 25.9 & 44.0 & 4.7 & 10.6 & 64.462 & 22.19  \\
120302080 & 147.0 & 25.2 & 7.7 & 122.5 & 29.7 & 22.15  \\
120624309 & 0.7 & -6.3 & 1.2 & 4.8 & 7.2376 & 14.14  \\
121128212 & 278.8 & 41.6 & 1.5 & 300.6 & 54.3 & 19.20 \\
\enddata

\end{deluxetable}

\begin{deluxetable}{ccccccc}
\tabletypesize{\scriptsize}  
\tablecaption{BALROG Systematic Location GRB Sample \label{tab:sys_bal}}
\tablewidth{0pt} 
\tablehead{\colhead{GRB} & \colhead{True RA} & \colhead{True Dec}  & \colhead{Balrog RA} & \colhead{BALROG Dec} & \colhead{BALROG Separation Angle} & \colhead{DOL Separation Angle}}
\startdata
080725435 & 121.7 & -14.0 & 114.95 : 125.30 & -26.62 : -5.93 & 5.49 & 9.2 \\
081101491 & 95.8 & -0.1 & 64.67 : 107.84 & -2.63 : 20.78 & 8.97 & 18.94 \\
090927422 & 344.0 & -71.0 & -82.49 : 110.39 & -87.40 : -55.35 & 22.9 & 27.38 \\
100116897 & 305.0 & 14.5 & 303.35 : 307.75 & 13.28 : 16.82 & 0.6 & 8.81 \\
100206563 & 47.2 & 13.2 & 34.00 : 51.68 & 7.08 : 17.12 & 1.74 & 16.25 \\
100414097 & 192.1 & 8.7 & 188.61 : 193.40 & 8.77 : 13.95 & 3.14 & 9.38 \\
110112934 & 10.6 & 64.462 & 10.82 : 62.79 & 51.54 : 72.23 & 13.74 & 22.19 \\
120302080 & 122.5 & 29.7 & 4.45 : 218.29 & -76.84 : 48.43 & 30.71 & 22.16 \\
120624309 & 4.8 & 7.2376 & -0.43 : 5.61 & 0.63 : 7.26 & 3.81 & 14.14 \\
121128212 & 300.6 & 54.3 & 299.38 : 304.32 & 51.57 : 54.59 & 1.38 & 19.21 \\
\enddata
\tablecomments{Separation angles for {\tt BALROG} are determined from the maximum likelihood point.}
\end{deluxetable}

\clearpage

\begin{deluxetable}{ccccccc}
\tabletypesize{\scriptsize}  
\tablecaption{BALROG Good Location GRB Sample \label{tab:good_bal}}
\tablewidth{0pt} 
\tablehead{\colhead{GRB} & \colhead{True RA} & \colhead{True Dec} & \colhead{Balrog RA} & \colhead{Balrog Dec} & \colhead{Balrog Separation} & \colhead{GBM Separation}}
\startdata
080723557 & 176.8 & -60.2 & 162.99 : 180.15 & -62.54 : -55.97 & 2.7 & 0.98 \\
091003191 & 251.5 & 36.6 & 250.89 : 251.80 & 36.02 : 36.80 & 0.25 & 0.74 \\
100704149 & 133.6 & -24.2 & 127.04 : 143.25 & -25.85 : -19.71 & 3.11 & 0.7 \\
100906576 & 28.7 & 55.63 & 7.46 : 39.41 & 39.42 : 60.38 & 5.87 & 0.59 \\
111103441 & 327.1 & -10.5 & 326.19 : 333.22 & -11.57 : 8.30 & 9.81 & 0.51 \\
120119170 & 120.0 & -9.1 & 118.86 : 120.97 & -11.02 : -8.56 & 0.77 & 0.99 \\
120512112 & 325.6 & 13.6 & 322.10 : 326.62 & 11.91 : 18.41 & 1.85 & 1.4 \\
120709883 & 318.4 & -50.0 & 305.01 : 335.24 & -60.05 : -48.08 & 4.33 & 0.96 \\
121125356 & 228.5 & 55.3 & 210.40 : 236.90 & 44.46 : 58.36 & 4.49 & 0.46 \\
130518580 & 355.7 & 47.5 & 353.63 : 356.70 & 46.10 : 47.68 & 0.72 & 0.64 \\
\enddata
\tablecomments{Separation angles for {\tt BALROG} are determined from the maximum likelihood point.}
\end{deluxetable}

\clearpage

\begin{deluxetable}{ccccccccc}
\tabletypesize{\scriptsize}  
\tablecaption{GRB 080916C Time-Integrated {\tt BALROG} Parameters (95\% HDP) \label{tab:080916C-int}}
\tablewidth{0pt} 
\tablehead{\colhead{Model} & \colhead{RA} & \colhead{Dec} & \colhead{$\Ep$} & \colhead{$\alpha$} & \colhead{$\beta$} & \colhead{$kT$} & \colhead{break scale} & \colhead{logz}}
\startdata
Band$^{1}$ & 132.8 : 136.5 & -51.3 : -48.7 & 276.6 : 370.2 & -0.91 : -0.82 & -2.24 : -1.94 & \nodata & \nodata & -2831.4 \\
Band & 120.9 : 144.8 & -59.5 : -36.0 & 296.4 : 416.7 & -0.95 : -0.86 & -2.28 : -1.94 & \nodata & \nodata & -1881.4 \\
Band+BB & 119.7 : 145.1 & -60.2 : -36.0 & 238.6 : 422.9 & -0.96 : -0.74 & -2.26 : -1.92 & 1.0 : 325.0 & \nodata & -1881.1 \\
SBPL & 114.8 : 144.3 & -62.2 : -37.0 & 136.2 : 194.3 & -1.07 : -1.00 & -2.07 : -1.82 & \nodata & 1.78E-03 : 3.51E-01 & -1878.8
\enddata
\tablenotetext{1}{Fit including NaI 6 and 7.}
\end{deluxetable}

\begin{deluxetable}{ccccccc}
\tabletypesize{\scriptsize}  
\tablecaption{GRB 080916C Time-Integrated Fixed DRM Parameters (95\% HDP) \label{tab:080916C-fix}}
\tablewidth{0pt} 
\tablehead{\colhead{Model}  & \colhead{$\Ep$} & \colhead{$\alpha$} & \colhead{$\beta$} & \colhead{$kT$} & \colhead{break scale} & \colhead{logz}}
\startdata
Band & 413.2 : 548.4 & -1.02 : -0.95 & -2.29 : -2.00 & \nodata & \nodata & -237.4 \\
Band+BB &  955.9 : 1545.8 & -1.28 : -1.16 & -2.89: -2.06 & 41.2 : 48.64 & \nodata & -231.2 \\
SBPL &  205.1: 249.3 & -1.09 : -1.11 & -2.13 : -1.88 & \nodata & 5.21E-03 : 3.98E-01 & -234.4
\enddata
\end{deluxetable}

\clearpage

\begin{deluxetable}{cccccccccc}
\tabletypesize{\scriptsize}  
\tablecaption{GRB 080916C Time-resolved parameters (95\% HDP) \label{tab:080916C}}
\tablewidth{0pt}  
\tablehead{\colhead{Time from T$_0$} & \colhead{Model} & \colhead{$\Ep$ (keV)} &  \colhead{$\alpha$} & \colhead{$\beta$} &   \colhead{$kT$ (keV)} &   \colhead{$\delta$} &  \colhead{RA} &  \colhead{Dec} &  \colhead{$\log Z$} }
\startdata 
-0.1 : 0.7 & Band & 154.6 : 618.9 & -0.97 : -0.41 & -4.86 : -1.89 & \nodata & \nodata & 82.5 : 157.9 & -76.1 : -9.9 & -346.0 \\
 & CPL+BB & 169.5 : 643.2 & -0.97 : -0.43 & \nodata & 1.00 : 314.79 & \nodata & 89.1 : 158.6 & -76.2 : -12.5 & -346.4 \\
 \vspace{.4cm}
 & CPL+BB+PL & 22.2 : 97694.0 & -0.7 (fix) & \nodata & 44.94 : 65.33 & -1.5 (fix) & 78.6 : 160.0 & -75.1 : -7.1 & -348.8 \\

0.7 : 1.2 & Band & 169.5 : 425.5 & -0.76 : -0.32 & -5.00 : -2.18 & \nodata & \nodata & 81.4 : 154.9 & -71.3 : -18.8 & -326.2 \\
 & CPL+BB & 183.9 : 419.7 & -0.76 : -0.33 & \nodata & 1.00 : 217.38 & \nodata & 81.6 : 155.6 & -71.4 : -17.5 & -326.5 \\
 \vspace{.4cm}
 & CPL+BB+PL & 16.4 : 82608.5 & -0.7 (fix) & \nodata & 49.67 : 66.65 & -1.5 (fix) & 75.7 : 158.9 & -71.3 : -6.2 & -332.6 \\

1.2 : 1.7 & Band & 625.8 : 1710.3 & -0.97 : -0.71 & -5.00 : -2.41 & \nodata & \nodata & 117.7 : 158.6 & -64.7 : -14.3 & -348.4 \\
 & CPL+BB & 169.9 : 2347.1 & -0.99 : -0.40 & \nodata & 6.77 : 797.42 & \nodata & 110.0 : 158.2 & -68.1 : -16.1 & -347.6 \\
 \vspace{.4cm}
 & CPL+BB+PL & 17.4 : 98087.7 & -0.7 (fix) & \nodata & 47.23 : 78.55 & -1.5 (fix) & 156.7 : 164.3 & -5.7 : 9.2 & -364.7 \\

1.7 : 2.3 & Band & 93.2 : 255.7 & -0.64 : -0.02 & -4.20 : -1.77 & \nodata & \nodata & 92.1 : 155.9 & -67.7 : -16.4 & -399.3 \\
 & CPL+BB & 119.0 : 378.0 & -0.71 : -0.19 & \nodata & 1.01 : 569.82 & \nodata & 96.1 : 154.7 & -67.1 : -17.8 & -400.1 \\
 \vspace{.4cm}
 & CPL+BB+PL & 15.0 : 88705.0 & -0.7 (fix) & \nodata & 40.66 : 52.89 & -1.5 (fix) & 86.8 : 160.4 & -66.6 : -0.8 & -407.0 \\

2.3 : 2.8 & Band & 270.2 : 634.1 & -0.89 : -0.58 & -5.00 : -2.45 & \nodata & \nodata & 100.2 : 154.5 & -70.4 : -26.7 & -336.0 \\
 & CPL+BB & 126.3 : 1513.9 & -1.07 : -0.40 & \nodata & 1.25 : 373.46 & \nodata & 105.1 : 154.6 & -69.5 : -26.5 & -335.6 \\
 \vspace{.4cm}
 & CPL+BB+PL & 10.1 : 53433.7 & -0.7 (fix) & \nodata & 47.46 : 60.70 & -1.5 (fix) & 74.7 : 153.8 & -72.9 : -20.8 & -346.8 \\

2.8 : 3.4 & Band & 238.9 : 549.0 & -0.86 : -0.54 & -5.00 : -2.12 & \nodata & \nodata & 110.3 : 154.9 & -66.2 : -21.2 & -396.5 \\
 & CPL+BB & 267.9 : 566.2 & -0.86 : -0.56 & \nodata & 1.00 : 263.49 & \nodata & 110.3 : 154.3 & -65.1 : -20.4 & -396.8 \\
 \vspace{.4cm}
 & CPL+BB+PL & 11.8 : 67813.5 & -0.7 (fix) & \nodata & 52.77 : 68.14 & -1.5 (fix) & 91.7 : 158.5 & -67.8 : -9.5 & -401.8 \\

3.4 : 3.8 & Band & 78.8 : 419.6 & -0.87 : 0.02 & -4.68 : -1.71 & \nodata & \nodata & 103.5 : 163.0 & -66.2 : 2.2 & -256.2 \\
 & CPL+BB & 86.2 : 767.5 & -0.96 : -0.25 & \nodata & 1.00 : 353.30 & \nodata & 113.5 : 163.0 & -63.0 : 0.4 & -256.6 \\
 \vspace{.4cm}
& CPL+BB+PL & 12.3 : 62089.7 & -0.7 (fix) & \nodata & 36.52 : 55.27 & -1.5 (fix) & 151.8 : 163.0 & -11.1 : 8.8 & -262.5 \\

3.8 : 4.3 & Band & 248.5 : 871.2 & -0.94 : -0.57 & -4.60 : -1.81 & \nodata & \nodata & 88.0 : 154.3 & -68.6 : -20.0 & -338.7 \\
 & CPL+BB & 227.9 : 2254.0 & -1.02 : -0.57 & \nodata & 1.22 : 808.97 & \nodata & 93.1 : 153.9 & -68.5 : -21.7 & -339.1 \\
 \vspace{.4cm}
 & CPL+BB+PL & 10.5 : 57624.8 & -0.7 (fix) & \nodata & 47.41 : 75.23 & -1.5 (fix) & 78.2 : 161.7 & -68.0 : 3.6 & -348.1
\enddata
\vspace{-0.6cm}

\end{deluxetable}

\bsp
\label{lastpage}
\end{document}